\documentclass[aps,twocolumn,prd,showpacs,showkeys,preprintnumbers,superscriptaddress,nobibnotes,floatfix,longbibliography,notitlepage,nofootinbib]{revtex4-2}

\pdfoutput=1
\usepackage{amsmath}
\usepackage{amsfonts}
\usepackage{amssymb}
\usepackage{mathrsfs}
\usepackage{graphicx}
\usepackage{color}
\usepackage{bbding}
\usepackage{tabularx}
\usepackage[usenames,dvipsnames,table]{xcolor}
\usepackage[normalem]{ulem}
\usepackage{makecell}

\usepackage{hyperref}
\usepackage{multirow}
\usepackage{makecell}
\usepackage{upgreek}
\usepackage[capitalise]{cleveref}

\newcommand{\ie}{{\it i.e.}}

\newcommand{\eg}{{\it e.g.}}

\newcommand{\eq}{Eq.}

\newcommand{\Fig}{Fig.}

\newcommand{\dcp}{\delta_{\rm CP}}

\graphicspath{{figures/}{figures_appendix/}}

\begin{document}

\title{The Future of High-Energy Astrophysical Neutrino Flavor Measurements}

\author{Ningqiang Song}
\email{ningqiang.song@queensu.ca}
\affiliation{Department of Physics, Engineering Physics and Astronomy, Queen's University, Kingston ON K7L 3N6, Canada}
\affiliation{Arthur B. McDonald Canadian Astroparticle Physics Research Institute,  Kingston ON K7L 3N6, Canada}
\affiliation{Perimeter Institute for Theoretical Physics, Waterloo ON N2L 2Y5, Canada}

\author{Shirley Weishi Li}
\email{shirleyl@fnal.gov}
\affiliation{SLAC National Accelerator Laboratory, 2575 Sand Hill Road, Menlo Park, CA, 94025, USA}
\affiliation{Theoretical Physics Department, Fermilab, P.O. Box 500, Batavia, IL 60510, USA}

\author{Carlos A. Arg{\"u}elles}
\email{carguelles@fas.harvard.edu}
\affiliation{Department of Physics \& Laboratory for Particle Physics and Cosmology, Harvard University, Cambridge, MA 02138, USA}

\author{Mauricio Bustamante}
\email{mbustamante@nbi.ku.dk}
\affiliation{Niels Bohr International Academy \& DARK, Niels Bohr Institute,\\University of Copenhagen, DK-2100 Copenhagen, Denmark}

\author{Aaron C. Vincent}
\email{aaron.vincent@queensu.ca}
\affiliation{Department of Physics, Engineering Physics and Astronomy, Queen's University, Kingston ON K7L 3N6, Canada}
\affiliation{Arthur B. McDonald Canadian Astroparticle Physics Research Institute,  Kingston ON K7L 3N6, Canada}
\affiliation{Perimeter Institute for Theoretical Physics, Waterloo ON N2L 2Y5, Canada}

\date{December 23, 2020}

\begin{abstract}
We critically examine the ability of future neutrino telescopes, including Baikal-GVD, KM3NeT, P-ONE, TAMBO, and IceCube-Gen2, to determine the flavor composition of high-energy astrophysical neutrinos, \ie, the relative number of $\nu_e$, $\nu_\mu$, and $\nu_\tau$, in light of improving measurements of the neutrino mixing parameters.
Starting in 2020, we show how measurements by JUNO, DUNE, and Hyper-Kamiokande will affect our ability to determine the regions of flavor composition at Earth that are allowed by neutrino oscillations under different assumptions of the flavor composition that is emitted by the astrophysical sources.
From 2020 to 2040, the error on inferring the flavor composition at the source will improve from $> 40\%$ to less than $6\%$.
By 2040, under the assumption that pion decay is the principal production mechanism of high-energy astrophysical neutrinos, a sub-dominant mechanism could be constrained to contribute less than 20\% of the flux at 99.7\% credibility.
These conclusions are robust in the nonstandard scenario where neutrino mixing is non-unitary, a scenario that is the target of next-generation experiments, in particular the IceCube-Upgrade.
Finally, to illustrate the improvement in using flavor composition to test beyond-the-Standard-Model physics, we examine the possibility of neutrino decay and find that, by 2040, combined neutrino telescope measurements will be able to limit the decay rate of the heavier neutrinos to below $1.8\times 10^{-5} (m/\mathrm{eV})$~s$^{-1}$, at 95\% credibility. 
\end{abstract}

\maketitle

%%%%%%%%%%%%%%%%%%%%%%%%%%%%%%%%%%%%%%%%%%%%%%%%%%%%%%%%
%%%%%%%%%%%%%%%%%%%%%%%%%%%%%%%%%%%%%%%%%%%%%%%%%%%%%%%%

\section{Introduction}
\label{sec:intro}

High-energy astrophysical neutrinos in the TeV--PeV energy range, discovered by the IceCube Neutrino Observatory~\cite{Aartsen:2013bka,Aartsen:2013jdh,Aartsen:2014gkd,Aartsen:2015rwa,Aartsen:2016xlq,Ahlers:2018fkn,Abbasi:2020jmh}, offer unprecedented insight into astrophysics~\cite{Anchordoqui:2013dnh,Ahlers:2018fkn,Halzen:2019qkf,Ackermann:2019ows,Halzen:2019lxc,Palladino:2020jol} and fundamental physics~\cite{Gaisser:1994yf,Ahlers:2018mkf,Ackermann:2019cxh,Arguelles:2019rbn}.
On the astrophysical side, they may reveal the identity of the most energetic non-thermal sources in the Universe, located at cosmological-scale distances away from us. 
These neutrinos attain energies well beyond the reach of terrestrial colliders, granting access to a variety of Standard Model and beyond-the-Standard-Model (BSM) physics scenarios.
Because of their small interaction cross sections, neutrinos are unlikely to interact en route to Earth, so the information they carry about distant sources and high-energy processes reaches us with little to no distortion.
Detecting these neutrinos and extracting that information is challenging for the same reason, requiring cubic kilometer or larger detectors to overcome their low detection rate~\cite{Markov:1960vja}.

The information is encoded in the energies, arrival directions, arrival times, and flavor composition of high-energy neutrinos, \ie, the proportion of $\nu_e$, $\nu_\mu$, and $\nu_\tau$ in the flux~\cite{Anchordoqui:2013dnh,Ahlers:2018fkn,Ackermann:2019cxh,Ackermann:2019ows,Palladino:2020jol}.
The flavor composition has long been regarded as a particularly versatile probe of astrophysics~\cite{Rachen:1998fd,Athar:2000yw,Crocker:2001zs,Barenboim:2003jm,Beacom:2003nh,Beacom:2004jb,Kashti:2005qa,Mena:2006eq,Kachelriess:2006fi,Lipari:2007su,Esmaili:2009dz,Choubey:2009jq,Hummer:2010ai,Palladino:2015zua,Bustamante:2015waa,Biehl:2016psj,Bustamante:2019sdb} and fundamental physics~\cite{Beacom:2002vi,Barenboim:2003jm,Beacom:2003nh,Beacom:2003eu,Beacom:2003zg,Serpico:2005bs,Mena:2006eq,Lipari:2007su,Pakvasa:2007dc,Esmaili:2009dz,Choubey:2009jq,Esmaili:2009fk,Bhattacharya:2009tx,Bhattacharya:2010xj,Bustamante:2010nq,Mehta:2011qb,Baerwald:2012kc,Fu:2012zr,Pakvasa:2012db,Chatterjee:2013tza,Xu:2014via,Aeikens:2014yga,Arguelles:2015dca,Bustamante:2015waa,Pagliaroli:2015rca,deSalas:2016svi,Gonzalez-Garcia:2016gpq,Bustamante:2016ciw,Rasmussen:2017ert,Dey:2017ede,Bustamante:2018mzu,Farzan:2018pnk,Ahlers:2018yom,Brdar:2018tce,Palladino:2019pid,Ahlers:2020miq,Karmakar:2020yzn,Fiorillo:2020gsb}.

The sources of the observed flux of high-energy astrophysical neutrinos---still unidentified today, save for two promising instances~\cite{IceCube:2018cha,Stein:2020xhk}---are presumably hadronic accelerators where high-energy protons and nuclei interact with surrounding matter and radiation~\cite{Margolis:1977wt,Stecker:1978ah,Waxman:1998yy,Mucke:1999yb,Kelner:2006tc,Hummer:2010vx} to make high-energy neutrinos.
Different neutrino production mechanisms yield different flavor compositions at the source, and during their journey to Earth over cosmological distances, neutrinos oscillate, \ie, they undergo flavor conversions~\cite{Pontecorvo:1967fh,Fukuda:1998mi,Ahmad:2002jz}.
The standard theory of neutrino oscillation allows us to map a given flavor composition at the source to an expected flavor composition at Earth.
Here, large-scale neutrino telescopes detect them; the flavor composition of the neutrino flux results from comparing the number of events with different morphologies, which roughly reflects the number of neutrinos of each flavor~\cite{Roberts:1992re,Learned:1994wg,Mena:2014sja,Palomares-ruiz:2015mka,Palladino:2015zua,Aartsen:2015ivb,Aartsen:2015knd,Vincent:2016nut,Li:2016kra,DAmico:2017dwq,Aartsen:2018vez,Abbasi:2020zmr}.
Additionally, if there is more than one mechanism of neutrino production, each producing neutrinos with a different flavor composition, constraining the average flavor composition amounts to asking how large the fractional contribution of each mechanism can be in order to be detected~\cite{Aartsen:2019mbc}.

At present, however, our ability to perform such a precise flavor reconstruction and recover the flavor composition at the source is hampered by two important yet surmountable limitations.
First, the prediction of how a given flavor composition at the source maps to a flavor composition at Earth relies on our knowledge of the values of the neutrino mixing parameters that drive the oscillations~\cite{Pakvasa:2007dc}.
Because these are not precisely known~\cite{Capozzi:2017ipn,deSalas:2020pgw,Esteban:2020cvm}, such predictions are uncertain.
Second, measuring the flavor composition in neutrino telescopes is challenging, and suffers from large statistical and systematic uncertainties~\cite{Mena:2014sja,Palomares-ruiz:2015mka,Aartsen:2015knd,Abbasi:2020zmr}.
This prevents us from distinguishing between predictions that are similar but based on different assumptions of neutrino production.

In this work, we show that these limitations will be overcome in the next two decades, thanks to new terrestrial and astrophysical neutrino experiments that are planned or in construction~\cite{Arguelles:2019xgp}.
Oscillation experiments that use terrestrial neutrinos---JUNO~\cite{An:2015jdp}, DUNE~\cite{Abi:2020wmh}, Hyper-Kamiokande (HK)~\cite{Abe:2018uyc}, and the IceCube-Upgrade~\cite{Ishihara:2019aao}---will reduce the uncertainties in the mixing parameters and put the standard oscillation framework to test.   Large-scale neutrino telescopes---Baikal-GVD~\cite{Avrorin:2019vfc}, IceCube-Gen2~\cite{Aartsen:2020fgd}, KM3NeT~\cite{Adrian-Martinez:2016fdl},  P-ONE~\cite{Agostini:2020aar}, and TAMBO~\cite{Romero-Wolf:2020pzh}---will detect more high-energy astrophysical neutrinos and improve the measurement of their flavor composition.

To show this, we make detailed, realistic projections of how the uncertainty in the predicted flavor composition at Earth of the isotropic flux of high-energy neutrinos and its measurement will evolve over the next two decades.
Our main finding is that, by 2040, we will be able to precisely infer the flavor composition at the sources, including possibly identifying the contribution of multiple neutrino-production mechanisms, even if oscillations are non-unitary~\cite{Xing:2008fg,Xu:2014via,Parke:2015goa,Brdar:2016thq,Ahlers:2018yom,Arguelles:2019tum,Ellis:2020hus,Ahlers:2020miq}.
Further, we illustrate the upcoming power of flavor measurements to probe BSM neutrino physics using neutrino decay~\cite{Beacom:2002vi,Meloni:2006gv,Maltoni:2008jr,Baerwald:2012kc,Pakvasa:2012db,Pagliaroli:2015rca,Huang:2015flc,Bustamante:2016ciw,Denton:2018aml,Bustamante:2020niz,Abdullahi:2020rge}.

%%%%%%%%%%%%%%%%%%%%%%%%%%%%%%
\begin{figure}[t]
\centering
\includegraphics[width=\columnwidth]{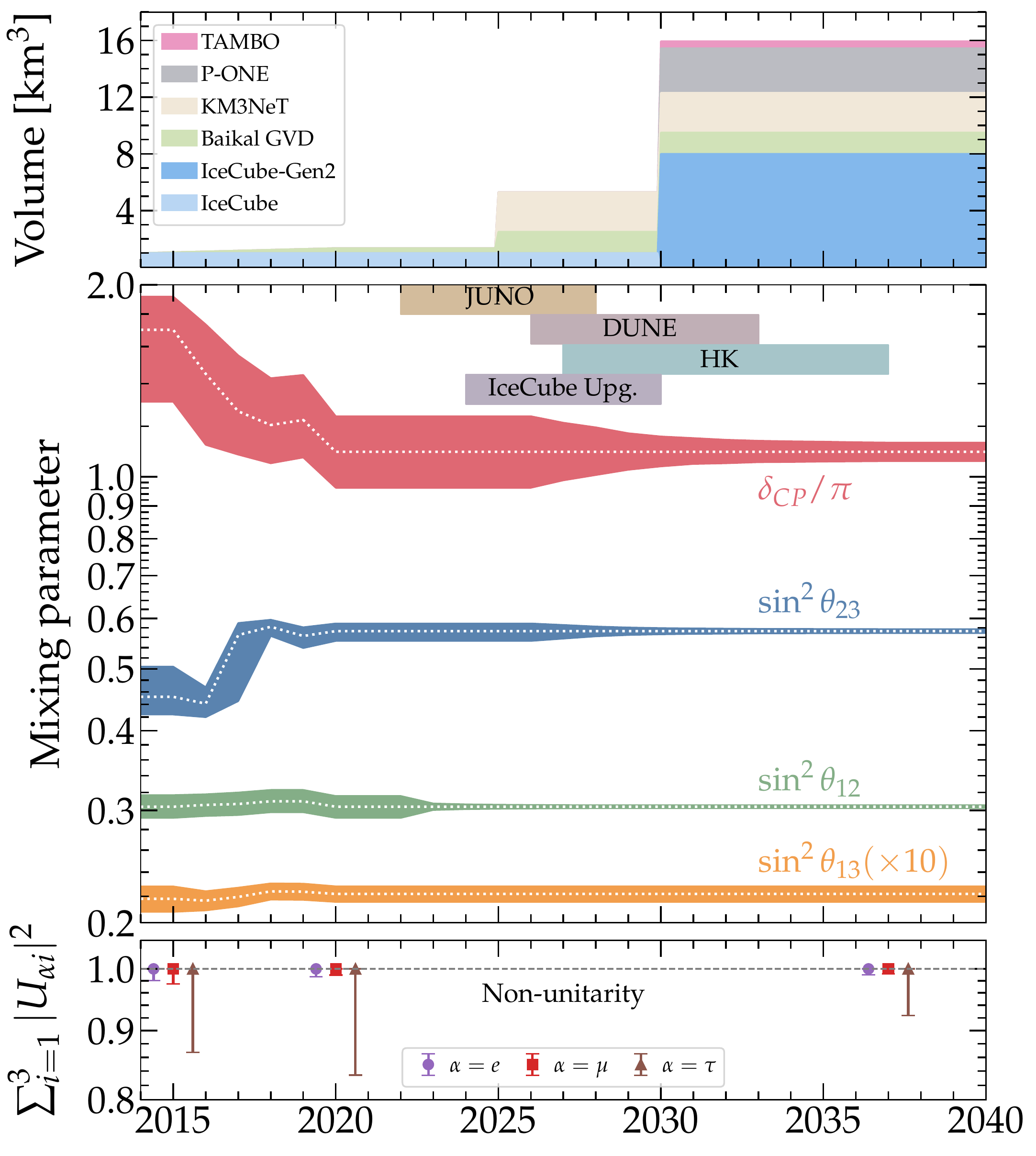}
\caption{\textbf{\textit{The future of neutrino telescopes, oscillation experiments, and flavor mixing measurements.}}
{\it Top:} Effective volume of future neutrino telescopes able to probe the flavor composition of astrophysical neutrinos.
{\it Center:} Time evolution of the oscillation parameters.
For each parameter, the dotted white line shows the best-fit value and the shaded region around it, the $1\sigma$ uncertainty.
Values up to and including the year 2020 are obtained from the NuFit global fit to oscillation data~\cite{Gonzalez-Garcia:2014bfa, Esteban:2016qun, Esteban:2018azc, Esteban:2020cvm}; NuFit 5.0~\cite{Esteban:2020cvm} is the latest fit.
Projections from 2020 to 2040 have best-fit values fixed at the current values from NuFit~5.0, but uncertainties reduced due to JUNO, DUNE, and Hyper-Kamiokande (HK) measurements, following our simulations.
The boxes at the top show the start and projected estimated running times for these experiments.
{\it Bottom:} Time evolution of the expected error on the unitarity of the neutrino flavor mixing matrix; values taken from Refs.~\cite{Parke:2015goa,Ellis:2020hus}.}
\label{fig:time_line}
\end{figure}
%%%%%%%%%%%%%%%%%%%%%%%%%%%%%%%%%

This article is organized as follows.
In Section~\ref{sec:theory} we revisit the basics of neutrino mixing, especially as it pertains to high-energy astrophysical neutrinos, and introduce the formalism of neutrino decay and non-unitary neutrino evolution.
In Section~\ref{sec:exp} we introduce the future neutrino experiments that we consider in our analysis and their measurement goals.
In Section~\ref{sec:stat} we present the statistical method that we use to produce the allowed regions of flavor composition at Earth.
In Section~\ref{sec:results} we present our results.
In Section~\ref{sec:conclusions}, we summarize and conclude.
In the appendices, we show additional analysis cases that we do not explore in the main text.

%%%%%%%%%%%%%%%%%%%%%%%%%%%%%%%%%%%%%%%%%%%%%%%%%%%%%%%%
%%%%%%%%%%%%%%%%%%%%%%%%%%%%%%%%%%%%%%%%%%%%%%%%%%%%%%%%

\section{Flavor composition of high-energy astrophysical neutrinos}
\label{sec:theory}

%%%%%%%%%%%%%%%%%%%%%%%%%%%%%%%%%%%%%%%%%%%%%%%%%%%%%%%%

\subsection{Flavor composition at the sources}
\label{subsection:ratios_source}

In astrophysical sites of hadronic acceleration, protons and heavier nuclei are accelerated to energies well beyond the PeV scale.  Likely candidate acceleration sites feature high particle densities, high baryon content, and matter that moves at relativistic bulk speeds, such as the jets of gamma-ray bursts and active galactic nuclei~\cite{Anchordoqui:2013dnh,Ahlers:2018fkn,Murase:2019tjj}.
There, high-energy protons interact with ambient matter and radiation~\cite{Margolis:1977wt,Stecker:1978ah,Mucke:1999yb,Kelner:2006tc,Hummer:2010vx}, generating secondary pions and kaons that decay into high-energy neutrinos.
The physical conditions at the sources determine what neutrino production channels are available and affect the maximum energy of the parent protons and the energy losses of the secondaries.
This, in turn, determines the relative number of neutrinos and anti-neutrinos produced---\ie, the flavor composition at the sources.

We parametrize the flavor composition at the source via the flavor ratios $(f_{e, {\rm S}}, f_{\mu, {\rm S}}, f_{\tau, {\rm S}})$, where $f_{\alpha, {\rm S}} \in [0,1]$ is the ratio of the flux of $\nu_\alpha$ and $\bar{\nu}_\alpha$, with $\alpha = e, \mu, {\rm or}~\tau$, to the total flux.
We do not separate neutrinos and anti-neutrinos because high-energy neutrino telescopes are unable to make this distinction on an event-by-event basis, with the exception of the Glashow resonance triggered by high-energy $\bar{\nu}_e$~\cite{Glashow:1960zz,Bhattacharya:2011qu,Bhattacharya:2012fh,Biehl:2016psj,Huang:2019hgs}. Neutrinos and anti-neutrinos may be distinguished statistically by measuring the inelasticity distribution~\cite{Gandhi:1995tf,Connolly:2011vc} of detected events; see Ref.~\cite{Aartsen:2018vez} for the first measurement of the flavor composition using this observable.
Henceforth, we use $\nu_\alpha$ to mean both neutrinos and anti-neutrinos of flavor $\alpha$.
Flavor ratios are normalized to one, \ie, $\sum_\alpha f_{\alpha, {\rm S}} = 1$, and if there are additional neutrino species, the sum also includes them; see Section~\ref{subsection:nonunit} for details.

%%%%%%%%%%%%%%%%%%%%%%%%%%%%%%%%%
\begin{figure*}[t!]
  \centering
  \includegraphics[trim=0 0.5cm 0 0, clip, width=0.49\textwidth]{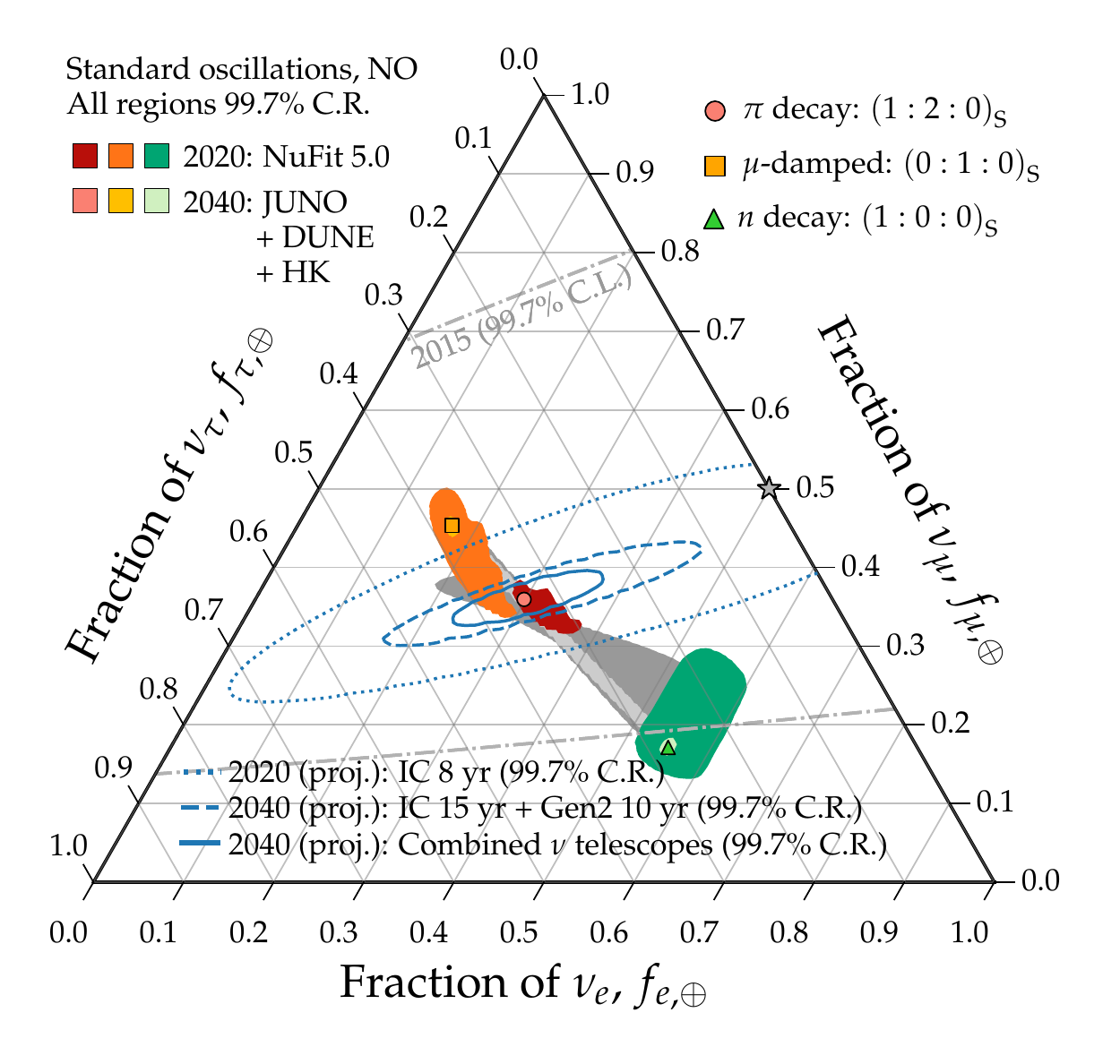}
  \includegraphics[trim=0 0.5cm 0 0, clip, width=0.49\textwidth]{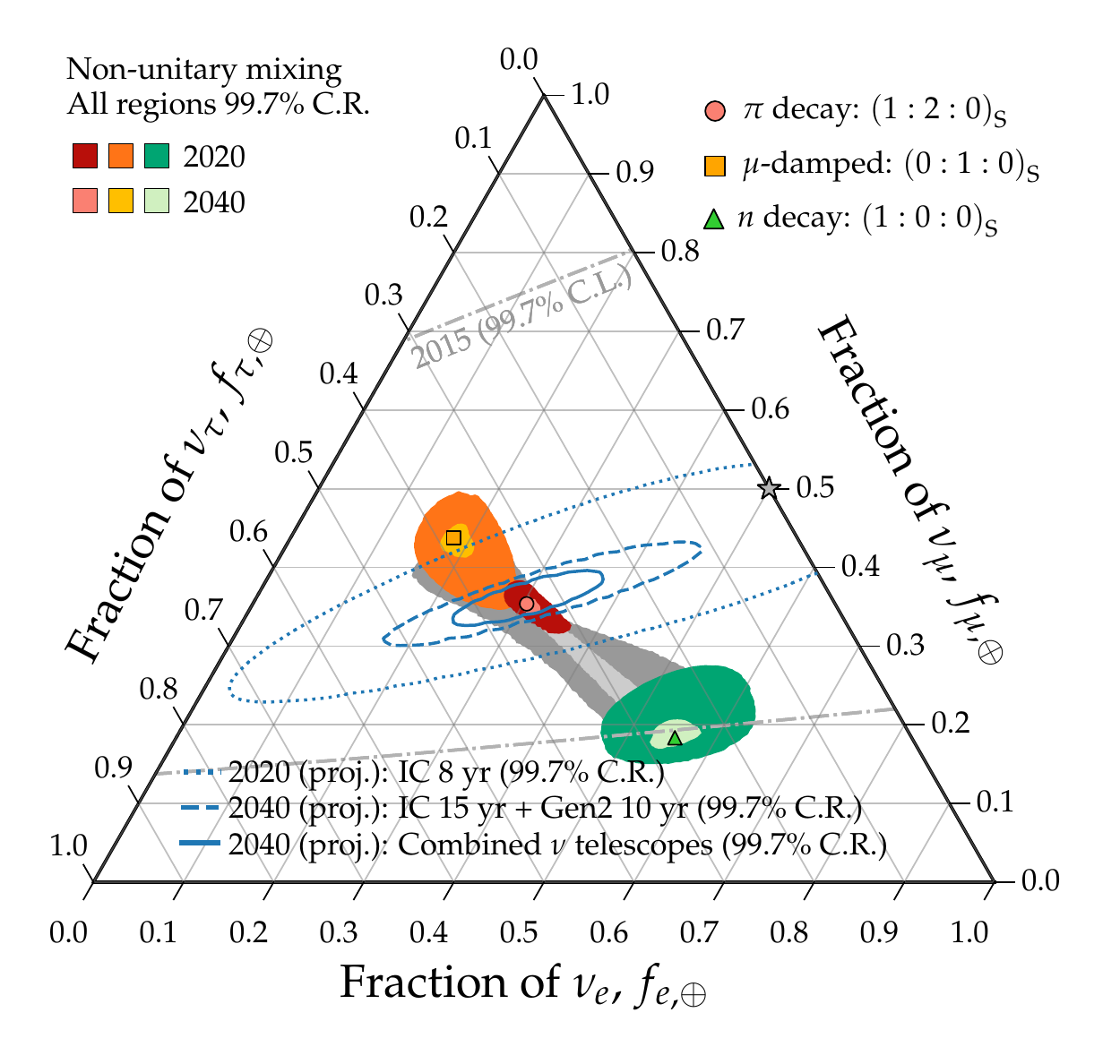}
  \caption{Comparison between the status of allowed regions of flavor composition at Earth in 2020 and 2040.
  Colored regions are computed separately for the three benchmark flavor compositions at the source $(f_e : f_\mu : f_\tau)_{\rm S}$ (pion decay, muon damping, neutron decay), and varying over all possible flavor compositions at the source (gray-shaded regions).
  Lines show the 99.7\% credibility regions (C.R.) of the astrophysical flux assuming a composition of $(0.3, 0.36, 0.34)_\oplus$ at Earth.
  The overlaid contours show the 2015 IceCube measurement of the flavor composition~\cite{Aartsen:2015knd} and projections for IceCube and IceCube-Gen2~\cite{Aartsen:2020fgd} and for the combination of all TeV--PeV neutrino telescopes available in 2040, derived here.
  {\it Left:} Assuming standard oscillations and unitarity in the 3$\times$3 flavor mixing matrix.   {\it Right:} Same as left panel, but without the assumption of unitarity.
  }
  \label{fig:triangle_sm_2020_vs_2040}
\end{figure*}
%%%%%%%%%%%%%%%%%%%%%%%%%%%%%%%%%

Presently, because the identity of the high-energy astrophysical neutrino sources is unknown, there is considerable uncertainty as to the dominant neutrino production mechanism and the physical conditions at production.
In addition, these may be different at different neutrino energies. However, because the flavor ratios reflect the neutrino production mechanism, we can use them---after accounting for oscillations en route to Earth as discussed in Section~\ref{subsection:std_osc}---to reveal the production mechanism and help identify the neutrino sources~\cite{Bustamante:2019sdb}.

In our analysis, we explore all possible flavor ratios at the sources, but showcase three physically motivated benchmark scenarios commonly discussed in the literature: full pion decay, muon damping, and neutron decay.  

In the full pion decay scenario, charged pions generate neutrinos via $\pi^+ \to \mu^+ + \nu_\mu$, followed by $\mu^+ \to \bar{\nu}_\mu + e^+ + \nu_e$, and their charge-conjugated processes. In this case, the flavor ratio is $\left( \frac{1}{3}, \frac{2}{3}, 0 \right)_{\rm S}$.
This is the canonical expectation for the flavor ratios at the sources.

In the muon-damped scenario, the intermediate muons cool via synchrotron radiation induced by strong magnetic fields harbored by the sources.
As a result, only the $\nu_\mu$ coming directly from pion decay have high energy.
In this case, the flavor composition is $\left( 0, 1, 0 \right)_{\rm S}$. The flavor composition may transition from the full pion decay scenario to the muon-damped scenario at an energy determined by the onset of synchrotron losses; see, \eg, Refs.~\cite{Kashti:2005qa,Lipari:2007su,Hummer:2010vx}.
Observing this transition would reveal the magnetic field strength of the sources and help identify them~\cite{Winter:2013cla, Bustamante:2020bxp}; this might be possible in IceCube-Gen2 if the transition occurs at PeV energies~\cite{Aartsen:2020fgd}.

In the neutron decay scenario, $\bar{\nu}_e$ exclusively are generated in the beta decay of neutrons or short-lived isotopes produced by spallation or photodisintegration of cosmic rays.
In this case, the flavor composition is $\left( 1, 0, 0 \right)_{\rm S}$.
This production scenario is unlikely, since neutrinos from beta decay are significantly less energetic than those from pion decay. Already, flavor-ratio measurements disfavor this production scenario at $\geq 2\sigma$~\cite{Aartsen:2015knd, Bustamante:2019sdb,Abbasi:2020jmh}; we keep it in our discussion because it remains a useful benchmark.

%%%%%%%%%%%%%%%%%%%%%%%%%%%%%%%%%%%%%%%%%%%%%%%%%%%%%%%%
\subsection{Standard neutrino oscillations}
\label{subsection:std_osc}

Because the neutrino flavor states, $|\nu_e\rangle$, $|\nu_\mu\rangle$, $|\nu_\tau\rangle$, and the energy eigenstates of the free-particle Hamiltonian, \ie, the {\it mass eigenstates} $|\nu_1\rangle$, $|\nu_2\rangle$, $|\nu_3\rangle$ are different, neutrinos change flavor, or {\it oscillate}, as they propagate from their sources to Earth.
Oscillations alter the neutrino flavor ratios that reach Earth. Below, we describe how this occurs within the standard oscillation scenario; for comprehensive reviews, see Refs.~\cite{Giunti:2007ry,Zyla:2020zbs}.
Later, in Sections~\ref{subsection:decay} and~\ref{subsection:nonunit}, we introduce alternative flavor-transition mechanisms.

In the standard oscillation scenario, the flavor and mass states are related via a unitary transformation, \ie,
\begin{equation}
  |\nu_\alpha\rangle  = \sum_{i=1}^3 U_{\alpha i}^* |\nu_i\rangle,
\end{equation}
where $\alpha = e, \mu, \tau$, and $U$ is the Pontecorvo-Maki-Nakagawa-Sakata (PMNS) lepton mixing matrix. 
We adopt the standard parametrization~\cite{Zyla:2020zbs} of $U$ as a  $3 \times 3$ complex ``rotation'' matrix, in terms of three mixing angles, $\theta_{12}$, $\theta_{23}$, and $\theta_{13}$, and one phase, $\dcp$.If neutrinos are Majorana fermions, $U$ contains two additional phases that do not affect oscillations.  
Neutrino oscillation experiments~\cite{Cleveland:1998nv,Hosaka:2005um,Cravens:2008aa,Bellini:2008mr,Abdurashitov:2009tn,Kaether:2010ag,Abe:2010hy,Bellini:2011rx,Aharmim:2011vm,Adamson:2013ue,Gando:2013nba,Adamson:2013whj,Bellini:2014uqa,Aartsen:2014yll,Honda:2015fha,An:2016srz,Vinyoles:2016djt,Abe:2017aap,Adey:2018zwh,ChoozNu2020,RENONu2020,T2KNu2020,NOvANu2020,SKNu2020} and global fits~\cite{Capozzi:2017ipn, deSalas:2020pgw, Esteban:2020cvm} to their data have determined the values of the mixing angles to the few-percent level and have started to corner the $CP$ phase. 
In this work, we use and build on the recent NuFit~5.0 global fit~\cite{Esteban:2020cvm,nufit5.0}, which uses data from 31 different analyses of solar, atmospheric, reactor, and accelerator neutrino experiments~\cite{Cleveland:1998nv,Hosaka:2005um,Cravens:2008aa,Bellini:2008mr,Abdurashitov:2009tn,Kaether:2010ag,Abe:2010hy,Bellini:2011rx,Aharmim:2011vm,Adamson:2013ue,Gando:2013nba,Adamson:2013whj,Bellini:2014uqa,Aartsen:2014yll,Honda:2015fha,An:2016srz,Vinyoles:2016djt,Abe:2017aap,Adey:2018zwh,ChoozNu2020,RENONu2020,T2KNu2020,NOvANu2020,SKNu2020}.

The characteristic neutrino oscillation length is $L_{\rm osc} = 4\pi E/\Delta m_{ij}^2$, where $E \gg m_i$ is the neutrino energy, and $\Delta m_{ij}^2 \equiv m_i^2 - m_j^2$ is the difference between squared masses of the mass eigenstates, with $(i,j = 1,2,3)$.
High-energy astrophysical neutrinos, with energies between 10~TeV and 10~PeV, have $L_{\rm osc} \ll 1$~pc.
Thus, compared to the cosmological-scale distances over which these neutrinos propagate and the energy resolutions of neutrino telescopes, the oscillations are rapid and cannot be resolved.
Instead, we are sensitive only to the average $\nu_\alpha \to \nu_\beta$ flavor-transition probability, \ie,
\begin{equation}
   \label{equ:prob_std}
   P_{\alpha\beta}^{\rm std} = \sum_{i=1}^{3} |U_{\alpha i}|^2 |U_{\beta i}|^2 \;.
\end{equation}
The average probability depends only on the mixing angles and the $CP$-violation phase.
We adopt this approximation in our standard oscillation analysis. Our choice is further motivated by the fact that the isotropic high-energy neutrino flux is the aggregated contribution of multiple unresolved sources, each located at a different distance, so that the individual oscillation patterns coming from each source are smeared by the spread of the distribution of distances, leaving the average flavor-transition probability as the only accessible quantity.

Because the complex phases in $U$ do not contribute to the average flavor-transition probability, Eq.~\eqref{equ:prob_std}, leptonic $CP$-violation does not affect the flavor composition at Earth~\cite{Giunti:2007ry}.
However, in the standard parameterization of $U$ that we use~\cite{Zyla:2020zbs,Denton:2020igp}, the value of $\dcp$ still impacts the probability, since
\begin{equation}
|U|^2 = \begin{pmatrix}
|U_{e 1}|^2  & |U_{e 2}|^2 & |U_{e 3}|^2  \\
\sqrt{ X_1 + Y \cos\dcp } & \sqrt{ X_2 - Y \cos\dcp }& |U_{\mu 3}|^2\\
\sqrt{ X_3 - Y \cos\dcp } & \sqrt{ X_4 + Y \cos\dcp } &  |U_{\tau 3}|^2
\end{pmatrix},
\end{equation}
where $X_1$, $X_2$, $X_3$, $X_4$, and $Y$ are computable functions of the mixing angles, but not of $\dcp$.
In other words, $\cos\dcp$ contributes to the content of $\nu_1$ and $\nu_2$ mass eigenstates in the $\nu_\mu$ and $\nu_\tau$ flavor states.
However, the effect of $\dcp$ on the flavor-transition probability is weak because it appears multiplied by $\sin^4 \theta_{13} \ll 1$.

Table~\ref{tab:NuFIT} shows the current best-fit values and uncertainties of the mixing parameters from NuFit~5.0. 
The values depend on the choice of the unknown neutrino mass ordering, which is labeled as \textit{normal}, if $\nu_1$ is the lightest, or \textit{inverted}, if $\nu_3$ is the lightest.
In the main text, our present-day results and projections are derived assuming the distributions of values of the mixing parameters under the normal mass ordering; see Section~\ref{sec:exp} for details.
Normal ordering has until recently been favored over inverted ordering at a significance of $\simeq$3$\sigma$~\cite{Capozzi:2017ipn,deSalas:2020pgw}, but such preference has weakened in light of the most recent data~\cite{Kelly:2020fkv,Esteban:2020cvm}.
Results of our analyses assuming an inverted ordering are very similar; we show them in Appendix~\ref{sec:appendix}. 

The ``solar'' mixing parameters $\theta_{12}$ and $\Delta m^2_{21}$ are measured in solar neutrino experiments~\cite{Aharmim:2011vm,Abe:2016nxk} and the reactor experiment KamLAND~\cite{Gando:2013nba}.
The angle $\theta_{13}$ is precisely measured in reactor experiments, \eg, Daya Bay~\cite{Adey:2018zwh}.
The ``atmospheric'' parameters $\theta_{23}$ and $\Delta m^2_{32}$ are measured in atmospheric and long-baseline accelerator experiments~\cite{Adamson:2013whj,Aartsen:2014yll,Abe:2017aap,Acero:2019ksn,Abe:2020vdv}.
The phase $\dcp$ is measured in long-baseline experiments~\cite{Acero:2019ksn,Abe:2019vii}.
Presently, the only significant correlation among the mixing angles and $\dcp$ is between $\theta_{23}$ and $\dcp$ (see Fig.~\ref{fig:correlation})~\cite{Esteban:2020cvm}, which we take into account below in our sampling of values of the mixing parameters.

%%%%%%%%%%%%%%%%%%%%%%%%%%%%%%%%%%%%%%%%%%%%%%%%%%%%%%%%

\subsection{Flavor composition at Earth:\\Standard oscillations}
\label{subsection:ratios_earth}

For a given flavor composition at the source $(f_{e,{\rm S}}, f_{\mu,{\rm S}}, f_{\tau,{\rm S}})$, the flavor composition of the neutrino flux that arrives at Earth, under standard oscillations, is 
\begin{equation}
    \label{equ:flavor_ratio_earth_std}
    f_{\beta,\oplus} = \sum\limits_{\alpha=e,\mu,\tau} P_{\alpha\beta}^{\rm std} f_{\alpha, {\rm S}} 
    \quad ({\rm std.~oscillations})
    \;.
\end{equation}
Using the best-fit values of the mixing parameters from Table~\ref{tab:NuFIT}, the expected standard flavor composition at Earth is approximately democratic for the full pion decay chain: $\left(0.3, 0.36, 0.34\right)_\oplus$. This becomes $(0.17,0.47,0.36)_\oplus$ for muon damping, and $(0.55,0.17,0.28)_\oplus$ for neutron decay; see \Fig~\ref{fig:triangle_sm_2020_vs_2040}.
Later, in Section \ref{sec:results}, we show that the present-day experimental uncertainties in the mixing parameters and in the measurement of the flavor ratios prevent us from distinguishing between these benchmark scenarios, but that future improvements will allow us to do so by 2040.

%%%%%%%%%%%%%%%%%%%%%%%%%%%%%%%%%%%%%%%%%%%%%%%%%%%%%%%%

\subsection{Flavor composition measurements in neutrino telescopes}
\label{subsection:ratios_icecube}

The flavor composition of a sample of events detected by a neutrino telescope is inferred from correlations in their energies, directions, and morphologies.
The morphology of an event, \ie, the spatial and temporal distribution of the collected light associated with it, correlates particularly strongly with the flavor of the neutrino that triggered it. 
In ice-based and water-based neutrino telescopes, the morphologies detected so far are showers (mainly from $\nu_e$ and $\nu_\tau$), tracks (mainly from $\nu_\mu$), and double bangs (from $\nu_\tau$).

\textbf{Showers}, also known as cascades, are generated by the charged-current (CC) deep-inelastic neutrino-nucleon scattering of a $\nu_e$ or a $\nu_\tau$ in the ice or water.
The scattering produces a particle shower in which charged particles emit Cherenkov radiation that is detected by photomultipliers embedded in the detector volume.
Neutral current (NC) interactions from all flavors also yield showers, though their contribution to the event rate is subdominant to that of CC interactions because the NC cross section is smaller and because, at a fixed shower energy, higher-energy neutrinos are required to make a NC shower than a CC shower.

\textbf{Tracks} are generated by the CC deep-inelastic scattering of a $\nu_\mu$.
This creates an energetic final-state muon that can travel several kilometers, leaving a visible track of Cherenkov light in its wake.
In addition, the momentum transferred to the nucleon produces a shower centered on the interaction vertex.
Tracks can also be produced by CC $\nu_\tau$ interactions where the tau promptly decays into a muon, which happens approximately 18\% of the time, and where the showers generated by the production and decay of the tau cannot be separated. 

\textbf{Double bangs}, or double cascades, are uniquely made in the CC interaction of $\nu_\tau$.  The neutrino-nucleon scattering triggers a first shower and produces a final-state tau that, if energetic enough, decays far from the first shower to trigger a second, identifiable shower~\cite{Learned:1994wg,Athar:2000rx}. 
The first double-bang events were only recently observed at IceCube~\cite{Abbasi:2020zmr}. 

There are other identifiable, but yet undetected, morphologies associated to $\nu_\tau$ CC interactions~\cite{Cowen:2007ny}; \eg, when the $\nu_\tau$ interacts outside the detector but the decay of the tau is visible.

Identifying flavor on an event-by-event basis is effectively unfeasible.  Showers generated by the CC interaction of $\nu_e$ and $\nu_\tau$ of the same energy look nearly identical---which leads to a degeneracy in measuring their flavor ratios---and so do the showers generated by the NC interaction of all flavors of neutrinos of the same energy.  Tracks may be made by final-state muons from $\nu_\mu$ CC interactions or by the decay into muons of final-state taus from $\nu_\tau$ CC interactions.
To address this limitation, future neutrino telescopes may be able to use timing information to distinguish $\nu_e$-induced electromagnetic showers, from hadronic showers originating mainly from $\nu_\tau$'s,  by using the difference in their late-time Cherenkov ``echoes'' from low-energy muons and neutrons~\cite{Li:2016kra}. 
This will require using photomultipliers with a low level of ``delayed pulses''~\cite{Steuer:2017tca,Kopketalk} that could mimic muon and neutron echoes. 
We do not include echoes in our analyses.

Thus, the flavor composition is reconstructed collectively for a sample of detected events, using statistical methods.
All flavor measurements use \textit{starting events}, where the neutrino interacts within the detector volume~\cite{Mena:2014sja,Palomares-ruiz:2015mka,Aartsen:2015ivb,Vincent:2016nut, Aartsen:2018vez,Abbasi:2020zmr} and all three morphologies are distinguishable. References~\cite{Aartsen:2015ivb,Aartsen:2018vez,Abbasi:2020zmr} reported IceCube measurements of the flavor composition based exclusively on starting events using 3, 5, and 7.5 years of data, respectively.
These analyses are statistically limited because of the low event rate of  $\sim$8 neutrinos per km$^3$ per year above 60~TeV~\cite{Abbasi:2020jmh}, including the background of atmospheric neutrinos.
Flavor measurements are improved by complementing them with \textit{through-going tracks}, which occur when $\nu_\mu$'s interact outside the instrumented volume, producing muons that cross part of the detector.
Because through-going tracks are more numerous, when combined with starting events they appreciably tighten the flavor measurements.  Reference~\cite{Aartsen:2015knd} reported the only IceCube measurement of the flavor composition of this type to date, based on 4 years of starting events and 2 years of through-going tracks.

Detailed analyses of the flavor-composition sensitivity that use through-going tracks require knowing the detector effective areas for these events, that, however, are not available outside the IceCube Collaboration.
Therefore, we base our analyses instead on the estimated projected IceCube (and IceCube-Gen2) sensitivities to flavor composition from Ref.~\cite{Aartsen:2020fgd}, using combined starting events and through-going tracks.

Figures~\ref{fig:triangle_sm_2020_vs_2040} and \ref{fig:triangle_decay_2020_vs_2040} show the present-day estimated IceCube sensitivity to flavor ratios~\cite{Aartsen:2020fgd}, based on a combination 8 years of starting events and through-going tracks.
The size of the sensitivity contour is representative of the present-day sensitivity of IceCube; it has been manually centered on the most likely best-fit composition assuming neutrino production in the full pion decay scenario. 

%%%%%%%%%%%%%%%%%%%%%%%%%%%%%%
\begin{table}[tbp]
\caption{Current best-fit values of the mixing parameters and their 1$\sigma$ uncertainties, taken from the global fit to oscillation data NuFit~5.0~\cite{Esteban:2020cvm,nufit5.0}, assuming normal or inverted neutrino mass ordering.  We include only the parameters that affect the average flavor-transition probabilities of high-energy astrophysical neutrinos: the mixing angles, $\theta_{12}$, $\theta_{23}$, $\theta_{13}$, and the phase $\dcp$.
}
\begin{center}  
\renewcommand{\arraystretch}{1.4}
\begin{tabular}{ccc} \hline \hline
   \textbf{Parameter} & \textbf{Normal ordering}  & \textbf{Inverted ordering} \\ \hline
$\sin^2\theta_{12}$ & $0.304^{+0.012}_{-0.012}$ & $0.304^{+0.013}_{-0.012}$  \\ 
% \hline
$\sin^2\theta_{23}$ & $0.573^{+0.016}_{-0.020}$ & $0.575^{+0.016}_{-0.019}$  \\
% \hline
$\sin^2\theta_{13}$ & $0.02219^{+0.00062}_{-0.00063}$ & $0.02238^{+0.00063}_{-0.00062}$  \\ 
% \hline
$\dcp~(^\circ)$ & $197^{+27}_{-24}$ & $282^{+26}_{-30}$  \\ \hline \hline
\end{tabular}
\end{center}
\label{tab:NuFIT}
\end{table}
%%%%%%%%%%%%%%%%%%%%%%%%%%%%%%

%%%%%%%%%%%%%%%%%%%%%%%%%%%%%%%%%%%%%%%%%%%%%%%%%%%%%%%%

\subsection{Flavor composition at Earth: Neutrino-decay-like new physics}
\label{subsection:decay}

As an example of physics beyond the standard oscillation picture, we consider the possibility that neutrinos decay~\cite{Beacom:2002vi, Beacom:2003nh,Maltoni:2008jr,Mehta:2011qb, Baerwald:2012kc,Pagliaroli:2015rca, Bustamante:2016ciw, Denton:2018aml, Bustamante:2020niz, Abdullahi:2020rge}.

Under decay, the flavor composition at Earth is determined by the flavor content of the surviving mass eigenstates, \ie,
\begin{equation}
    f_{\beta, \oplus} = \sum\limits_{i=1}^3 |U_{\beta i}|^2 f_{i,\oplus} 
    \label{equ:flavor_ratio_earth_decay}
\end{equation}
where $f_{i,\oplus}$ is the fraction of surviving $\nu_i$ in the flux that reaches Earth, and depends on the neutrino lifetimes, energies, and traveled distances.
By comparing the flavor composition at Earth under decay to the flavor composition measured in neutrino telescopes, we constrain the lifetime of the decaying neutrinos.

As illustration, we explore the case of invisible neutrino decay~\cite{Beacom:2002vi,Barenboim:2003jm,Maltoni:2008jr,Mehta:2011qb,Baerwald:2012kc,Pagliaroli:2015rca,Denton:2018aml,Barenboim:2020vrr}, in which the two heaviest mass eigenstates decay to species that are undetectable in neutrino telescopes, \eg, into a sterile neutrino or into a low-energy active neutrino.
For example, if the mass ordering is normal and neutrinos have a Dirac mass, $\nu_2$ and $\nu_3$ could decay into a right-handed $\nu_1$ and a new scalar; if it is inverted, $\nu_1$ and $\nu_2$ could decay into a right-handed $\nu_3$ and a new scalar.
In our discussion, we focus only on decay in the normal ordering and we take the lightest neutrino, $\nu_1$, to be stable.

Figure~\ref{fig:triangle_decay_2020_vs_2040} shows the flavor content $\vert U_{\alpha i} \vert^2$ of the mass eigenstates.
In the extreme case of complete decay, all unstable neutrinos have decayed upon reaching Earth, and the flavor composition of the flux is determined by the flavor content of $\nu_1$, \ie, $f_{\beta , \oplus} = \vert U_{\beta 1} \vert^2$.
If a fraction of the unstable neutrinos survive, the flavor composition is a combination of their flavor contents, Eq.~\eqref{equ:flavor_ratio_earth_decay}.

In order to estimate bounds on the neutrino lifetime, we turn to a concrete model, in which we  assume that $\nu_2$ and $\nu_3$ have the same lifetime-to-mass ratio $\tau/m$ and only $\nu_1$ is stable. We calculate the diffuse flux of high-energy neutrinos produced by a nondescript population of extragalactic sources, including the effect of neutrino decay during propagation, following Ref.~\cite{Bustamante:2016ciw}. 
We adopt the formalism of invisible decay from Refs.~\cite{Baerwald:2012kc,Bustamante:2016ciw}.

We assume that each neutrino source produces neutrinos with the same power-law energy spectrum, $J_{\nu}(E) \equiv E^2 dN_\nu/dE \propto E^{2-\gamma}$, where the value of the spectral index $\gamma$ is common to neutrinos and anti-neutrinos of all flavors. We assume $\gamma=2.5$ corresponding to the neutrino flux adopted to produce the IceCube projections of the sensitivity to flavor composition that we use~\cite{Aartsen:2015knd}.
For the number density of the neutrino sources at redshift $z$, we use the generic parametrization from Ref.~\cite{vanVliet:2019nse}, \ie, 
\begin{equation}
    \rho(z) 
    \propto 
    \left\{
    \begin{array}{ll}
    (1+z)^n,   & z < z_c\\
    (1+z_c)^n, & z \geq z_c 
    \end{array} 
    \right. \;,
    \label{eq:sourcerho}
\end{equation}
where different values of $n$ describe different candidate source populations and $z_c$ is a critical redshift above which their evolution is flat. We take $n = 1.5$ and $z_c = 1.5$, which roughly corresponds to the expected distribution of active galactic nuclei sources~\cite{vanVliet:2019nse}.

The diffuse flux of $\nu_\beta$ with energy $E$ detected at Earth is the sum of the contributions from all sources~\cite{Bustamante:2016ciw}, \ie,
\begin{eqnarray}
    E^2\frac{d\phi_{\beta,\oplus}}{dE} &=& \frac{1}{4\pi}\int_0^{z_{\max}} dz \sum\limits_\alpha P_{\alpha\beta}^\mathrm{decay}(E,z)f_{\alpha, {\rm S}} \nonumber\\
    &&\times\frac{\rho(z)}{(1+z)^2 H(z)}J_{\nu}(E(1+z),z) \;,
\end{eqnarray}
where $H(z) = H_0\sqrt{\Omega_\Lambda + \Omega_m (1+z)^3}$ is the Hubble parameter, $H_0 = 67.4$ km\,s$^{-1}$\,Mpc$^{-1}$ is the Hubble constant, $\Omega_m = 0.315$ is the energy density of matter, and $\Omega_\Lambda = 1- \Omega_m$ is the energy density of vacuum~\cite{aghanim:2018eyx}. {We integrate over the neutrino sources up to $z_{\max}=4$ beyond which we expect negligible contribution to the neutrino flux.} The neutrino flavor oscillation probability considering invisible decay is 
\begin{equation}
    \label{equ:prob_decay}
    P_{\alpha\beta}^\mathrm{decay}(E,z)=\sum\limits_i |U_{\alpha i}|^2|U_{\beta i}|^2\mathcal{Z}_i(z)^{-\frac{m_i}{\tau_i}\frac{1}{H_0 E}} \,,
\end{equation}
where $\mathcal{Z}_i$ is the redshift-dependent decay suppression factor introduced in Ref.~\cite{Baerwald:2012kc}.
Since the neutrino mass $m_i$ and lifetime $\tau_i$ appear together in Eq.~(\ref{equ:prob_decay}), we perform our analysis in terms of the ratio $m_i/\tau_i$.
Because $\nu_1$ is stable, $\mathcal{Z}_{1}=1$, while, for $\nu_2$ and $\nu_3$,
\begin{equation}
    \mathcal{Z}_{2,3}(z) \simeq a+be^{-cz}\,,
\end{equation}
where $a\simeq 1.67$, $b=1-a$, and $c\simeq 1.43$ for our choice of values of the cosmological parameters.

Under decay, the flavor composition changes with neutrino energy (see, \eg, Refs.~\cite{Beacom:2002vi,Mehta:2011qb,Baerwald:2012kc,Bustamante:2016ciw}).  In our analysis, we compute the average flavor composition at Earth over the energy interval from $E_{\min} = 60$~TeV to $E_{\max} = 10$~PeV, \ie,
\begin{equation}
    f_{\beta,\oplus}
    =
    \frac{\int_{E_{\min}}^{E_{\max}}dE \frac{d\phi_{\beta,\oplus}}{dE}}
    {\sum\limits_\alpha \int_{E_{\min}}^{E_{\max}} dE \frac{d\phi_{\alpha,\oplus}}{dE}} 
    \quad ({\rm neutrino~decay}) \; .
    \label{eq:decayflavor}
\end{equation}
%

%%%%%%%%%%%%%%%%%%%%%%%%%%%%%%%%%%%%%%%%%%%%%%%%%%%%%%%%

\subsection{Flavor composition at Earth:\\Non-unitary mixing}
\label{subsection:nonunit}

So far, we have assumed that the $3 \times 3$ mixing matrix $U$ is unitary, \ie, that the flavor states $\nu_\alpha$, and thus also the mass eigenstates $\nu_i$, form a complete basis.
The assumption of unitarity imposes constraints on the elements of $U$.  
However, the ``true'' mixing matrix could be larger than $3 \times 3$, as a result of the three active neutrinos mixing with additional states, such as a fourth, ``sterile'' neutrino.
In this case, $U$ is a $3 \times 3$ submatrix of the larger, true mixing matrix.
Relaxing the assumption of the unitarity of $U$ leads to a broader range of allowed flavor composition at Earth due to the active neutrinos mixing with the new states~\cite{Antusch:2006vwa,Xing:2008fg,Xu:2014via,Parke:2015goa,Brdar:2016thq,Arguelles:2019tum,Ellis:2020ehi,Ellis:2020hus,Hu:2020oba,Ahlers:2020miq}.
This is true even if the new states are too massive to be kinematically accessible.

We will examine how much the prediction of the allowed flavor composition at Earth relies on the assumption of the unitarity of neutrino mixing, and how much it affects the ability of future neutrino telescopes to infer the flavor composition at the source.  In the case of non-unitary mixing, a flavor state can be written as~\cite{Giunti:2004zf,Antusch:2006vwa,Ellis:2020hus}
\begin{eqnarray}
    |\nu_\alpha\rangle
    =
    \frac{1}{\sqrt{N_\alpha}} \sum\limits_{i=1}^3 U_{\alpha i}^* |\nu_i\rangle \;, 
    \label{eq:flavor_def_uv}
\end{eqnarray}
where the normalization $N_\alpha \equiv \sum_{i=1}^3 |U_{\alpha i}|^2$
ensures that $|\nu_\alpha\rangle$ is a properly normalized state, \ie, that $\langle\nu_\alpha|\nu_\alpha\rangle = 1$. 
The non-unitary (NU) average flavor-transition probability $|\langle\nu_\beta|\nu_\alpha\rangle|^2$ is
\begin{eqnarray}
P_{\alpha\beta}^{\rm NU} =\frac{1}{N_\alpha N_\beta} \sum\limits_{i=1}^3|U_{\alpha i}|^2|U_{\beta i}|^2 \;. 
\end{eqnarray}

The flavor ratios at Earth are computed in analogy to  \eq~(\ref{equ:flavor_ratio_earth_std}), \ie, for a given flavor composition at the source, they are
\begin{equation}
    \label{equ:flavor_ratio_earth_nu}
    f_{\beta,\oplus} = \sum\limits_{\alpha=e,\mu,\tau} P_{\alpha\beta}^{\rm NU} f_{\alpha, {\rm S}} 
    \quad (\text{non-unitary})
    \;.
\end{equation}
However, because some active neutrinos oscillate away into sterile states, the sum over active flavors at Earth is no longer unity, \ie, $f_{e, \oplus} + f_{\mu,\oplus} + f_{\tau, \oplus} < 1$.
Since neutrino telescopes can only measure the flavor composition of the flux of active neutrinos, we renormalize the flavor ratios as $\tilde{f}_{\alpha, \oplus} = f_{\alpha, \oplus} / \sum_{\beta=e,\mu,\tau} f_{\beta,\oplus}$.  Below, we show our results in the case of non-unitarity exclusively in terms of these renormalized flavor ratios.  To lighten the notation, below we refer to them simply as $f_{\alpha, \oplus}$.

%%%%%%%%%%%%%%%%%%%%%%%%%%%%%%%%%%%%%%%%%%%%%%%%%%%%%%%%
%%%%%%%%%%%%%%%%%%%%%%%%%%%%%%%%%%%%%%%%%%%%%%%%%%%%%%%%

\section{Next-Generation Experiments}
\label{sec:exp}

In the next two decades, oscillation experiments that use terrestrial neutrinos will significantly improve the precision of mixing parameters.
In parallel, future neutrino telescopes will precisely measure the flavor composition of astrophysical neutrinos.
Combined, they will provide the opportunity to pinpoint the flavor composition at the sources and thus help identify the origin of the high-energy astrophysical neutrinos.
In this section, we describe how we model these future experiments.

%%%%%%%%%%%%%%%%%%%%%%%%%%%%%%%%%%%%%%%%%%%%%%%%%%%%%%%%
\subsection{Future oscillation experiments}
\label{subsection:future_osc_exp}

Figure~\ref{fig:time_line} summarizes our projected evolution of the measurement precision of the mixing parameters using a combination of next-generation terrestrial neutrino experiments\footnote{A similar version of the central panel of this figure, showing the evolution up to 2020 only, was first shown in Ref.~\cite{denton-aps}.}.

Presently, as seen in Table~\ref{tab:NuFIT},  $\sin^2 \theta_{12}$ and $\sin^2 \theta_{23}$ are known to within $\simeq$4\%, from the NuFit~5.0 global fit~\cite{Esteban:2020cvm, nufit5.0}.
We consider the future measurement of $\sin^2\theta_{12}$ by JUNO, and of $\sin^2\theta_{23}$ and $\dcp$ by HK and DUNE.
We assume there will be no improvement on $\sin^2\theta_{13}$.
Presently, $\sin^2\theta_{13}$ is measured to $\sim$3\% by Daya Bay.
This is because, while JUNO, HK, and DUNE are sensitive to $\sin^2\theta_{13}$, no single one of them is expected to achieve better precision than Daya Bay, assuming their nominal exposures~\cite{An:2015jdp,Abe:2018uyc,Abi:2020evt}.
For example, DUNE will only reach 7\% resolution with its nominal exposure~\cite{Abi:2020evt}.
Below we describe the oscillation experiments that we use in our predictions.

{\bf JUNO}, the Jiangmen Underground Neutrino Observatory~\cite{An:2015jdp}, will be a 20-kt liquid scintillator detector, located in Guangdong, China.
It will measure the oscillation probability $P(\bar\nu_e\rightarrow \bar\nu_e)$ of 2--8-MeV reactor neutrinos at a baseline of $\simeq$53~km.
JUNO seeks to determine the neutrino mass ordering and precisely measure $\sin^2\theta_{12}$ and $\Delta m^2_{21}$.
Its nominal sensitivity on $\sin^2\theta_{12}$ is 0.54\% after 6 years of data-taking~\cite{An:2015jdp}, which is the value we adopt in this work.
JUNO is under construction and will start taking data in 2022~\cite{JUNONu2020}.

To simulate the time evolution of the sensitivity to $\sin^2\theta_{12}$, we simulate JUNO following Ref.~\cite{An:2015jdp}.
We take the reactor neutrino flux from Refs.~\cite{Mueller:2011nm,Huber:2011wv,An:2013uza,Capozzi:2013psa} and the inverse-beta-decay cross sections from Ref.~\cite{Strumia:2003zx}.
We include a correlated flux uncertainty of 2\%, an uncorrelated flux uncertainty of 0.8\%, a spectrum shape uncertainty of 1\%, and an energy scale uncertainty of 1\%~\cite{An:2015jdp}.
We do not include matter effects in the computation of the oscillation probability because they only shift the central value of $\sin^2\theta_{12}$ and not its sensitivity, and we do not consider backgrounds.  
With 6 years of collected data, our simulated sensitivity in $\sin^2\theta_{12}$ is 0.46\%.
We then take the time evolution of our simulated sensitivity and scale it by a factor of 1.17, so that our 6-year sensitivity matches that of Ref.~\cite{An:2015jdp}. 

{\bf DUNE}, the Deep Underground Neutrino Experiment~\cite{Acciarri:2015uup,Abi:2020wmh}, is a long-baseline neutrino oscillation experiment made out of large liquid argon time projection chambers.
It will measure the appearance and disappearance probabilities, $P(\nu_\mu\rightarrow\nu_e)$ and $P(\nu_\mu\rightarrow\nu_\mu)$, in the 0.5--5-GeV range using accelerator neutrinos, in neutrino and antineutrino modes~\cite{Abi:2020evt}.
DUNE seeks to determine the mass ordering and measure $\dcp$ and $\sin^2\theta_{23}$ precisely.
We use the official analysis framework released with the DUNE Conceptual Design Report~\cite{Acciarri:2015uup,Alion:2016uaj}, which is comparable to the DUNE Technical Design Report.

DUNE will start taking data in 2026~\cite{DUNENu2018} using a staged approach~\cite{Abi:2020evt}.
At the start of the beam run, its far detector will have two modules with a total fiducial volume of 20~kton, with a 1.2~MW beam.
After one year, an additional detector will be deployed, and then after 3 years of running, the last detector will be installed, totaling 40 kton of liquid argon.
Then, the beam will be upgraded to 2.4~MW.
We follow this timeline and assume an equal running time for neutrino and antineutrino modes to simulate the time evolution of the mixing parameter measurements.
At completion, the nominal projected sensitivity of DUNE envisions 300~kt$\cdot$MW$\cdot$year exposure, which corresponds to 7 years of collected data.

{\bf HK} is the multipurpose water Cherenkov successor to Super-K, with a fiducial mass of 187~kt, under construction in Kamioka, Japan.
This long-baseline experiment will measure the appearance and disappearance probabilities of accelerator neutrinos~\cite{Abe:2018uyc}.
It operates at slightly lower energies ($\simeq$0.6~GeV) and with a shorter baseline (295~km) than DUNE.
Like DUNE, HK will also measure $\dcp$ and $\sin^2\theta_{23}$ precisely.
It will start operation in 2027~\cite{HKNu2020} with a projected nominal exposure of 10 years using one Cherenkov tank as far detector~\cite{Abe:2018uyc}. 

Our simulations of HK are modified from that of Ref.~\cite{Huber:2002mx}, which follows Refs.~\cite{Itow:2001ee,Ishitsuka:2005qi}.
We adjusted the systematic errors on signal and background normalizations to match the official expected sensitivities on $\sin^2\theta_{23}$, $\Delta m^2_{32}$, and $\dcp$.
Figure~\ref{fig:correlation} shows our projected DUNE and HK sensitivities on $\sin^2\theta_{23}$ and $\dcp$, using their nominal exposures.

Beyond the mixing parameters, we will examine how robust are our results to oscillations being non-unitary.
The lower panel of Fig.~\ref{fig:time_line} shows the global limits on non-unitarity from current and future experiments by quantifying the deviation from $N_\alpha = 1$, for the $\alpha = e, \mu, \tau$ rows.
The 2015 values are taken from Ref.~\cite{Parke:2015goa}, and the 2020 and 2038 values are from Ref.~\cite{Ellis:2020hus}.
While future experiments may limit the non-unitarity in the $e$ and $\mu$ rows to the $\mathcal{O}(1\%)$ level, the non-unitarity in the $\tau$ row will remain relatively unchanged from its present value of 17\%.
The {\bf IceCube-Upgrade} will extend the current IceCube detector by 2025, with the addition of seven new closely-packed strings~\cite{Ishihara:2019aao}, including a number of calibration devices and sensors designed to help improve ice modeling~\cite{Ishihara:2019uei,Nagai:2019uaz}.
The sensitivity of the IceCube-Upgrade to $\nu_\tau$ appearance will play a major role in constraining $\sum|U_{\tau i}|^2$.
The ORCA subdetector of KM3NeT~\cite{Akindinov:2019flp}, to be deployed in the Mediterranean sea, is expected to perform similar measurements. 

%%%%%%%%%%%%%%%%%%%%%%%%%%%%%%%%%%%%%%%%%%%%%%%%%%%%%%%%

\subsection{Neutrino telescopes}
\label{sec:telescopes}

\textbf{IceCube} is an in-ice Cherenkov neutrino observatory that has been in operation for nearly a decade~\cite{Aartsen:2016nxy}. 
The experiment comprises a cubic kilometer of clear Antarctic ice, instrumented with 86 vertical strings, each of which is equipped with 60 digital optical modules (DOMs) to detect Cherenkov light from neutrino-nucleon interactions.
After 7.5 years of data-taking, IceCube has seen 103 High-Energy Starting Events (HESEs), of which 48.4 are above 60 TeV and expected to be of astrophysical origin~\cite{Abbasi:2020jmh}.
In 10 years, IceCube has seen 100--150 through-going tracks per year of astrophysical origin above 1~TeV~\cite{Stettner:2019tok,Abbasi:2020jmh}.

As mentioned in Section~\ref{subsection:ratios_icecube}, we base our analysis on \textit{projections} of the measurement of flavor composition in IceCube (and IceCube-Gen2), shown originally in Ref.~\cite{Aartsen:2020fgd}, that estimate the sensitivity obtained by combining starting events and through-going tracks collected over 8 or 15 years (in the latter case, combined with 10 years of IceCube-Gen2), as such an analysis has not been performed on real data yet. 
Figures~\ref{fig:triangle_sm_2020_vs_2040} and \ref{fig:triangle_decay_2020_vs_2040} show the 99.7\%~ credible regions (C.R.) 8-year IceCube contour from Ref.~\cite{Aartsen:2020fgd}.  

\textbf{IceCube-Gen2} is the planned extension of IceCube~\cite{Aartsen:2019swn,Aartsen:2020fgd}.
It will add 120 new strings to the existing experiment, leading to an instrumented volume of 7.9~km$^3$ and an effective area that varies from 7 to 8.5 times that of IceCube between 100~TeV and 1~PeV.
Here, we assume a full array effective start date of  2030.  Figures~\ref{fig:triangle_sm_2020_vs_2040} and \ref{fig:triangle_decay_2020_vs_2040} show the 99.7\%~C.R. 15-year IceCube plus 10-year IceCube-Gen2  contour from Ref.~\cite{Aartsen:2020fgd}.  
Later we detail how we use the IceCube and IceCube-Gen2 projections to estimate projections also for the other neutrino telescopes.

\textbf{KM3NeT}~\cite{Adrian-Martinez:2016fdl} is the successor to ANTARES~\cite{Collaboration:2011nsa}, located in the Mediterranean Sea.
The high-energy component, called KM3Net/ARCA, will be deployed as two 115-string arrays with 18 DOMs each, 100~km off the coast of Sicily, and should be complete by 2024~\cite{Aiello:2018usb}.
Based on a projected event rate of 15.6 cosmic neutrino-induced cascades per year~\cite{Adrian-Martinez:2016fdl}, we estimate the exposure of KM3Net to be $\simeq$2.4 times that of IceCube.   

\textbf{Baikal-GVD}~\cite{Safronov:2020dtw} is a gigaton volume detector that expands on the existing NT-200 detector~\cite{Belolaptikov:1997ry} in lake Baikal, Siberia.
The first modules are already installed, and the detector has been operating since 2018 with an effective volume of 0.35~km$^3$.
This will rise to 1.5~km$^3$ in 2025 when the detector is complete, consisting of 90 strings, with 12 DOMs each.
Baikal-GVD has already seen at least one candidate neutrino cascade event with reconstructed energy of 91 TeV~\cite{Zaborov:2020idc}.  

\textbf{P-ONE}~\cite{Agostini:2020aar}, the Pacific Ocean Neutrino Experiment, is a planned water Cherenkov experiment, to be deployed in the Cascadia basin off Vancouver Island, using Ocean Networks Canada infrastructure that is already in place. P-ONE is expected to be complete in 2030 and will include 70 strings, with 20 DOMs each, deployed in a modular array covering a cylindrical volume with a 1~km height and 1~km radius. 

\textbf{TAMBO}~\cite{Wissel:2019alx,Romero-Wolf:2020pzh}
the Tau Air-Shower Mountain-Based Observatory, is a proposed array of water-Cherenkov tanks to be located in a deep canyon in Peru.
TAMBO will search for Earth-skimming $\nu_\tau$ in the 1--100~PeV range.  It is expected to detect approximately 7 $\nu_\tau$ per year in the energy range considered here.
Because it is sensitive to a single flavor, TAMBO will be particularly helpful in breaking the $\nu_e$-$\nu_\tau$ degeneracy in measuring flavor composition.
Unlike the other future neutrino telescopes, whose projected sensitivity we obtain by scaling the IceCube sensitivity (see below), we model the contribution of TAMBO to the projected flavor likelihood in 2040 as
\begin{equation}
    \label{equ:ll_tambo}
     -2 \ln \mathcal{L}_{\rm TAMBO} = \frac{(N_{\nu_\tau} - \bar N_{\nu_\tau})^2}{ \bar N_{\nu_\tau}} \;,
\end{equation}
where $ \bar N_{\nu_\tau} \simeq 70$ is the expected number of $\nu_\tau$ detected between 2030 and 2040 and $N_{\nu_\tau}$ is the number of $\nu_\tau$ events if $f_{\tau,\oplus}$ deviates from the assumed true value of 0.34.

Table~\ref{tab:flavoruncertainty} shows what neutrino telescopes are expected to contribute to flavor measurements in 2020, 2030, and 2040, and their combined exposures.
Reference~\cite{Aartsen:2020fgd} presented the projected sensitivity of 8 and 15 years of IceCube, and 15 years of IceCube plus 10 years of IceCube-Gen2, in the form of iso-contours of posterior density in the plane of flavor compositions at Earth.
We use these as likelihood functions $\mathcal{L}(f_{\alpha, \oplus})$ that represent the sensitivity of the flavor measurements, \ie, $\mathcal{L}_{\rm IC8}$ and $\mathcal{L}_{\rm IC15}$ for 8 and 15 years of IceCube, which we use for our 2020 and 2030 projections, and $\mathcal{L}_{\rm IC+Gen2}$ for 15 years of IceCube plus 10 years of IceCube-Gen2, which we use for our 2040 projections.

We are interested in assessing the flavor sensitivity achieved by combining all of the available neutrino telescopes in 2040.
However, with the exception of IceCube and IceCube-Gen2~\cite{Aartsen:2020fgd}, detailed projections for the sensitivity to flavor composition in upcoming neutrino telescopes are unavailable.
Therefore, we estimate 2040 projections for the other neutrino telescopes ourselves based on the projections for IceCube-Gen2.
First, we single out the contribution of 10 years of IceCube-Gen2 via $\ln\mathcal{L}_{\rm Gen2} \equiv \ln\mathcal{L}_{\rm IC+Gen2} - \ln\mathcal{L}_{\rm IC15}$.
Second, we estimate the combined sensitivity of Baikal-GVD, KM3NeT, and P-ONE by rescaling the IceCube-Gen2 contribution by the exposures $\Xi$ of these telescopes in 2040 (see Table~\ref{tab:flavoruncertainty}).   
Third, we add to that the contribution of 15 years of IceCube and of TAMBO, Eq.~\eqref{equ:ll_tambo}.
Thus, in 2040, we calculate the flavor sensitivity as 
\begin{equation}
\ln\mathcal{L}_{\rm comb} = \Xi_S\ln\mathcal{L}_{\rm Gen2}   + \ln\mathcal{L}_{\rm IC15} + \ln\mathcal{L}_{\rm TAMBO},
\end{equation}
where $\Xi_S$ is the effective IceCube-Gen2-equivalent exposure defined by
\begin{equation}
\Xi_S = \frac{\Xi_{\rm Gen2}+\Xi_{\rm GVD}+\Xi_{\rm KM3NeT} + \Xi_{\rm P-ONE}}{\Xi_{\rm Gen2}}.
\end{equation}

Based on the projections presented above, the estimated exposures by 2040 for individual experiments are: $\Xi_{\rm Gen2} = 81.6$~km$^3$ yr for IceCube-Gen2, $\Xi_{\rm KM3NeT} = 42.1$~km$^3$ yr for KM3NeT, $\Xi_{\rm GVD} = 24.3$~km$^3$ yr for Baikal-GVD, and $\Xi_{\rm P-ONE} = 31.6$~km$^3$ yr for P-ONE.
Figures~\ref{fig:triangle_sm_2020_vs_2040} and \ref{fig:triangle_decay_2020_vs_2040} show our 99.7\%~C.R. contour for all neutrino telescopes combined in 2040.

All of the projected contours of flavor-composition sensitivity in our analysis are centered on the flavor composition at Earth corresponding to the full pion decay chain computed using the best-fit values of the mixing parameters from NuFit~5.0, \ie,  $\left(0.30, 0.36, 0.34\right)_\oplus$.
While the position on which the contours are centered may be different in reality, their size is representative of the sensitivity of IceCube in 2020, the combination of IceCube and IceCube-Gen2 in 2030 and 2040, and the combination of all available neutrino telescopes in 2040.
Later, in Section~\ref{sec:results_sources}, we use these likelihoods to infer the sensitivity to flavor composition at the sources based on the flavor composition measured at Earth.

%%%%%%%%%%%%%%%%%%%%%%%%%%%%%%
\begin{table*}[!htb]
    \centering
    \setlength\extrarowheight{7pt}
    \begin{tabular}{c  c  c c c c c c c}
    % \begin{tabular}{c | c | c| c c c| c c c}
         \hline \hline
         \multirow{3}{*}{\textbf{Year}} & \multirow{3}{*}{\textbf{Neutrino telescopes}} & \multirow{3}{*}{\textbf{Oscillation parameters}} &\multicolumn{3}{c}{\textbf{Flavor ratios at Earth}} & & \multicolumn{2}{c}{\makecell{\textbf{Flavor ratios at source}\\{\textbf{(assuming $f_{\tau,{\rm S}} = 0$)}}}} \\[2ex]
         \cline{4-6} \cline{8-9}
         &  &  & $f_{e,\oplus}$ & $f_{\mu,\oplus}$ & $f_{\tau,\oplus}$ & & $f_{e,\rm S}$ & $f_{\mu,\rm S}$ \\[1ex] \hline
         2020  & IC 8 yr & NuFit 5.0 & $0.30^{+0.13}_{-0.11}$                 & $0.36^{+0.059}_{-0.053}$                  & $0.34^{+0.16}_{-0.18}$  & & $0.31^{+0.08}_{-0.13}$ & $0.69_{-0.08}^{+0.13}$ \\[1ex]
        %  2030  & IC 15 yr & NuFit+JUNO & $0.30^{+0.088}_{-0.079}$                & $0.36^{+0.042}_{-0.038}$                  & $0.34^{+0.11}_{-0.12}$  & & & \\[1ex]
         2040 & IC 15 yr+Gen2 10 yr & NuFit+JUNO+DUNE+HK & $0.30^{+0.039}_{-0.037}$                 & $0.36^{+0.017}_{-0.016}$                  & $0.34^{+0.049}_{-0.050}$ & & $0.33^{+0.02}_{-0.02}$ & $0.67^{+0.02}_{-0.02}$\\[1ex]
         2040 & \makecell{IC+Gen2+KM3NeT\\+GVD+P-ONE+TAMBO} & NuFit+JUNO+DUNE+HK &  $0.30^{+0.030}_{-0.027}$                 & $0.36^{+0.011}_{-0.011}$                  & $0.34^{+0.037}_{-0.039}$ & & $0.33^{+0.02}_{-0.01}$ & $0.67^{+0.01}_{-0.02}$\\
         \hline \hline
    \end{tabular}    
    % \internallinenumbers
    \caption{Projected 68\%~C.R. uncertainties on the allowed flavor ratios at Earth, $f_{\alpha,\oplus}$ ($\alpha=e,\mu,\tau$), and on the inferred flavor ratios at the astrophysical sources, $f_{\alpha,{\rm S}}$.  For this table, we assume standard oscillations and set the true value of the flavor ratios at the sources to $\left(\frac{1}{3}, \frac{2}{3}, 0\right)_{\rm S}$, coming from the full pion decay chain (see Section~\ref{subsection:ratios_earth}).  
    }   
    \label{tab:flavoruncertainty}
\end{table*}
%%%%%%%%%%%%%%%%%%%%%%%%%%%%%%

%%%%%%%%%%%%%%%%%%%%%%%%%%%%%%%%%%%%%%%%%%%%%%%%%%%%%%%%
%%%%%%%%%%%%%%%%%%%%%%%%%%%%%%%%%%%%%%%%%%%%%%%%%%%%%%%%

\section{Statistical methods}
\label{sec:stat}

We present results in a Bayesian framework, as 68\% or 99.7\% credible regions (C.R.) or intervals.
These represent the iso-posterior contours within which 68\% or 99.7\%  of the marginalized---integrated-over nuisance parameters---posterior mass is located.

In this section, we describe in detail how we obtain the regions of the neutrino flavor composition at Earth, $\pmb{f}_\oplus \equiv ( f_{e,\oplus}, f_{\mu,\oplus}, f_{\tau,\oplus} )$, given different assumptions about the flavor composition at the sources, $\pmb{f}_{\rm S} \equiv ( f_{e,{\rm S}}, f_{\mu,{\rm S}}, f_{\tau,{\rm S}} )$, and under the three flavor-transition scenarios introduced in Section~\ref{sec:theory}: standard oscillations, non-unitary mixing, and neutrino decay.  

We assess the compatibility of a given flavor composition $\pmb{f}_\oplus$ with the probability distribution of mixing parameters, either today or in the future, and with our prior belief about what the flavor composition at the source is.
To do this, we adopt the Bayesian approach first introduced in Ref.~\cite{Gonzalez-Garcia:2016gpq}.
The posterior probability of $\pmb{f}_\oplus$ is 
\begin{eqnarray}
    \mathcal{P}(\pmb{f}_\oplus)
    &=&
    \int d\pmb{\vartheta}
    \int d\pmb{{f}}_{\rm S}~
    {\rm det}(\pmb{J}(\pmb{{f}}_{\rm S}, \pmb{\vartheta}))
    \nonumber \\
    && \quad 
    \times
    \delta( \pmb{f}_\oplus - \pmb{\tilde{f}}_\oplus(\pmb{{f}}_{\rm S}, \pmb{\vartheta}) )
    \mathcal{L}(\pmb{\vartheta}) 
    \pi(\pmb{\vartheta}) 
    \pi(\pmb{f}_{\rm S})
     \;,
     \label{equ:posterior}
\end{eqnarray}
where $\pmb{J} \equiv ( \partial^2 \pmb{f}_\oplus / \partial \pmb{f}_{\rm S} \partial \pmb{\vartheta} )^{-1}$ is the Jacobian matrix.
Here, $\mathcal{L}$ is the likelihood function, defined as the probability of obtaining a particular set of measurements $\pmb{\mathcal{E}}$ in oscillation experiments conditional on the mixing parameters being $\pmb{\vartheta} \equiv (\sin^2 \theta_{12}, \sin^2 \theta_{23}, \sin^2 \theta_{13}, \dcp)$.
The prior on the mixing parameters is $\pi(\pmb{\vartheta})$, and the prior on the flavor ratios at the source is $\pi(\pmb{f}_{\rm S})$ .
Below, we describe how to compute these functions.
The Dirac delta ensures that we account for all combinations of $\pmb{f}_{\rm S}$ and $\pmb{\vartheta}$ that produce the specific flavor ratios $\pmb{{f}}_\oplus$ at Earth.
Inside the Dirac delta, the flavor ratios at Earth, $\pmb{\tilde{f}}_\oplus(\pmb{f}_{\rm S}, \pmb{\vartheta})$, are computed using \eq~\eqref{equ:flavor_ratio_earth_std} for standard oscillations, \eq~\eqref{equ:flavor_ratio_earth_decay} for neutrino decay, and \eq~\eqref{equ:flavor_ratio_earth_nu} for non-unitary mixing.
In the case of neutrino decay, $\pmb f_\oplus$ also depends on the $\nu_i$ fractions, $f_{i,\oplus}$ (see Section~\ref{subsection:decay}), while in the case of non-unitary mixing, $\pmb{\vartheta}$ represents the elements of the non-unitary mixing matrix instead of the standard mixing parameters (see Section~\ref{subsection:nonunit}).

To compute the likelihood $\mathcal{L}$ in~\eq~\eqref{equ:posterior}, we construct a $\chi^2$ test-statistic that incorporates the combined information from future oscillation experiments---JUNO, DUNE, HK---on the mixing parameters $\pmb{\vartheta}$.
We fix the best-fit values of the mixing parameters to the current NuFit~5.0 best fit (see Table~\ref{tab:NuFIT}); for these, we assume normal mass ordering in the main text and inverted mass ordering in our appendices.
In our projections, we assume that the measurement of each mixing parameter $\vartheta_i$ will have a normal distribution, and that the measurement of different mixing parameters will be uncorrelated except for $\delta_{\rm CP}$ and $\sin^2\theta_{23}$.
With this, the sensitivity associated to each experiment $\mathcal{E}$ is
\begin{equation}
    \chi_{\mathcal{E}}^2 =\sum_{i,j}(\vartheta_i-\bar \vartheta_i)\Sigma^{-1}_{\mathcal{E},ij}(\vartheta_j-\bar \vartheta_j) \;,
    \label{equ:chi2}
\end{equation}
where $\Sigma_{ij}$ is the covariance matrix for parameters $\vartheta_i, \vartheta_j$.
The likelihood of the combined set of future experiments is
\begin{equation}
 -2\ln \mathcal{L}(\pmb{\vartheta}) = \sum_\mathcal{E} \chi_{\mathcal{E} }^2 \;,    
 \label{equ:likelihood}
\end{equation}
where the sum runs over NuFit~5.0~\cite{Esteban:2020cvm,nufit5.0} and each of the relevant experiments described above.
For the prior $\pi(\pmb{\vartheta})$ in \eq~\eqref{equ:posterior}, we sample uniformly from $\sin^2\theta_{12}$, $\sin^2\theta_{13}$ and $\sin^2\theta_{23}$.
% we use the Haar measure of SU(3)~\cite{Haba:2000be}; see Appendix of Ref.~\cite{Arguelles:2019tum} for details.

For the prior on the flavor composition at the source, $\pi(\pmb{f}_{\rm S})$ in \eq~\eqref{equ:posterior}, we explore two alternatives separately.
In both, we ensure that the prior is normalized by demanding that 
\begin{equation}
   \int_0^1 df_{e,{\rm S}} \int_0^{1-f_{e,{\rm S}}} df_{\mu,{\rm S}} \pi(\pmb{f}_{\rm S}) = 1 \;. 
\end{equation}
We only need to integrate over $f_{e,{\rm S}}$ and $f_{\mu,{\rm S}}$ because $f_{\tau,{\rm S}} = 1-f_{e,{\rm S}}-f_{\mu,{\rm S}}$. 
The two alternatives are:
\begin{enumerate}
    \item Every flavor composition at the source is equally likely, and we let it vary over all the possibilities.
    In this case, $\pi(\pmb{f}_{\rm S}) = 2$. 
    \item The flavor composition is fixed to one of the three benchmark scenarios: pion decay ($\pmb{f}_{\rm S}^\pi \equiv (\frac{1}{3}, \frac{2}{3}, 0)$), muon-damped ($\pmb{f}_{\rm S}^\mu \equiv (0, 1, 0)$), or neutron decay ($\pmb{f}_{\rm S}^n \equiv (1, 0, 0)$).
    In this case, $\pi(\pmb{f}_{\rm S}) = \delta(\pmb{f}_{\rm S}-\pmb{f}_{\rm S}^\pi)$ for pion decay, and similarly for the other benchmarks.
\end{enumerate}

In practice, we build the posterior function, \eq~\eqref{equ:posterior}, by randomly sampling values of $\pmb{\vartheta}$ and $\pmb f_{\rm S}$ from their respective priors, computing the corresponding value of $\pmb{\tilde{f}}_\oplus(\pmb f_{\rm S}, \pmb{\vartheta})$, and assigning it a weight $\mathcal{L}(\pmb{\vartheta})$.
Using the sampled values of $\pmb{\tilde{f}}_\oplus$, we build a kernel density estimator that is proportional to the posterior distribution.

%%%%%%%%%%%%%%%%%%%%%%%%%%%%%%%%%%%%%%%%%%%%%%%%%%%%%%%%
%%%%%%%%%%%%%%%%%%%%%%%%%%%%%%%%%%%%%%%%%%%%%%%%%%%%%%%%

\section{Results and Discussion}
\label{sec:results}

%%%%%%%%%%%%%%%%%%%%%%%%%%%%%%%%%%%%%%%%%%%%%%%%%%%%%%%%

\subsection{Finding the sources of the high-energy astrophysical neutrinos}
\label{sec:results_sources}

{\it Allowed regions of flavor composition at Earth.---}  The left panel of Figure~\ref{fig:triangle_sm_2020_vs_2040} shows the 99.7\% C.R. of flavor composition at Earth for the years 2020 and 2040, assuming standard oscillations, obtained using the statistical method outlined above.
The larger gray regions are sampled from a flat prior in source composition, while each of the colored regions assumes 100\% pion decay (red), muon-damped $\pi$ decay (orange), or neutron decay (green).
Table~\ref{tab:flavoruncertainty} shows the 68\%~C.R.~sensitivity to each of the flavor ratios for the different combinations of neutrino telescopes. 
These are shown for the year 2020---using the distribution of mixing parameters from NuFit~5.0---and for the years 2030 and 2040---using the projected sensitivity to the mixing parameters of the combined JUNO, DUNE, and HK, with their true values fixed at the best-fit values of NuFit~5.0.

Figure~\ref{fig:triangle_sm_2020_vs_2040} shows that, for a given flavor composition at the source, the allowed region of flavor composition at Earth shrinks approximately by a factor of ten between 2020 and 2040.
When allowing the flavor composition at the source to vary over all possible combinations instead, the allowed flavor region at Earth shrinks approximately by a factor of 5 between 2020 and 2040.
In this case, the improvement is smaller because the prior volume is larger since, in addition to sampling over the mixing parameters, we sample also over all possible values of $\pmb{f}_{\rm S}$.

The reduction in the size of the allowed flavor regions from 2020 to 2040 stems mainly from the improved measurement of $\sin^2 \theta_{12}$ by JUNO, which shrinks the regions along the $f_{e,\oplus}$ direction, and of $\sin^2 \theta_{23}$ by DUNE and HK, which shrinks the regions along the $f_{\tau,\oplus}$ direction.
We keep the uncertainty on $\sin^2 \theta_{13}$ fixed at its current value (see Section~\ref{sec:exp}). 
While we account for improvements in the measurement of $\dcp$ over time, the effect of $\dcp$ on flavor transitions is weak (see Section~\ref{subsection:std_osc}). 

In Fig.~\ref{fig:triangle_sm_2020_vs_2040}, we also include the estimated 2020 IceCube 8-year flavor sensitivity, the projected 2040 IceCube 15-year + IceCube-Gen2 10-year flavor sensitivity, and an ``all-telescope'' sensitivity that additionally includes the contributions of Baikal-GVD, KM3NeT, P-ONE, and TAMBO.
These contours are produced under the assumption of a ``true'' flavor ratio at Earth of about $(0.30, 0.36, 0.34)_\oplus$ coming from the full pion decay chain; see Section~\ref{sec:telescopes} for details.
The uncertainty in flavor measurement shrinks by roughly a factor of 2 between 2020 and 2040.
This improvement stems from the larger event sample size and, to a lesser extent, the inclusion of TAMBO, which measures the $\nu_\tau$-only neutrino flux.
For the remaining neutrino telescopes, which are sensitive to all the neutrino flavors, these projections use the same morphology confusion matrix as recent IceCube analyses~\cite{Abbasi:2020jmh,Abbasi:2020zmr}.
This is a conservative assumption, as these rates are expected to improve for IceCube thanks to the calibration devices of the IceCube-Upgrade~\cite{Aartsen:2014oha,Ishihara:2019aao} and are expected to be better in Baikal-GVD, KM3NeT, and P-ONE due to the reduced scattering of Cherenkov photons in water compared to ice.  Additional improvement may come from combining all of the available neutrino telescopes in a global observatory~\cite{resconi2019}.

The change from 2020 to 2040 is most striking when we focus on the two most likely neutrino production scenarios: full pion decay and muon damping.
In 2020, their 99.7\% C.R. overlap, which makes it challenging to distinguish between them, especially because of the large uncertainty with which IceCube currently measures flavor composition.
In contrast, by 2040, their flavor regions will be well separated, at the level of many standard deviations.
This, combined with the roughly factor-of-two reduction in the uncertainty of flavor measurements, will allow IceCube-Gen2 to unequivocally distinguish between the full-pion-decay and muon-damped scenarios, and realistically help identify the population of sources at the origin of the high-energy astrophysical neutrinos, as we will discuss in Sec.~\ref{sec:results_sources}.

%%%%%%%%%%%%%%%%%%%%%%%%%%%%%%

{\it Robustness against non-unitary mixing.---}  The right panel of Fig.~\ref{fig:triangle_sm_2020_vs_2040} shows that our conclusions 
hold even if neutrino mixing is non-unitary. 
The allowed regions with and without unitarity in the mixing---in the left vs.~right panels of Fig.~\ref{fig:triangle_sm_2020_vs_2040}---have approximately the same size.
This means that our ability to pinpoint the dominant mechanism of neutrino production is not affected by the existence of additional neutrino mass states.

Our analysis of non-unitarity assumes that the new mass eigenstates are too heavy to be produced in weak interactions~\cite{Ellis:2020hus}.
This is not the case for additional neutrinos motivated by the short-baseline oscillation anomalies~\cite{Athanassopoulos:1996wc,Athanassopoulos:1997pv,Dentler:2018sju,Aguilar-Arevalo:2018gpe,Diaz:2019fwt}, in which case large deviations from the allowed standard-oscillation flavor regions are possible~\cite{Brdar:2016thq,Arguelles:2019tum}, because light ($< 100$~MeV) sterile neutrinos could be produced at the sources, and so the sum in Eq.~\eqref{eq:flavor_def_uv} would then be over all mass states, active and sterile, with masses smaller than the mass of the parent pion.

%%%%%%%%%%%%%%%%%%%%%%%%%%%%%%

%%%%%%%%%%%%%%%%%%%%%%%%%%%%%%
\begin{figure}
 \centering
 \includegraphics[width=\columnwidth]{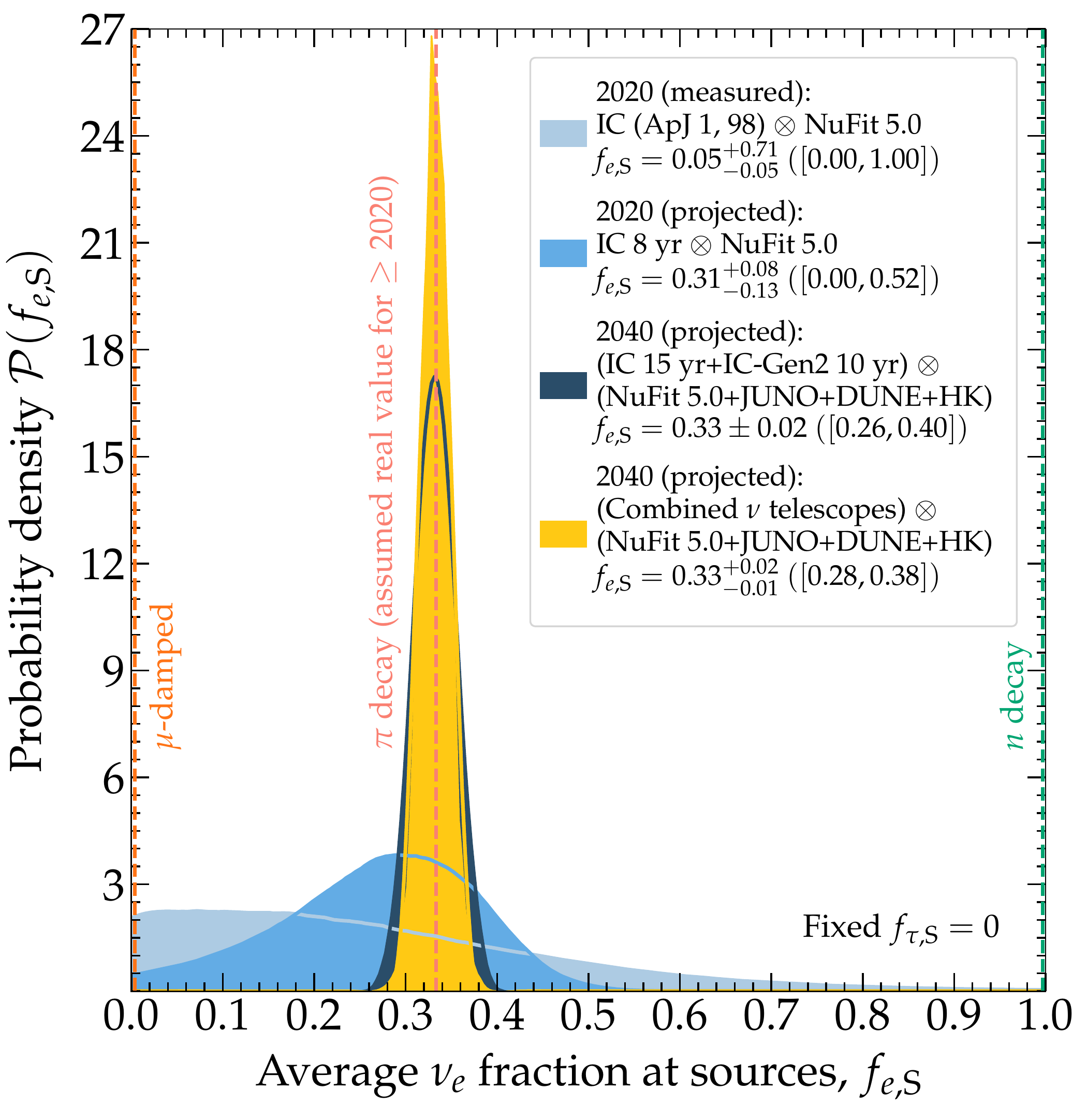}
 \caption{Fraction of $\nu_e$ produced in astrophysical sources, inferred from the flavor composition measured at Earth, in IceCube (IC), IceCube-Gen2 (IC-Gen2), and future neutrino telescopes combined, and accounting for the uncertainties in the mixing parameters.  In each case, we show the best-fit value of $f_{e, {\rm S}}$, its 68\%~C.R. interval and, in parentheses, its 99.7\%~C.R. interval.  The  2020 (measured) curve is based on the measurement of flavor composition $\mathcal{L}(\pmb{f}_\oplus)$ reported by IceCube in Ref.\ \cite{Aartsen:2015knd} (following Ref.\ \cite{Bustamante:2019sdb}, we convert the frequentist likelihood reported therein into a probability density) and mixing-parameter likelihood $\mathcal{L}(\pmb{\theta})$ from NuFit~5.0\ \cite{Esteban:2020cvm}.  The curves for 2020 (projected) and 2040 are based on projections of $\mathcal{L}_{\rm exp}$ from Ref.\ \cite{Aartsen:2020fgd}, and $\mathcal{L}(\pmb{\theta})$ built by combining projections of different oscillation experiments, as detailed in Section\ \ref{sec:stat}.  For the 2020 and 2040 curves, we assume that the real value of $f_{e, {\rm S}} = 1/3$, coming from the full pion decay chain.  We fix $f_{\tau, {\rm S}} = 0$, \ie, we assume that sources do not produce $\nu_\tau$.
 }
 \label{fig:recon_feS_pdf}
\end{figure}
%%%%%%%%%%%%%%%%%%%%%%%%%%%%%%

{\it Inferring the flavor composition at the sources.---} 
Ultimately, we are interested in learning about the identity of the sources of high-energy neutrinos and the physical conditions that govern them.

To illustrate the improvement over time in the reconstruction of the flavor composition at the source, we compute the posterior probability of $\pmb{f}_{\rm S}$ as
\begin{eqnarray}
    \mathcal{P}(\pmb{f}_{\rm S}) = 
    \int d\pmb{\vartheta}
    \mathcal{L}(\pmb{\vartheta}) 
    \mathcal{L}(\pmb{f}_\oplus(\pmb{f}_{\rm S},\vartheta))
    \pi(\pmb{\vartheta}) 
    \pi(\pmb{f}_{\rm S})
     \label{equ:ps},
\end{eqnarray} 
where $\mathcal{L}(\pmb{f}_\oplus(\pmb{f}_S,\vartheta))$ is the (projected) constraint on the flavor composition at Earth from neutrino telescope observations,
$\pi(\pmb{f}_S)$ is the prior on the flavor composition at the source. We assume $f_{\tau,\rm S}=0$ and put a uniform prior on $f_{e,\rm S}$.
  
%%%%%%%%%%%%%%%%%%%%%%%%%%%%%%
\begin{figure*}[t!]
 \centering
 \includegraphics[width=\columnwidth]{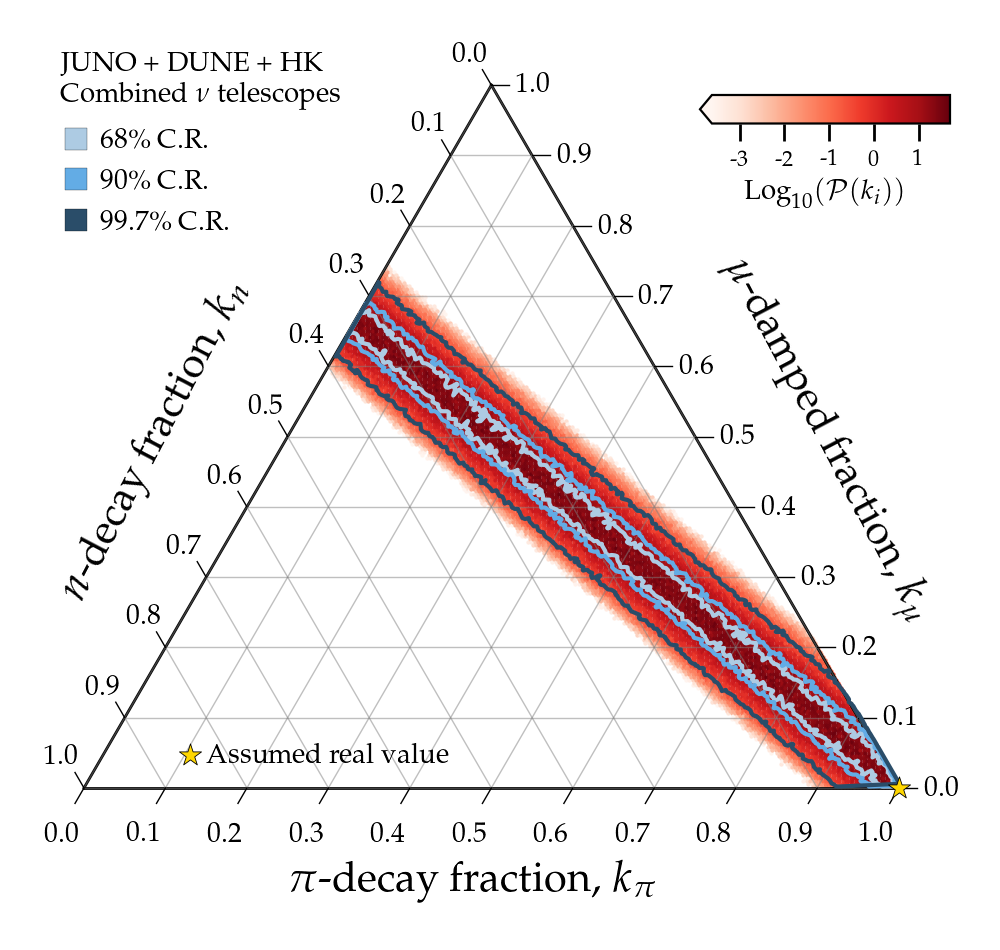}
 \includegraphics[width=\columnwidth]{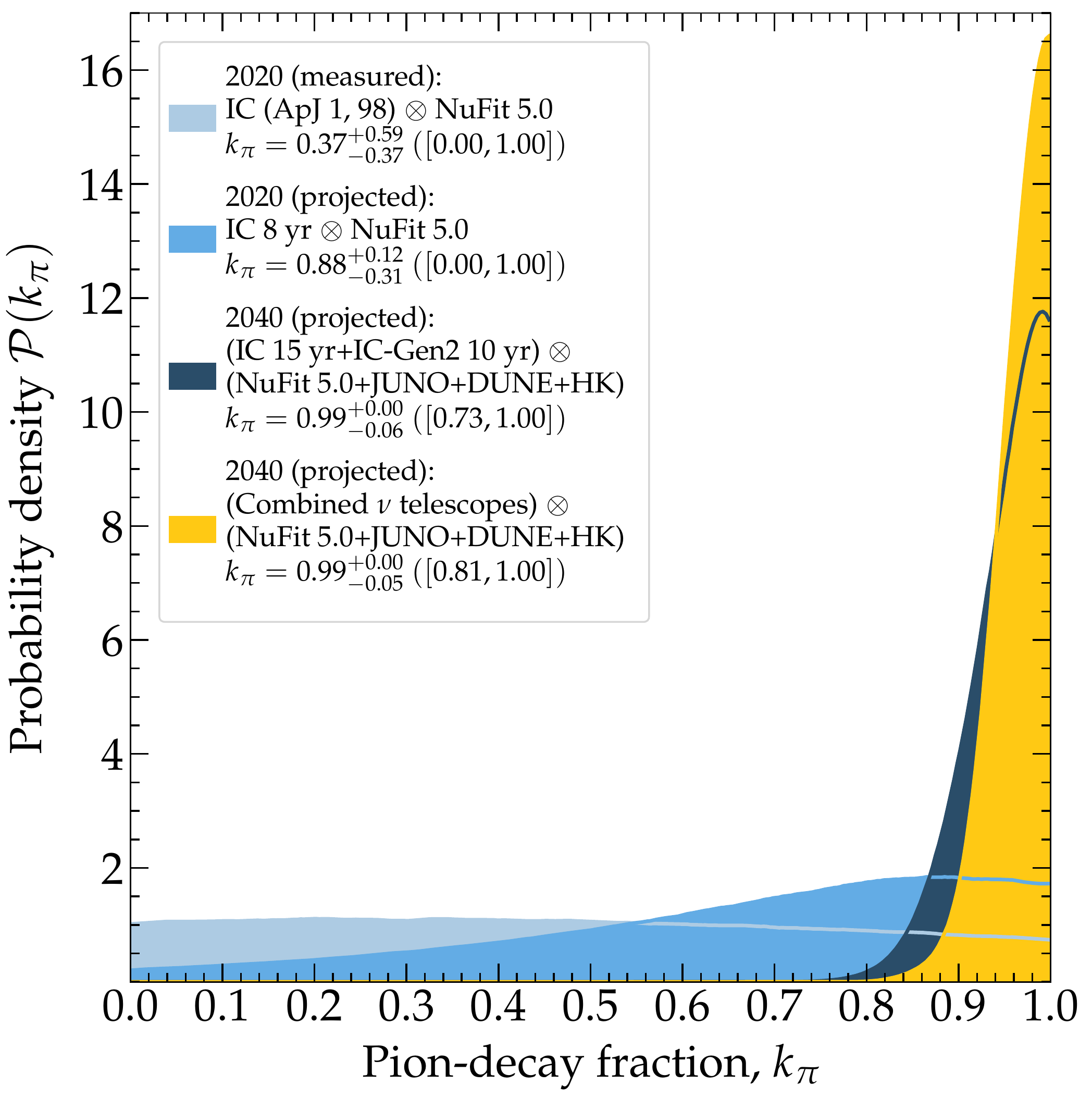}
%%  \internallinenumbers
 \caption{Sensitivity to the fraction of the diffuse flux of high-energy neutrinos that is contributed by the three benchmark scenarios.  The real value is assumed to be $k_\pi = 1$, \ie, production only via full pion decay. {\it Left:} Allowing for production via the three benchmark scenarios.  {\it Right:} Allowing for production only via the full pion decay and muon-damped scenarios, in IceCube (IC), IceCube-Gen2 (IC-Gen2), and future neutrino telescopes combined, and accounting for the uncertainties in the mixing parameters.  In each case, we show the best-fit value of $k_\pi$, its 68\%~C.R. interval and, in parentheses, its 99.7\%~C.R interval.}
 \label{fig:subcomponent_fractions}
\end{figure*}
%%%%%%%%%%%%%%%%%%%%%%%%%%%%%%

Our results update those from Ref.~\cite{Bustamante:2019sdb}, by improving in four different ways.
First, for the 2015 and 2020 results, we use $\mathcal{L}(\pmb{\theta})$ taken directly from the NuFit~5.0 $\chi^2$ profiles, which include two-parameter correlations, compared to Ref.~\cite{Bustamante:2019sdb}, which assumed Gaussian, uncorrelated likelihoods centered around the NuFit~3.2~\cite{Esteban:2016qun,nufit3.2} best-fit values.
Second, for the 2020 and 2040 projections, we use more recent and accurate projections of $\mathcal{L}(\pmb{f}_\oplus)$ for IceCube and IceCube-Gen2, from Ref.~\cite{Aartsen:2020fgd}, instead of the early estimate from Ref.~\cite{Aartsen:2014njl} used in Ref.~\cite{Bustamante:2019sdb}.
Third, for the 2040 projections, we build detailed projected likelihoods $\mathcal{L}(\pmb{\theta})$ by combining the results of simulating different oscillation experiments (see Section~\ref{subsection:future_osc_exp}), versus Ref.~\cite{Bustamante:2019sdb}, which assumed an estimated reduction in the parameter uncertainties in the near future and perfect knowledge of the parameters in the far future.
Finally, we now include in our projection not only IceCube-Gen2, as in Ref.~\cite{Bustamante:2019sdb}, but also the combination of all upcoming TeV--PeV neutrino telescopes.

Figure~\ref{fig:recon_feS_pdf} shows our results. We assume that $\nu_\tau$ are not produced in the sources, \ie, that $f_{\tau,{\rm S}} = 0$, as in the full-pion-decay and muon-damped scenarios, since that would require producing rare mesons in the sources, like $D_s^\pm$.
Using the 2015 IceCube measurements of flavor composition~\cite{Aartsen:2015knd}, the preferred value is $f_{e, {\rm S}} \simeq 0$, favoring muon-damped production, as was first reported in Ref.~\cite{Bustamante:2019sdb}.
To produce our 2020 and 2040 projections, we assume that the true flavor composition at Earth is that from the full pion decay chain (see Section~\ref{sec:telescopes}), and attempt to recover it.
Figure~\ref{fig:recon_feS_pdf} shows that, by 2040, using the projected sensitivity to flavor composition in 15 years IceCube plus 10 years of IceCube-Gen2, and the projected reduction in the uncertainty in mixing parameters, we should be able to recover the true value of $f_{e,{\rm S}}$, to within $2\%$ at 68\%~C.R., or within 21\% at 99.7\%~C.R.
By combining all of the available TeV--PeV neutrino telescopes in 2040, $f_{e,{\rm S}}$ could be measured to within 15\% at the 99.7\%~C.R.
The improvement in the precision of $f_{e,{\rm S}}$ is driven by larger sample size of the future neutrino telescopes as discussed in Section~\ref{sec:telescopes}.

%%%%%%%%%%%%%%%%%%%%%%%%%%%%%%

{\it Revealing multiple production mechanisms.---}
It is conceivable that the diffuse flux of high-energy astrophysical neutrinos is due to more than one population of sources and that each population generates neutrinos with a different flavor composition.
Alternatively, even if there is a single population of neutrino sources, each one could produce neutrinos via multiple mechanisms, each yielding its own flavor composition.
Given the expected improvements in the precision of the mixing parameters and flavor measurements, we study whether we can identify subdominant neutrino production mechanisms by measuring the flavor composition.

The left panel of Fig.~\ref{fig:subcomponent_fractions} shows the 2040 projected sensitivity to the fractions of the diffuse flux that can be attributed to each of the three benchmark production scenarios: full pion decay ($k_\pi$), muon-damped ($k_\mu$), and neutron decay ($k_n$), where $k_\pi + k_\mu + k_n = 1$.
The flavor composition at the source combining all these three contributions is $\pmb{f}_{\rm S}=k_\pi \pmb{f}_{\rm S}^\pi+k_\mu \pmb{f}_{\rm S}^\mu+k_n \pmb{f}_{\rm S}^n$.
To produce Fig.~\ref{fig:subcomponent_fractions}, we assume that $k_\pi = 1$, and compute how well we can recover that value, given the projected combined  sensitivity $\mathcal{L}(\pmb{f}_{\oplus})$ of all the neutrino telescopes, and the projected combined likelihood $\mathcal{L}(\pmb{\theta})$ of all the oscillation experiments. 
The posterior probability of the fractions $\pmb{k}=(k_\pi,k_\mu,k_n)$ at the source is 
\begin{equation}
    \mathcal{P}(\pmb{k})
    =
    \int d\pmb{\vartheta}
    \mathcal{L}(\pmb{\vartheta})
    \mathcal{L}(\pmb{f}_\oplus(\pmb{f}_{\rm S}(\pmb{k}), \pmb{\vartheta}))\pi(\pmb{\vartheta})\pi(\pmb{k}) \;,
\end{equation}
where $\pi(\pmb{k})$ is a uniform prior in $\pmb{k}$.

The left panel of Fig.~\ref{fig:subcomponent_fractions} shows that, while the ``true'' value of $k_\pi = 1$ is within the favored region, lower values of $k_\pi$ are also allowed, with the same significance, at the cost of increasing the contribution of muon-damped and neutron-decay production.
The value of $k_\pi$ is anti-correlated with the values of $k_\mu$ and $k_n$:
lowering the contribution of pion-decay production to $k_\pi < 1$ decreases $f_{e,{\rm S}}$ and $f_{\mu,{\rm S}}$, but the former is compensated by the correlated increase in $k_n$ and the latter, by the correlated increase in $k_\mu$.
Remarkably, the contribution of neutron-decay production cannot be larger than 40\%.

In some astrophysical sources, especially the ones that do not accelerate hadrons past PeV energies, the production of TeV--PeV neutrinos via neutron decay might be strongly suppressed, since beta decay yields neutrinos of lower energy than pion decay.
Below we explore the sensitivity to $\pmb{k}$ in the limit of no neutrino production via neutron decay.

The right panel of Fig.~\ref{fig:subcomponent_fractions} shows our results if we restrict production to only the pion decay and muon-damped scenarios, \ie, to $k_\pi$ and $k_\mu=1-k_\pi$. 
At present, using the 2015 IceCube measurements of flavor composition~\cite{Aartsen:2015knd} and the NuFit~5.0 measurements of mixing parameters~\cite{Esteban:2020cvm,nufit5.0}, the entire range of $k_\pi$ is allowed even at 68\%~C.R.  By 2040, the constraints are significantly stronger: $k_\pi$ can be measured to within 5\% at the 68\%~C.R. and to within 20\% at the 99.7\%~C.R.

In practice, searches for the neutrino production mechanism will use not only the flavor composition but also the energy spectrum.
In the muon-damped scenario, the synchrotron losses of the muons would leave features in the energy spectrum that are not expected in the full pion decay scenario, and which may indicate the strength of the magnetic field of the sources~\cite{Winter:2013cla,Bustamante:2020bxp}.
Presently, there is little sensitivity to these features in the energy spectrum~\cite{Bustamante:2020bxp}, but improved future sensitivity may help break the degeneracy between $k_\pi$ and $k_\mu$.

%%%%%%%%%%%%%%%%%%%%%%%%%%%%%%%%%%%%%%%%%%%%%%%%%%%%%%%%

\subsection{Testing new neutrino physics: neutrino decay}

%%%%%%%%%%%%%%%%%%%%%%%%%%%%%%
\begin{figure}[t!]
  \centering
  \includegraphics[trim=0 0.5cm 0 0, clip, width=0.49\textwidth]{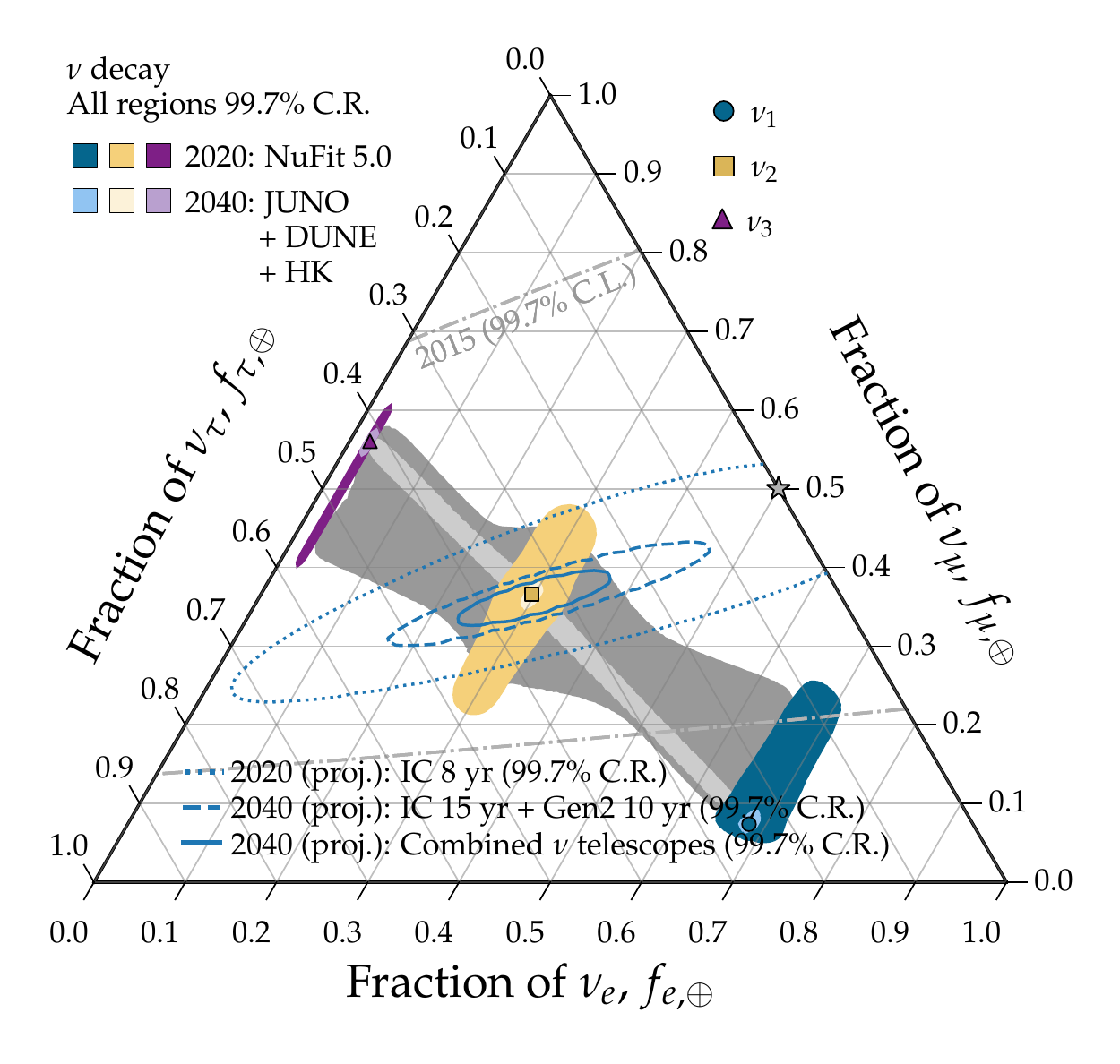}
  \caption{{Comparison of the regions of flavor content $\vert U_{\alpha i} \vert^2$ ($\alpha = e, \mu, \tau$) of the neutrino mass eigenstates $\nu_1$ (blue), $\nu_2$ (mustard) and $\nu_3$ (aubergine) in 2020 and 2040.  The overlaid contours denote the sensitivity to flavor measurement, as in \Fig\ \ref{fig:triangle_sm_2020_vs_2040}.  If all but one eigenstate decays completely while propagating to Earth, the allowed flavor composition at Earth matches the flavor content of the one remaining eigenstate.  Otherwise, the flavor composition is a combination (gray) of the flavor contents of the surviving $\nu_1$, $\nu_2$, and $\nu_3$.}}
  \label{fig:triangle_decay_2020_vs_2040}
\end{figure}
%%%%%%%%%%%%%%%%%%%%%%%%%%%%%%

Figure~\ref{fig:triangle_decay_2020_vs_2040} shows that, by 2040, the higher precision to which we will know the mixing parameters will also allow us to perform more precise tests of new physics, which we illustrate by considering the case neutrino decay (see Section~\ref{subsection:decay})~\cite{Beacom:2002vi,Meloni:2006gv,Maltoni:2008jr,Baerwald:2012kc,Pakvasa:2012db,Pagliaroli:2015rca,Huang:2015flc,Bustamante:2016ciw,Denton:2018aml,Bustamante:2020niz,Abdullahi:2020rge}.
The flavor contents $\vert U_{\alpha i} \vert^2$ of the mass eigenstates $\nu_i$ are required to compute the flavor composition at the Earth under decay, Eq.~\eqref{equ:flavor_ratio_earth_decay}.
Figure~\ref{fig:triangle_decay_2020_vs_2040} shows the uncertainty in them, in 2020 and 2040.
If all the eigenstates but one decay completely en route to Earth, the allowed flavor composition at Earth matches the flavor content of the one remaining eigenstate.
If multiple eigenstates survive, the flavor composition is a combination of the flavor contents of the surviving eigenstates.
Figure~\ref{fig:triangle_decay_2020_vs_2040} shows the allowed region of flavor composition that results from all possible combinations $k_1 \vert U_{\alpha 1} \vert^2 + k_2 \vert U_{\alpha 2} \vert^2 + k_3 \vert U_{\alpha 3} \vert^2$, where $k_1 + k_2 + k_3 = 1$ and each $k_i \in [0,1]$.
Reference~\cite{Bustamante:2015waa} showed an earlier version this region, generated using the 2015 uncertainties of the mixing parameters from Ref.~\cite{Gonzalez-Garcia:2014bfa}.

Under the assumption that $\nu_2$ and $\nu_3$ decay into invisible products with the same decay rate $m/\tau$ (see Section~\ref{subsection:decay}), we estimate upper limits on their common decay rate,  or, equivalently, lower limits on their common lifetime, for the years 2020 with IceCube 2015 measurement or with projected 8 year IceCube data, and 2040 using IceCube data or the flavor measurement at all future neutrino telescopes.
To do this, we compare the expected flavor composition at Earth computed for different values of the decay rate to a ``no decay'' scenario, where the flavor composition is computed under standard oscillations under different choices of the flavor composition at the source.
We use the likelihood of the mixing parameters, $\mathcal{L}(\pmb{\vartheta})$, and the likelihood of flavor measurements in neutrino telescope, $\mathcal{L}(\pmb{f}_\oplus)$, to translate any decay-induced deviation of $\pmb{f}_\oplus$ away from the ``no decay'' scenario into a bound on the decay rate.
The posterior probability of the decay rate $m/\tau$ is 
\begin{equation}
    \mathcal{P}\left(\frac{m}{\tau}\right)
    =
    \int d\pmb{\vartheta}
    \mathcal{L}(\pmb{\vartheta})
    \mathcal{L}\left(\pmb{f}_\oplus\left(\frac{m}{\tau}, \pmb{\vartheta}\right)\right)
    \pi(\pmb{\vartheta})
    \pi\left(\frac{m}{\tau}\right) \;,
\end{equation}
where $\pi(m/\tau)$ is a uniform prior on the decay rate and the flavor composition at Earth is computed following Eq.~\eqref{eq:decayflavor}.

%%%%%%%%%%%%%%%%%%%%%%%%%%%%%%
\begin{figure*}[t!]
 \centering
 \includegraphics[width=1.0\columnwidth]{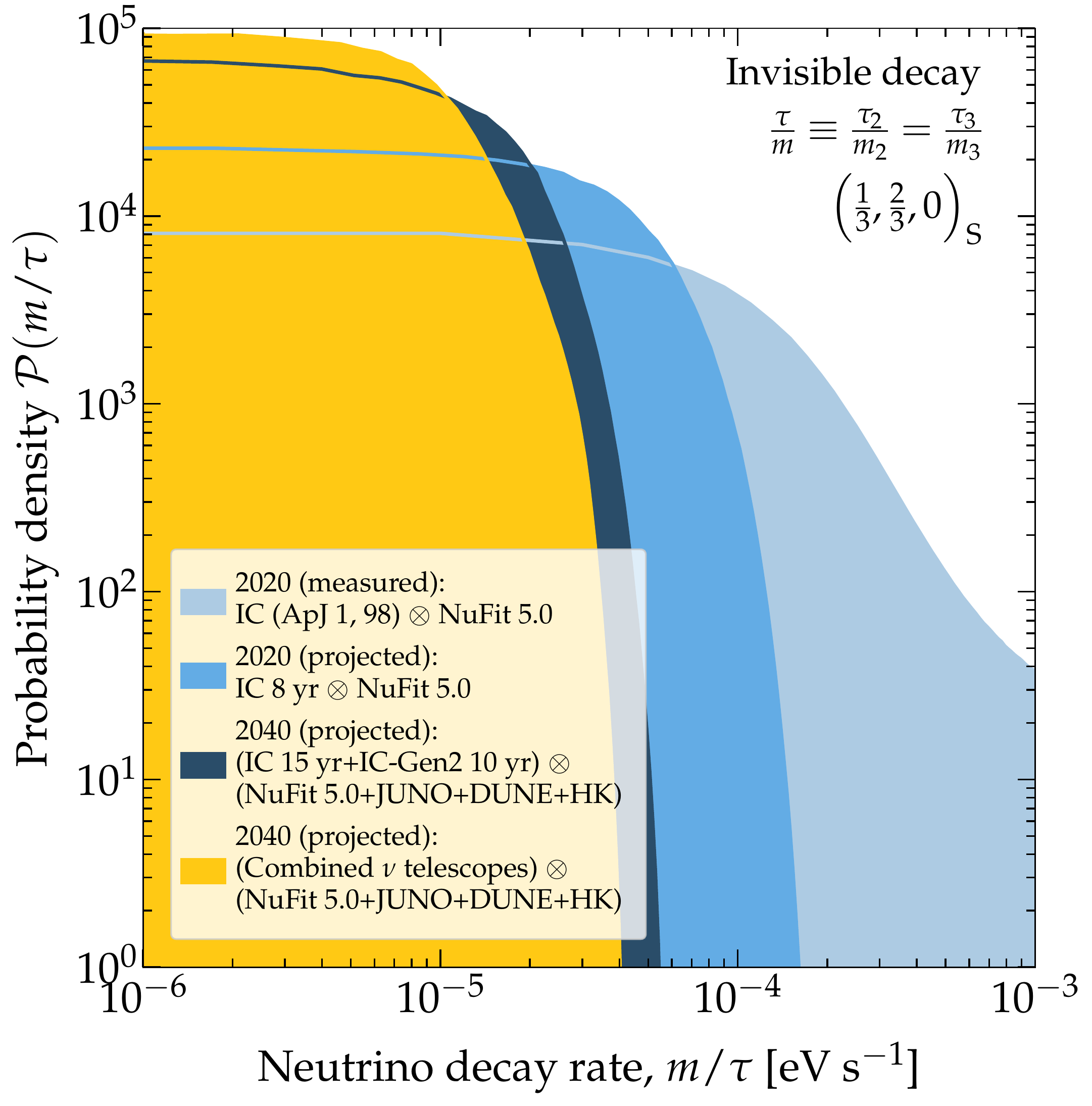}
 \includegraphics[width=1.0\columnwidth]{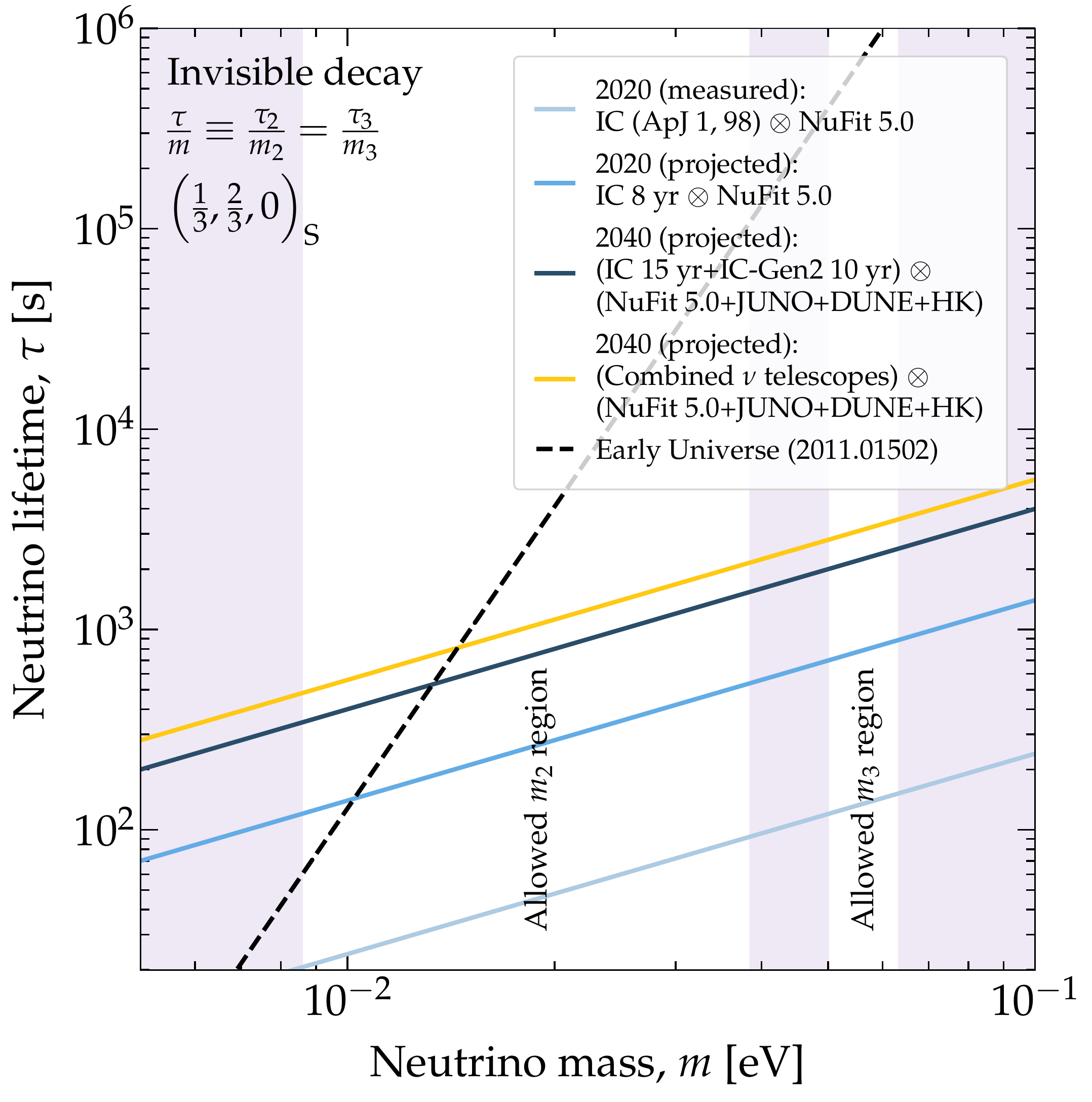}
 \caption{{\it Left:} Posterior probability density of the neutrino lifetime $\tau/m \equiv \tau_2/m_2 = \tau_3/m_3$, extracted assuming invisible decay, and a fixed flavor composition at the source of $(\frac{1}{3}, \frac{2}{3}, 0)_{\rm S}$, with the diffuse flux calculated as described in Section~\ref{subsection:decay}.  {\it Right:}  Comparison of the estimated 95\%~C.R.~lower limits on the lifetime derived here.  Low masses are excluded by the measurement of $\Delta m_{i1}^2 \equiv m_i^2 - m_1^2$ ($i=2,3$) in oscillation experiments\ \cite{deSalas:2020pgw, nufit5.0}; high masses, by cosmological limits on the sum of neutrino masses~\cite{Stocker:2020nsx} (see Ref. \cite{Bustamante:2016ciw} for details), where we have assumed normal ordering. The early-Universe limit of $\tau > 4 \times 10^5~{\rm s}~(m/50~{\rm meV})^5$~\cite{Barenboim:2020vrr} is from CMB and LSS constraints.}
 \label{fig:decay1d}
\end{figure*}
%%%%%%%%%%%%%%%%%%%%%%%%%%%%%%

The left panel of Fig.~\ref{fig:decay1d} shows the resulting posterior distributions computed assuming that the flavor composition at the source is $\pmb{f}_{\rm S} = \pmb{f}_{\rm S}^\pi \equiv \left(\frac{1}{3}:\frac{2}{3}:0\right)_{\rm S}$.
The posteriors reach their peak as $m/\tau \rightarrow 0$, favoring longer lifetimes; we thus place upper limits on the decay rates.
These become more constraining over time, as $\mathcal{L}(\pmb{\vartheta})$ and $\mathcal{L}(\pmb{f}_\oplus)$ become narrower.
They translate into lower limits on the lifetimes of $\tau/m \geq 2.4 \times 10^{3} (\mathrm{eV}/m)$~s, using 2015 data, to $5.6\times 10^{5} (\mathrm{eV}/m)$~s in 2040.
The right panel of Fig.~\ref{fig:decay1d} shows the corresponding lower limits on the lifetime as a function of neutrino mass.
We have highlighted the allowed interval of masses assuming normal ordering by shading out the regions that are respectively disfavored for each of the mass eigenstates due to constraints on the mass splitting from oscillation experiments, and limits on the sum of the masses from by the latest global fit including cosmological observations and terrestrial experiments~\cite{Stocker:2020nsx}.

A realistic analysis needs to take into account the uncertainties on the flavor compositions. 
To this end, we explore two alternative choices of the flavor composition at the source: varying over all possible values of $\pmb{f}_{\rm S}$ (``$\pmb{f}_{\rm S}$ free''); and production via full pion decay, but allowing its contribution to the neutrino flux to vary below its nominal value of 100\%, with a half-Gaussian prior with a 10\% width (``$\pmb{f}_{\rm S}$ constr.'') and the rest of the flux comes from the muon damped scenario.

Table~\ref{tab:decayrates} shows the 95\%~C.R.~upper limits on the decay rate for the three cases.
In the most conservative case, ``$\pmb{f}_{\rm S}$ free,'' we see the same decay rate limit with 2020 and 2040 data, $m/\tau \simeq 2\times10^{-4}$.
This corresponds to a transition energy between fully-decay and no-decay at $E\simeq m/\tau H_0 \simeq 100$~TeV, close to the lower limit of our energy window.
For any smaller decay rates, only a small fraction (exponentially suppressed, see Eq.~\eqref{equ:prob_decay}) of neutrinos in the energy window would have decayed during the propagation, thus causing negligible changes to the flavor composition integrated over energy.
This leads to strong degeneracy between the flavor composition at the source and the decay rate.  
By choosing instead the ``$\pmb{f}_{\rm S}$ constr.'', the degeneracy is largely lifted.
This illustrates that any future bounds for neutrino decay will need to be carefully weighed against our understanding of the flavor composition at the source.
However, note that we only use the flavor information to test decay.
If there are indeed hints for neutrino decay, the measured energy spectrum will also provide crucial information~\cite{Denton:2018aml}.

%%%%%%%%%%%%%%%%%%%%%%%%%%%%%%
\begin{table*}[t!]
    \centering
    \begin{tabular}{c c c c c c}
         \hline \hline \\[-1.5ex]
         \multirow{2}{*}{\textbf{Year}} & \multirow{2}{*}{\textbf{Neutrino telescopes}} & \multirow{2}{*}{\textbf{Oscillation parameters}} & \multicolumn{3}{c}{\textbf{Upper limit $m/\tau$ [eV s$^{-1}$] (95\%~C.R.)}} \\[1ex] 
         & & & $\pmb{f}_{\rm S} = \left(\frac{1}{3},\frac{2}{3},0\right)$ & $\pmb{f}_{\rm S}$ \textbf{free} & $\pmb{f}_{\rm S}$ \textbf{constr.} \\[1ex] \hline \\[-1.5ex]
         {{2020 (measured)}}       & {{IC 2015}  }             & NuFit~5.0 (2020)                  & $4.1\times 10^{-4}$ &$4.4\times 10^{-2}$ & $6.4\times 10^{-3}$\\[1ex]
         2020 (projected)      & IC 8 yr            & NuFit~5.0 (2020)                   & $7.4\times 10^{-5}$ &$2.2\times 10^{-4}$ & $9.1\times 10^{-5}$\\[1ex]
         2040 (projected)      & IC 15 yr + IC-Gen2 10 yr & NuFit + JUNO + DUNE + HK & $2.5\times 10^{-5}$ &$2.4\times 10^{-4}$ & $5.3\times 10^{-5}$\\[1ex]
         2040 (projected)      & Combined $\nu$ telescopes     & NuFit + JUNO + DUNE + HK & $1.8\times 10^{-5}$ &$2.4\times 10^{-4}$ & $4.6\times 10^{-5}$ \\[1ex]
         \hline \hline
    \end{tabular}
    \caption{Estimated upper limits on the common decay rate of $\nu_2$ and $\nu_3$ into an invisible $\nu_1$, assuming a population of sources evolving in redshift with $m = 1.5$ (see Eq.~\eqref{eq:sourcerho}) and producing neutrinos assuming spectral index of $\gamma = 2.5$, via the full pion decay chain (fourth column, $\pmb{f}_{\rm S} = \left(\frac{1}{3},\frac{2}{3},0\right)$), and allowing the source flavor composition to vary freely (fifth column, $\pmb{f}_{\rm S}$ free). The last column assumes a pion decay fraction of 100\%, with 10\% (half) Gaussian uncertainty at the source, with the remaining neutrino flux from the muon-damped scenario.}
    \label{tab:decayrates}
\end{table*}
%%%%%%%%%%%%%%%%%%%%%%%%%%%%%%

The limits that we find are for the case of invisible decays and are, therefore, more conservative than the case of visible decay.
For visible decays~\cite{Beacom:2002vi,Bustamante:2016ciw, Abdullahi:2020rge}, the heavier mass eigenstates decay into the lightest one and can still be detected in neutrino telescopes.
In the normal mass ordering, where $\nu_1$ is the lightest neutrino, visible decay leads to a larger surviving fraction of $\nu_1$, moving the flavor composition further away from the flavor composition expected from full pion decay, and potentially strengthening the limits on the decay rate.  
However, by 2040, and assuming that the measured flavor composition is centered on $\pmb{f}_\oplus^\pi$---as in the projected measurement contours in Fig.~\ref{fig:triangle_decay_2020_vs_2040}---then only decays that leave $\nu_2$ as the dominant surviving neutrino in the flux will still be allowed.
For a detailed treatment of the nuances of visible decay, see Ref.~\cite{Abdullahi:2020rge}.
 
The right panel of Fig.~\ref{fig:decay1d} shows that our lower limits on the neutrino lifetime are far from the lower limit stemming from early-Universe constraints~\cite{Barenboim:2020vrr}.  Although those limits assume a scalar-mediated decay from heavier to lighter mass eigenstates, decays to completely invisible products should not produce appreciably weaker bounds~\cite{Escudero:2019gfk} owing to the self-interactions induced by such a new mediator. Our limits are independent of early-Universe cosmology and are thus not susceptible to modifications to $\Lambda$CDM nucleosynthesis or recombination. For example, models in which a late-time phase transition leads to neutrino decay \cite{Dvali:2016uhn,Funcke:2019grs} easily evade the cosmological limits, making our constraints dominant.

%%%%%%%%%%%%%%%%%%%%%%%%%%%%%%%%%%%%%%%%%%%%%%%%%%%%%%%%
%%%%%%%%%%%%%%%%%%%%%%%%%%%%%%%%%%%%%%%%%%%%%%%%%%%%%%%%

\section{Summary and Conclusions}
\label{sec:conclusions}

The flavor composition of TeV--PeV astrophysical neutrinos, \ie, the proportion of $\nu_e$, $\nu_\mu$, and $\nu_\tau$ in the neutrino flux, has long been regarded as a versatile tool to learn about high-energy astrophysics and test fundamental physics.
However, in practice, present-day uncertainties in the neutrino mixing parameters and in the measurement of flavor composition in neutrino telescopes limit its reach.
Fortunately, this situation will change over the next two decades, thanks to the significant progress that is expected from terrestrial neutrino experiments.
We have found that the full potential of flavor composition will finally be fulfilled over the next 20 years, thanks to a host of new neutrino oscillation experiments that will improve the precision of the mixing parameters using terrestrial neutrinos and neutrino telescopes that will improve the measurement of the flavor composition of high-energy neutrinos.

Regarding neutrino mixing parameters, by 2040, improved measurements of $\theta_{12}$ by JUNO~\cite{An:2015jdp} and of $\theta_{23}$ by DUNE~\cite{Abi:2020wmh} and Hyper-Kamiokande~\cite{Abe:2018uyc} will reduce the size of the allowed flavor regions at Earth predicted by standard oscillations by a factor of 5--10 compared to today.
Additionally, the IceCube-Upgrade, together with the previously mentioned experiments, will provide improved constraints on non-unitarity of the PMNS matrix.
This will clearly separate the flavor composition predicted by different neutrino production mechanisms, at a credibility level well in excess of 99.7\%, and will also sharpen the distinction between expectations from standard and nonstandard oscillations.

Regarding the measurement of the flavor composition of high-energy neutrinos, the deployment of new neutrino telescopes will increase the precision of the measurement of flavor composition thanks to the larger sample of high-energy neutrinos that they will detect.
Beyond the continuing operation of IceCube, Baikal-GVD~\cite{Safronov:2020dtw} and KM3NeT/ARCA should already be in operation by 2025~\cite{Adrian-Martinez:2016fdl},  P-ONE~\cite{Agostini:2020aar} and IceCube-Gen2~\cite{Aartsen:2020fgd} by 2030---at which point the combined effective volume of neutrino telescopes exceeds the present one by more than an order of magnitude---and TAMBO~\cite{Romero-Wolf:2020pzh,tambo_loi}, dedicated to measuring the $\nu_\tau$ flux.
From their combined measurements, the uncertainty in flavor composition is expected to shrink by a factor of 2 from 2020 to 2040.
Our projections are conservative: they rely mainly on statistical improvements due to larger exposures and the inclusion of TAMBO as a dedicated tau-neutrino experiment.
Any improvement in the methods use to reconstruct flavor, which we have not considered, will only improve the projections further.

Combining these two improvements, by 2040 we will be able to distinguish with high confidence between similar predictions of the flavor composition at Earth expected from different neutrino production mechanisms. 
Notably, we will be able to robustly differentiate the flavor composition expected from neutrino production due to full pion decay from the composition expected from muon-damped pion decay, the two most likely production scenarios.
The combined effect of smaller allowed flavor regions and more precise flavor measurements anticipate that progress in using flavor measurements to identify the still-unknown sources of the bulk of the high-energy diffuse neutrino flux will be not merely incremental but transformative.

Further, by 2040 we will be able to use the measured flavor composition, and our precise knowledge of the mixing parameters, to infer the flavor composition at the source with high precision. 
In particular, the average $\nu_e$ fraction at the source will be known to within 6\%, a marked improvement over the 42\% precision to which it is known today (see Table~\ref{tab:flavoruncertainty}).
Moreover, if high-energy neutrinos are produced by a variety of production mechanisms, each yielding a different flavor composition, we will be able to identify the dominant and sub-dominant mechanisms.
We find that if production via pion decay is the dominant mechanism, this constrains the contribution from production via neutron decay to be smaller than 40\%.
If production only via pion decay and muon-damped decay are allowed, the dominant production mechanism can be pinned down to less than 20\% at 99.7\% credible level.

The presence of new physics effects, specifically non-unitarity in the PMNS mixing matrix, only modestly affects the flavor triangle: by 2040, all three canonical source compositions will be distinguishable even in the presence of non-unitary mixing.

We explore neutrino decay into invisible products to illustrate the improvement that we will achieve in testing beyond-the-Standard-Model neutrino physics using the flavor composition.
Complete neutrino decay to $\nu_3$ or $\nu_1$ is strongly disfavored today, and will be excluded at more than 5$\sigma$ by 2040.
Under certain conservative assumptions, we have shown that future observations will be able to constrain the lifetime of the heavier neutrinos to nearly $\sim 10^{5} (\mathrm{eV}/m)$~s if only $\nu_1$ is stable.
This is nearly eight orders of magnitude stronger than the limits set by solar neutrino observations~\cite{Berryman:2014qha}, and competitive with bounds that could be obtained from observing a Galactic supernova~\cite{Bustamante:2016ciw}; however they are significantly weaker than the constraints for early universe observables~\cite{Barenboim:2020vrr}

Approximately fifty years have passed since the original proposal by Markov to build large detectors to observe high-energy neutrinos. The last ten years have brought us the discovery of the diffuse high-energy astrophysical neutrino flux by IceCube, the discovery of the potential first few astrophysical sources of high-energy neutrinos, and first measurements of the flavor composition.
We have shown that these efforts will come to dramatic fruition in the next two decades, yielding a more complete picture of the Universe as seen with high-energy neutrinos.
The future is bright for neutrino hunters. \\

%%%%%%%%%%%%%%%%%%%%%%%%%%%%%%%%%%%%%%%%%%%%%%%%%%%%%%%%
%%%%%%%%%%%%%%%%%%%%%%%%%%%%%%%%%%%%%%%%%%%%%%%%%%%%%%%%

\section*{Acknowledgements}

We thank Nikita Blinov, Francesco Capozzi, Gwenhael De Wasseige, Miguel Escudero, Kevin Kelly, Qinrui Liu, Thomas Schwetz, V\'ictor Valera, and Jakob van Santen for their valuable input.
MB is supported by the {\sc Villum Fonden} under project nos.~13164 and 29388.
SWL is supported by the U.S. Department of Energy under Contract No. DE-AC02- 76SF00515 and later by Fermi Research Alliance, LLC under contract DE-AC02-07CH11359 with the U.S. Department of Energy.
NS and ACV are supported by the Arthur B. McDonald Canadian Astroparticle Physics Research Institute,  with equipment funded by the Canada Foundation for Innovation and the Province of Ontario and housed at the Queen's Centre for Advanced Computing.
Research at Perimeter Institute is supported by the Government of Canada through the Department of Innovation, Science, and Economic Development, and by the Province of Ontario.
CA is additionally supported by the Faculty of Arts and Sciences of Harvard University.

The authors dedicate this paper to those we lost this year and in particular to the memory of \href{https://inspirehep.net/literature?sort=mostrecent&size=25&page=1&q=a\%20S.Pakvasa.1\%20and\%20ac\%201-\%3E9&ui-citation-summary=true}{Sandip Pakvasa}, an outstanding neutrino physicist who was the first to tackle many topics covered in this work~\cite{Learned:1994wg}.
Additionally, CA dedicates this work to the memory of Lilian Arg\"{u}elles, Rafael Com, and Enrique Delgado, who were loving, supporting, and always-inspiring family members.

%%%%%%%%%%%%%%%%%%%%%%%%%%%%%%%%%%%%%%%%%%%%%%%%%%%%%%%%
%%%%%%%%%%%%%%%%%%%%%%%%%%%%%%%%%%%%%%%%%%%%%%%%%%%%%%%%

%\pagebreak
\bibliographystyle{apsrev4-1}

\bibliography{references.bib}

%%%%%%%%%% supplemental materials %%%%%%%%%%
\pagebreak
\clearpage

\onecolumngrid
\appendix

%%%%%%%%%%%%%%%%%%%%%%%%%%%%%%%%%%%%%%%%%%%%%%%%%%%%%%%%
%%%%%%%%%%%%%%%%%%%%%%%%%%%%%%%%%%%%%%%%%%%%%%%%%%%%%%%%
\newpage

\renewcommand{\theequation}{A\arabic{equation}}
\clearpage

%%%%%%%%%%%%%%%%%%%%%%%%%%%%%%%%%%%%%%%%%%%%%%%%%%%%%%%%
%%%%%%%%%%%%%%%%%%%%%%%%%%%%%%%%%%%%%%%%%%%%%%%%%%%%%%%%

\section{Supplementary figures}
\label{sec:appendix}

Here we present supplementary figures to the ones discussed in the main text.
\begin{itemize}
\item[$\clubsuit$] Fig. \ref{fig:correlation} shows the correlation between the mixing parameter $\sin^2 \theta_{23}$ and the phase $\dcp$ from the NuFit 5.0, as well as projected correlations that we use for DUNE and HK. \\

\item[$\heartsuit$]  Fig. \ref{fig:triangle_sm_upper_io} Shows the allowed oscillation regions in the case of inverted ordering (IO), if any source flavor ratio is allowed, for various combinations of experimental results. Fig. \ref{fig:triangle_sm_upper_io_fixed} is the same, but highlights the allowed regions if the flavor composition at the source is fixed by pion decay, muon-damping, or neutron decay. \\ 

\item[$\spadesuit$] Figs.~\ref{fig:triangle_sm_upper_no} and  \ref{fig:triangle_sm_upper_no_fixed} are the same as Figs.~\ref{fig:triangle_sm_upper_io} and  \ref{fig:triangle_sm_upper_io_fixed}, respectively, but under normal ordering (NO). \\ 

\item[$\diamondsuit$] Figs.~\ref{fig:triangle_sm_upper_no} and \ref{fig:triangle_sm_lower_no_fixed} show a subset of these results for NO, but under the assumption that $\theta_{23}$ is in the lower octant. 

\end{itemize}

%%%%%%%%%%%%%%%%%%%%%%%%%%%%%%%%%%%%%%%%%%%%%%%%%%%%%%%%
%%%%%%%%%%%%%%%%%%%%%%%%%%%%%%%%%%%%%%%%%%%%%%%%%%%%%%%%

%%%%%%%%%%%%%%%%%%%%%%%%%%%%%%%%
\renewcommand{\thefigure}{A\arabic{figure}}
\setcounter{figure}{0}
\begin{figure}[t]
\centering
\includegraphics[width=0.32\textwidth]{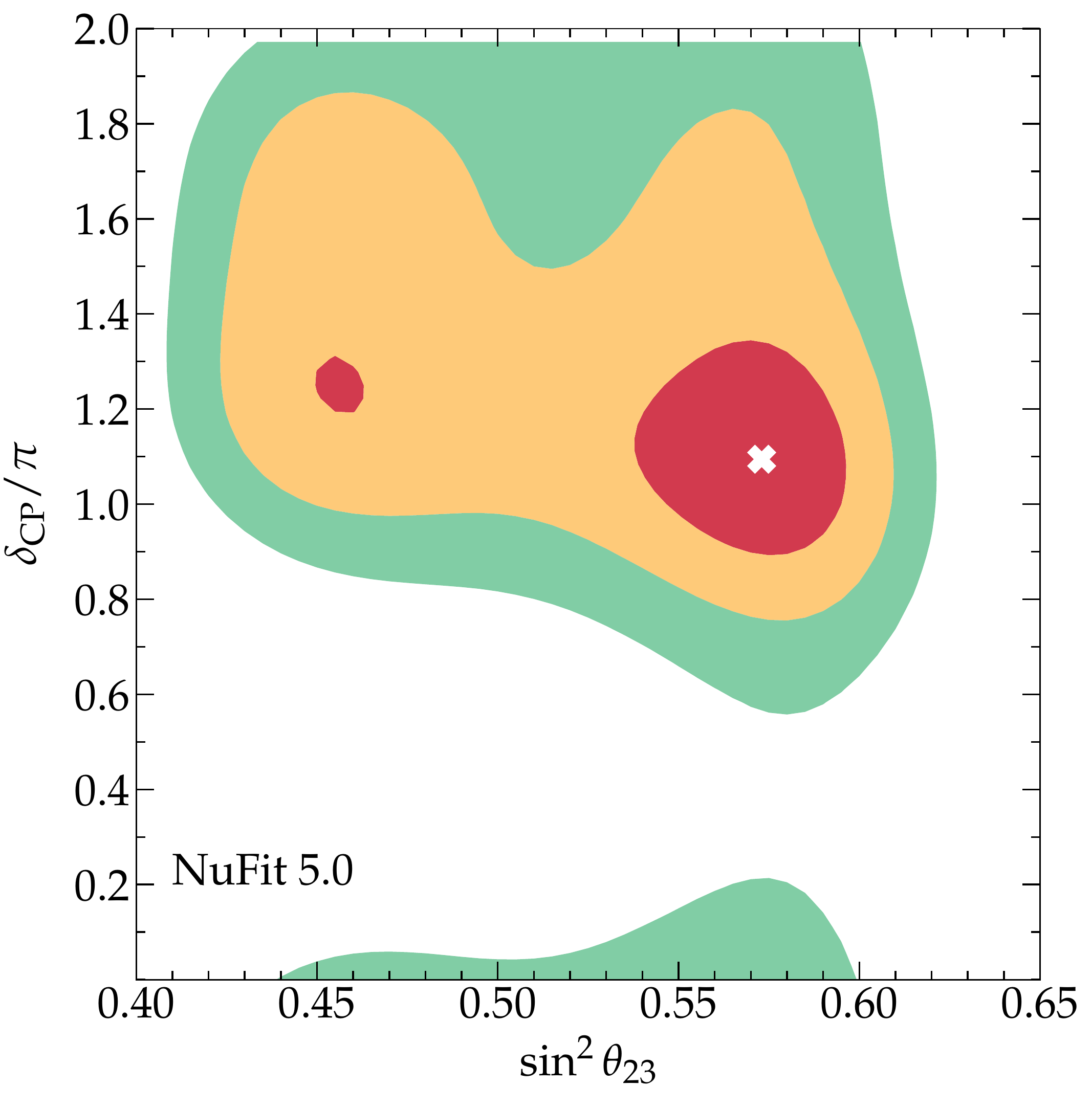}
\includegraphics[width=0.32\textwidth]{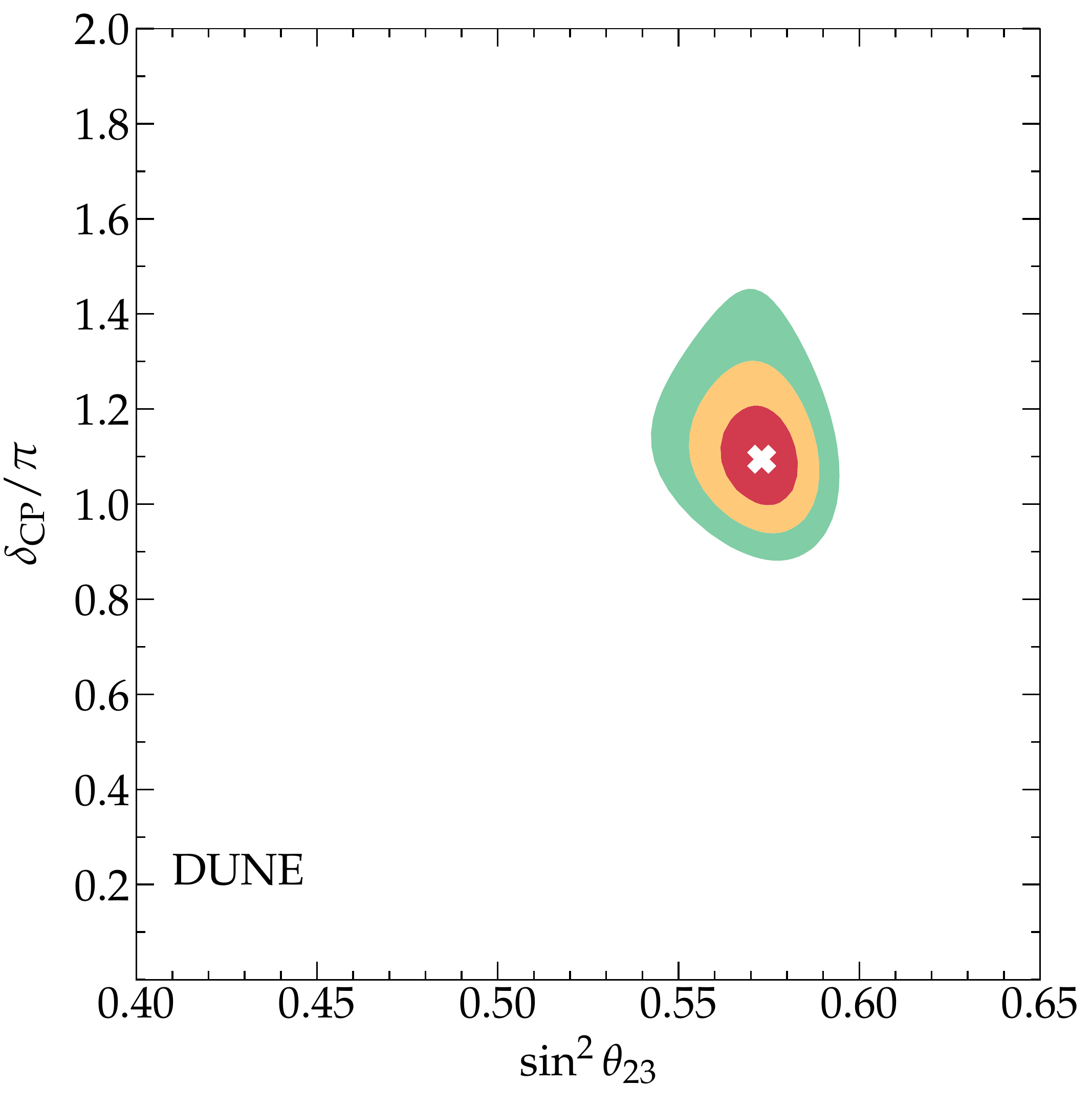}
\includegraphics[width=0.32\textwidth]{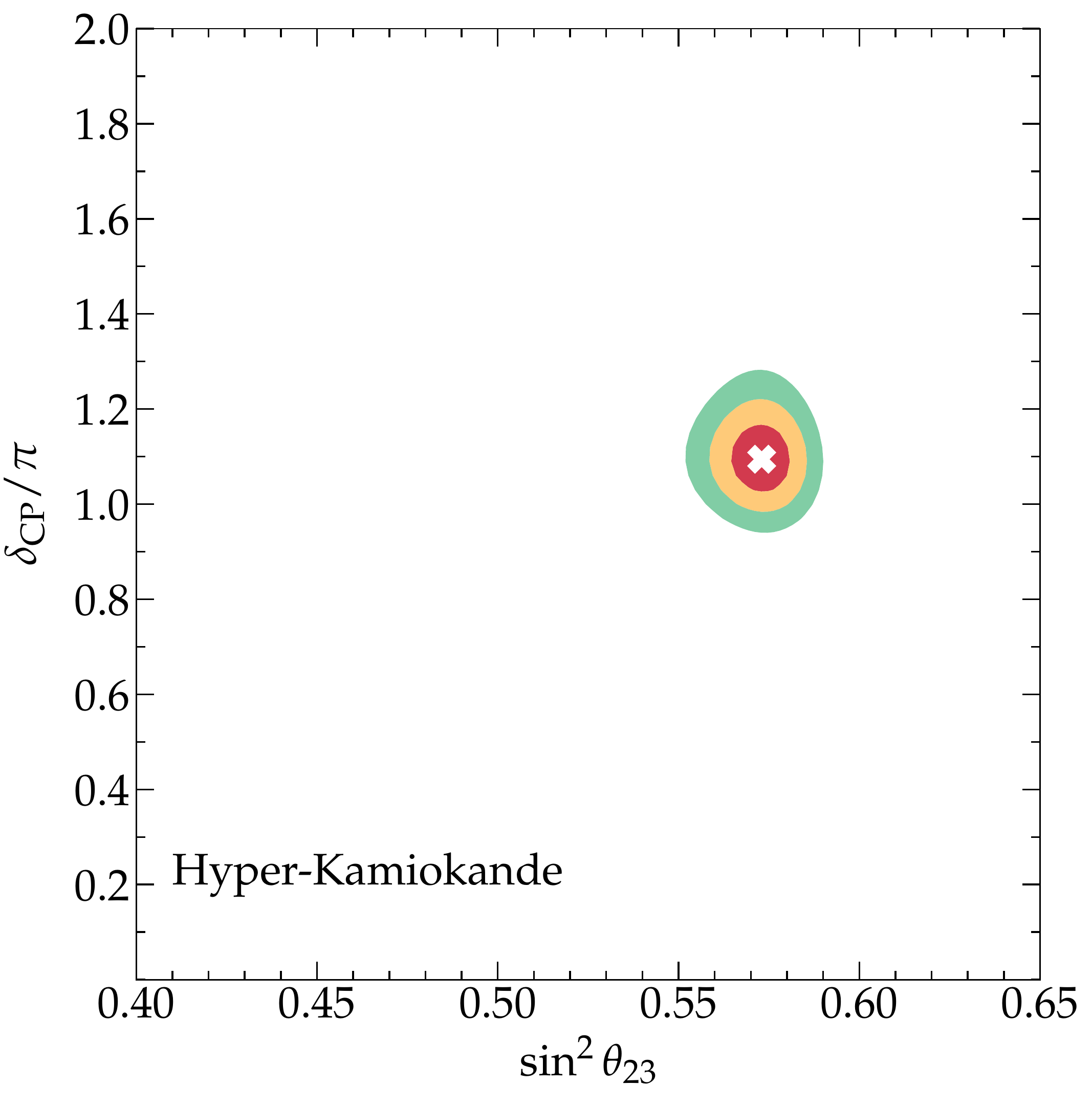}
%
% \internallinenumbers
\caption{Correlation between $\sin^2\theta_{23}$ and $\dcp$ from the NuFit~5.0 global oscillation fit and for projected sensitivities for DUNE and Hyper-Kamiokande (HK).  The shaded regions indicate $1\sigma$ (red), $2\sigma$ (orange), and $3\sigma$ (green) allowed regions of the parameters.}
\label{fig:correlation}
\end{figure}
%%%%%%%%%%%%%%%%%%%%%%%%%%%%%%

%%%%%%%%%%%%%%%%%%%%%%%%%%%%%%%%%%%%%%%%%%%%%%%%%%%%%%%%
%%%%%%%%%%%%%%%%%%%%%%%%%%%%%%%%%%%%%%%%%%%%%%%%%%%%%%%%

% SM, IO, upper th23 octant, all fS
\begin{figure*}
  \centering
  \includegraphics[trim=0 0.5cm 0 0, clip, width=0.49\textwidth]{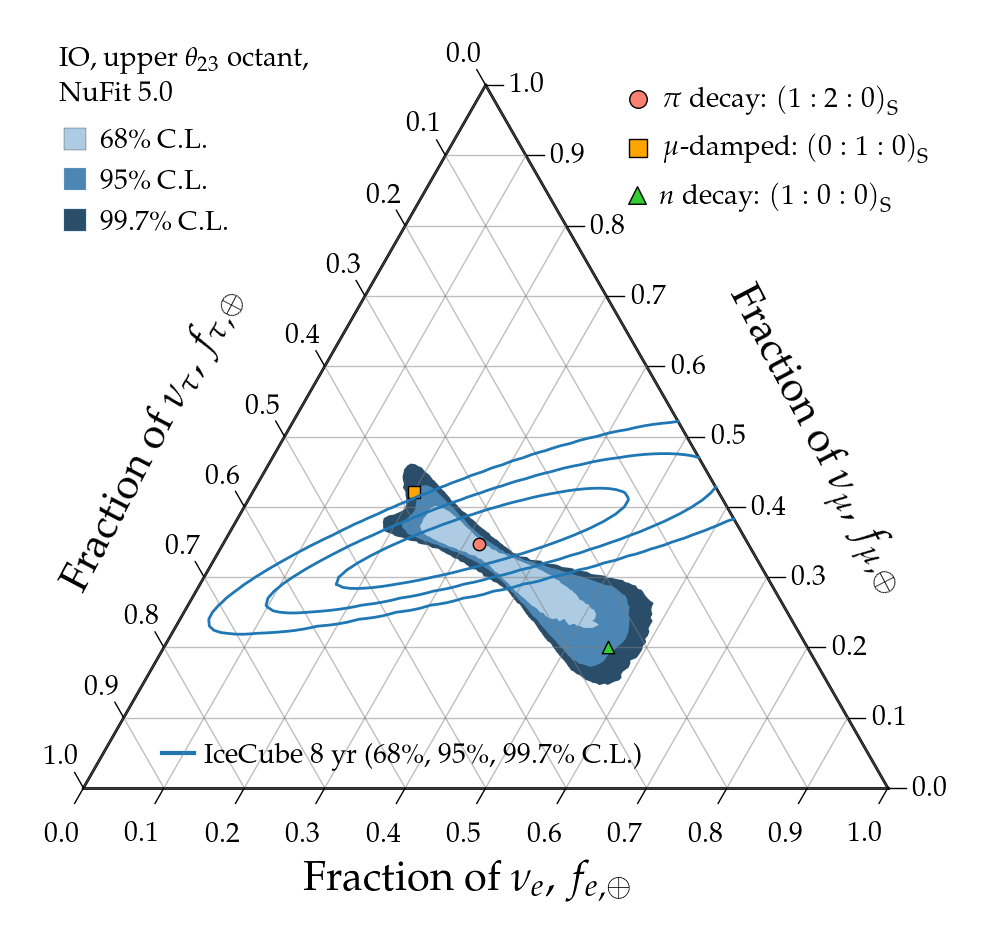}
  \includegraphics[trim=0 0.5cm 0 0, clip, width=0.49\textwidth]{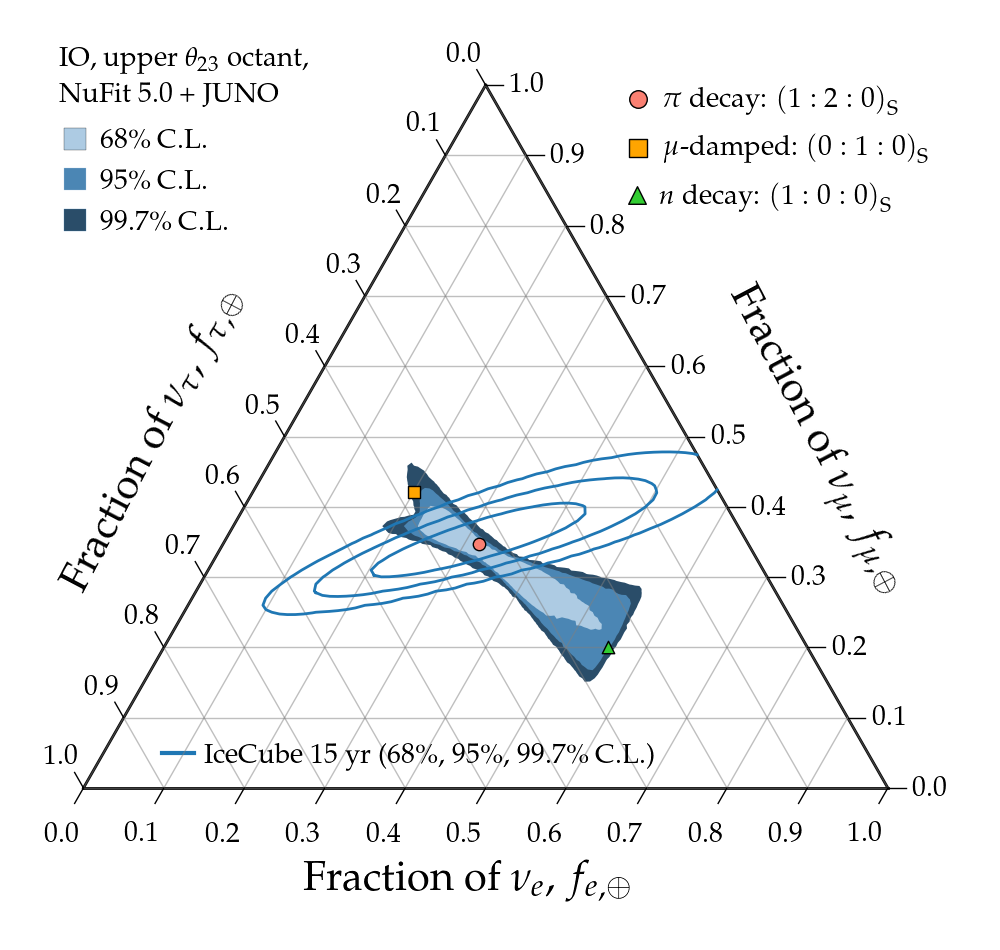}
  \includegraphics[trim=0 0.5cm 0 0, clip, width=0.49\textwidth]{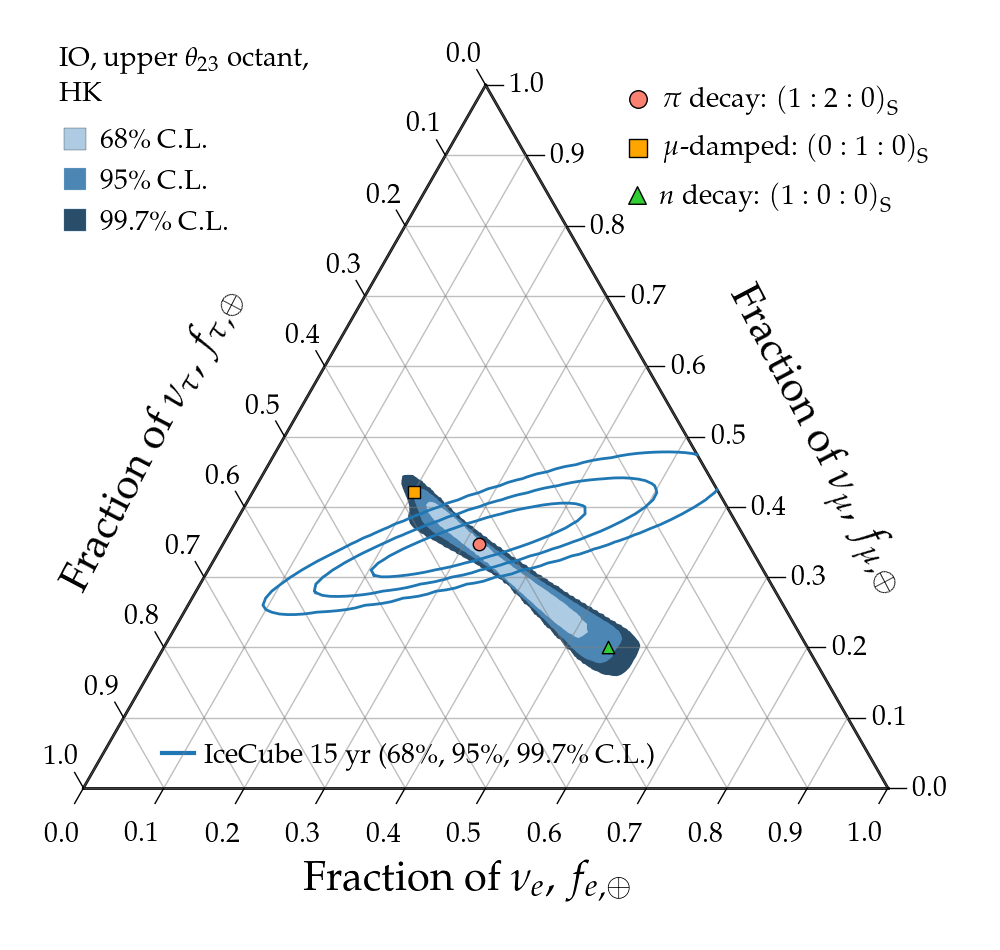}
  \includegraphics[trim=0 0.5cm 0 0, clip, width=0.49\textwidth]{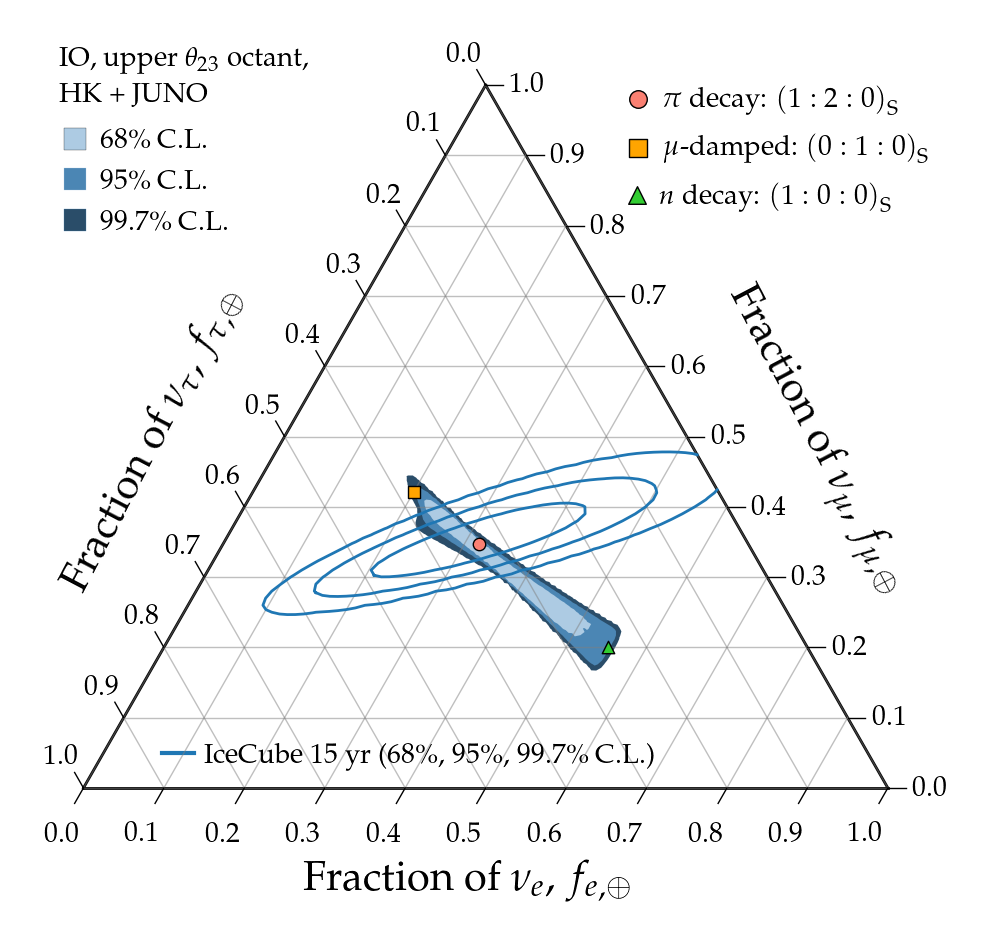}
  \includegraphics[trim=0 0.5cm 0 0, clip, width=0.49\textwidth]{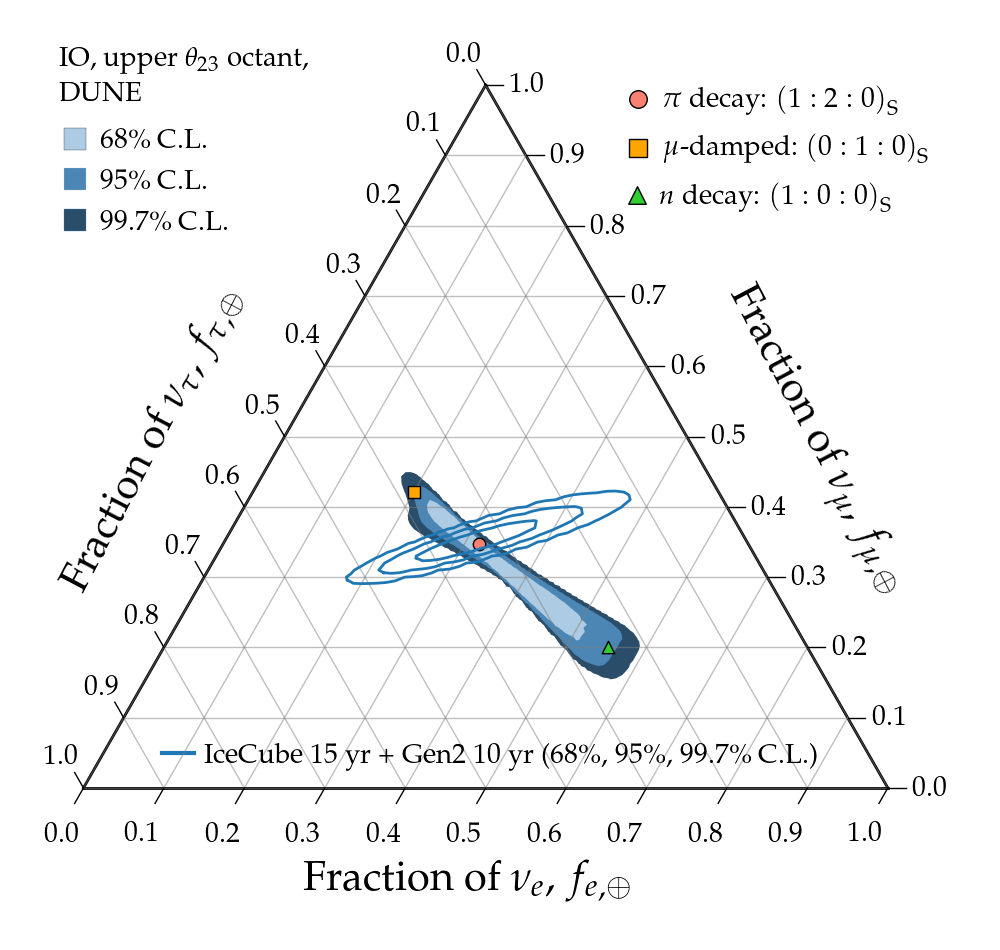}
  \includegraphics[trim=0 0.5cm 0 0, clip, width=0.49\textwidth]{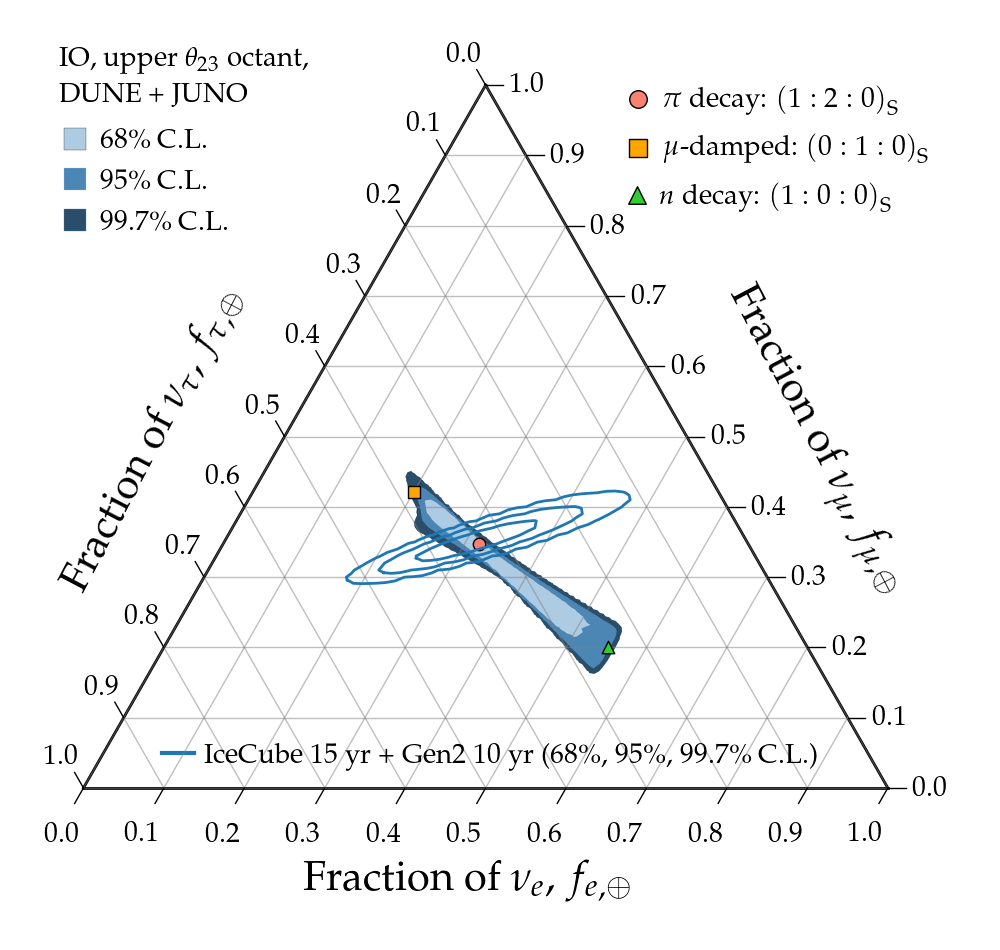}
%  \internallinenumbers
  \caption{Standard oscillation regions, varying over all flavor compositions $f_{\alpha, {\rm S}}$ at the source: inverted ordering (IO), upper $\theta_{23}$ octant.}
  \label{fig:triangle_sm_upper_io}
\end{figure*}

% SM, IO, upper th23 octant, fixed fS
\begin{figure*}
  \centering
  \includegraphics[trim=0 0.5cm 0 0, clip, width=0.49\textwidth]{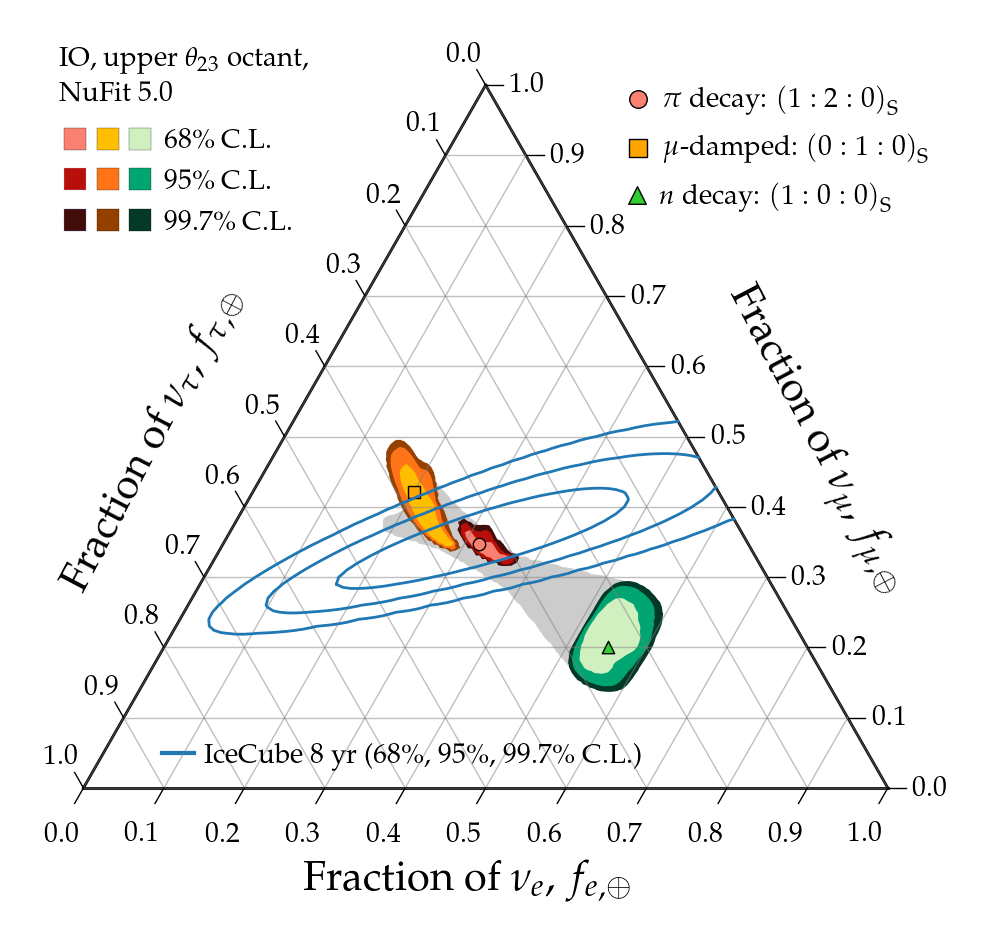}
  \includegraphics[trim=0 0.5cm 0 0, clip, width=0.49\textwidth]{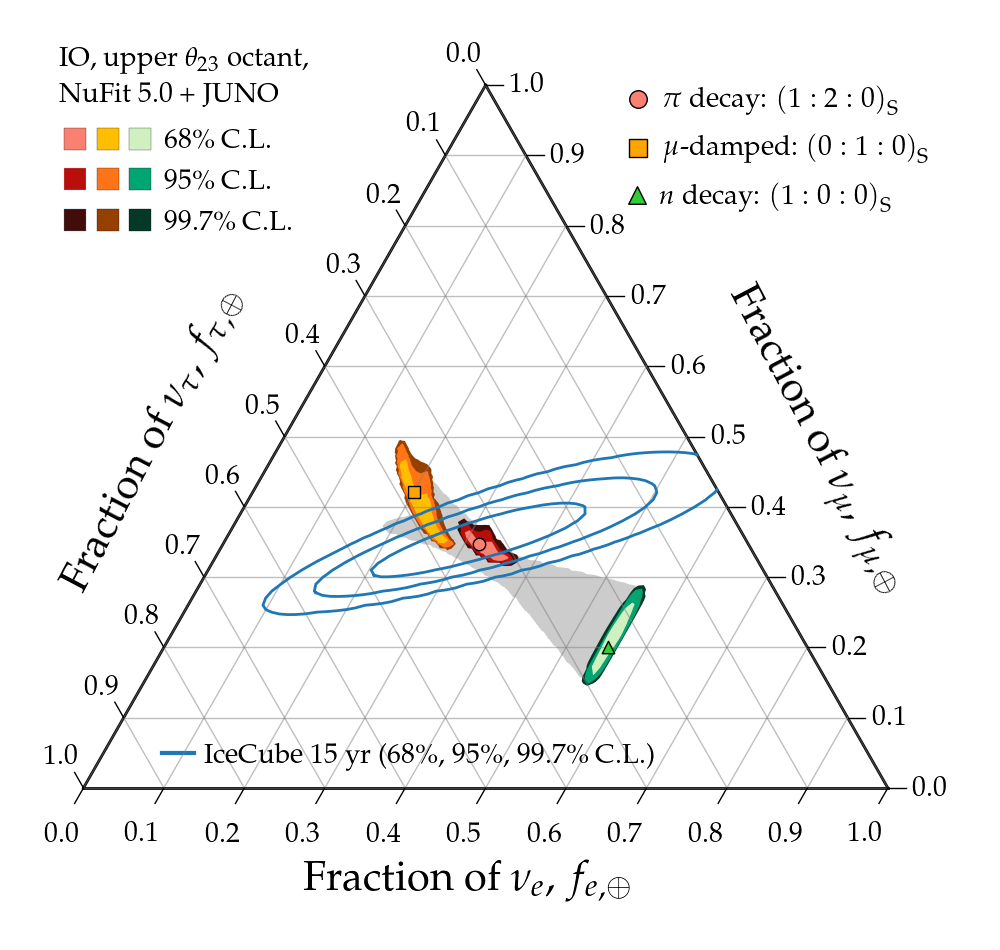}
  \includegraphics[trim=0 0.5cm 0 0, clip, width=0.49\textwidth]{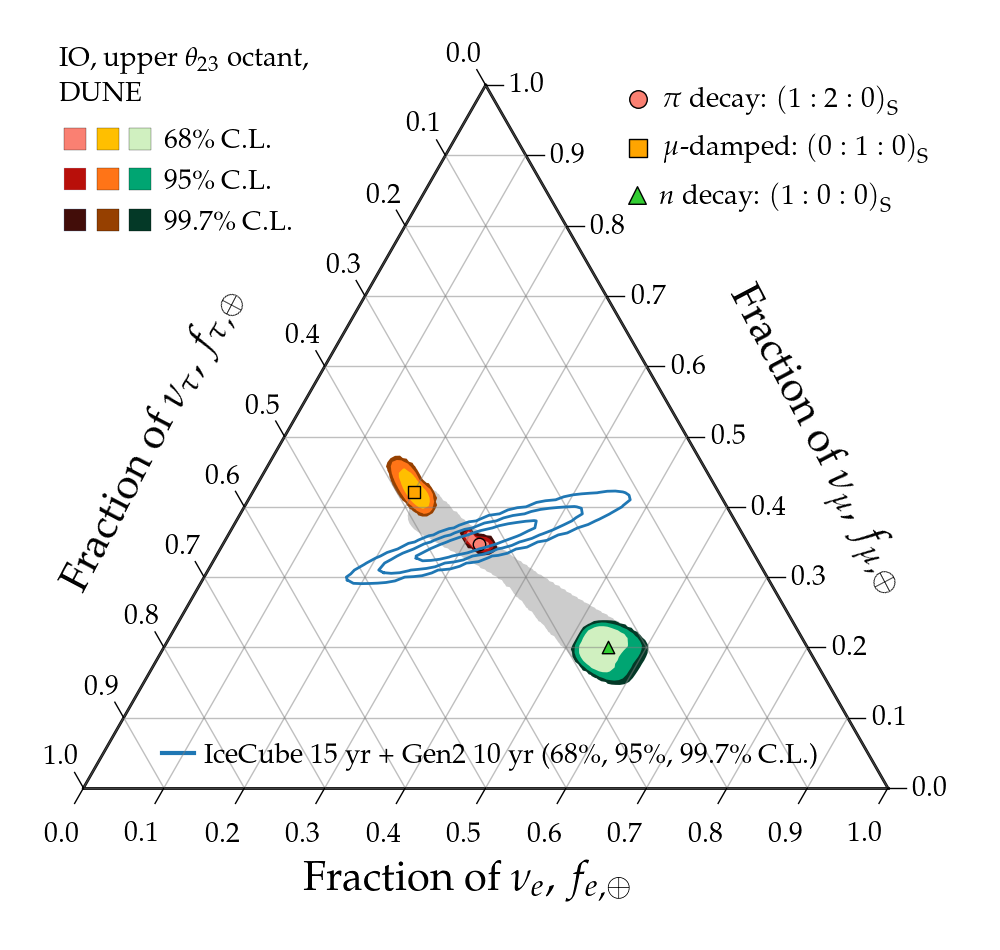}
  \includegraphics[trim=0 0.5cm 0 0, clip, width=0.49\textwidth]{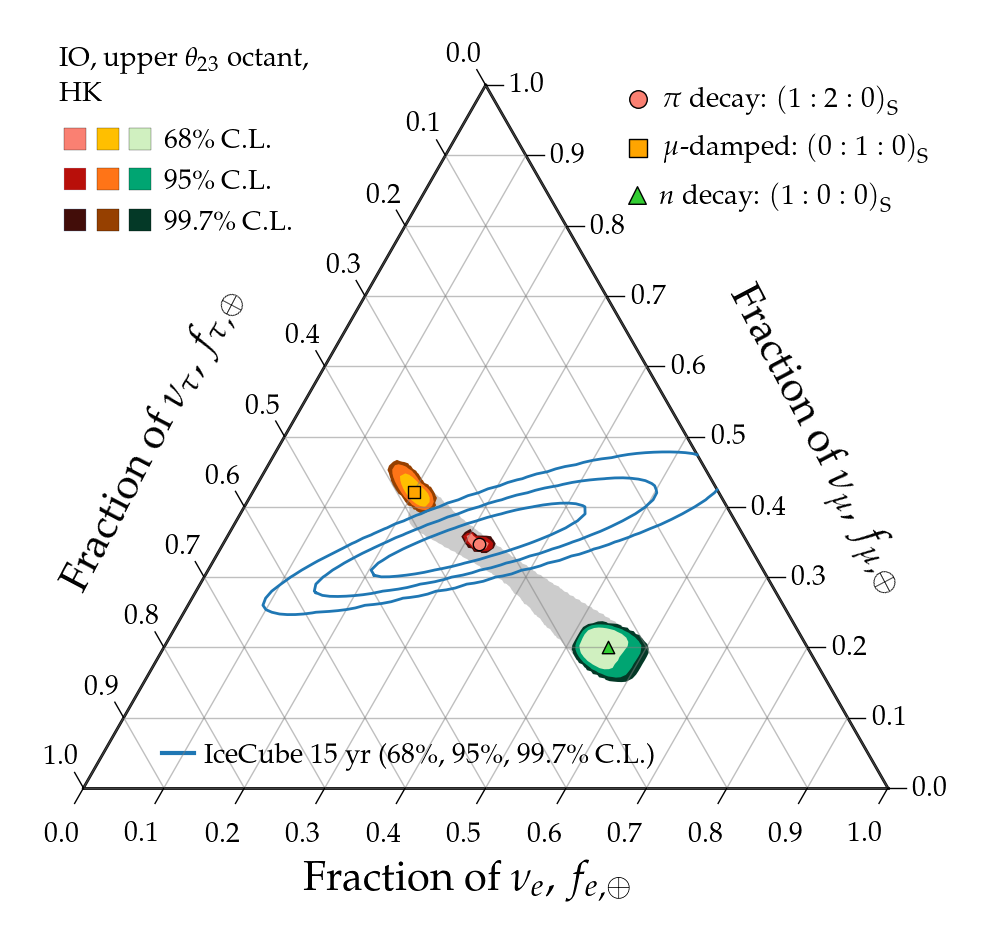}
  \includegraphics[trim=0 0.5cm 0 0, clip, width=0.49\textwidth]{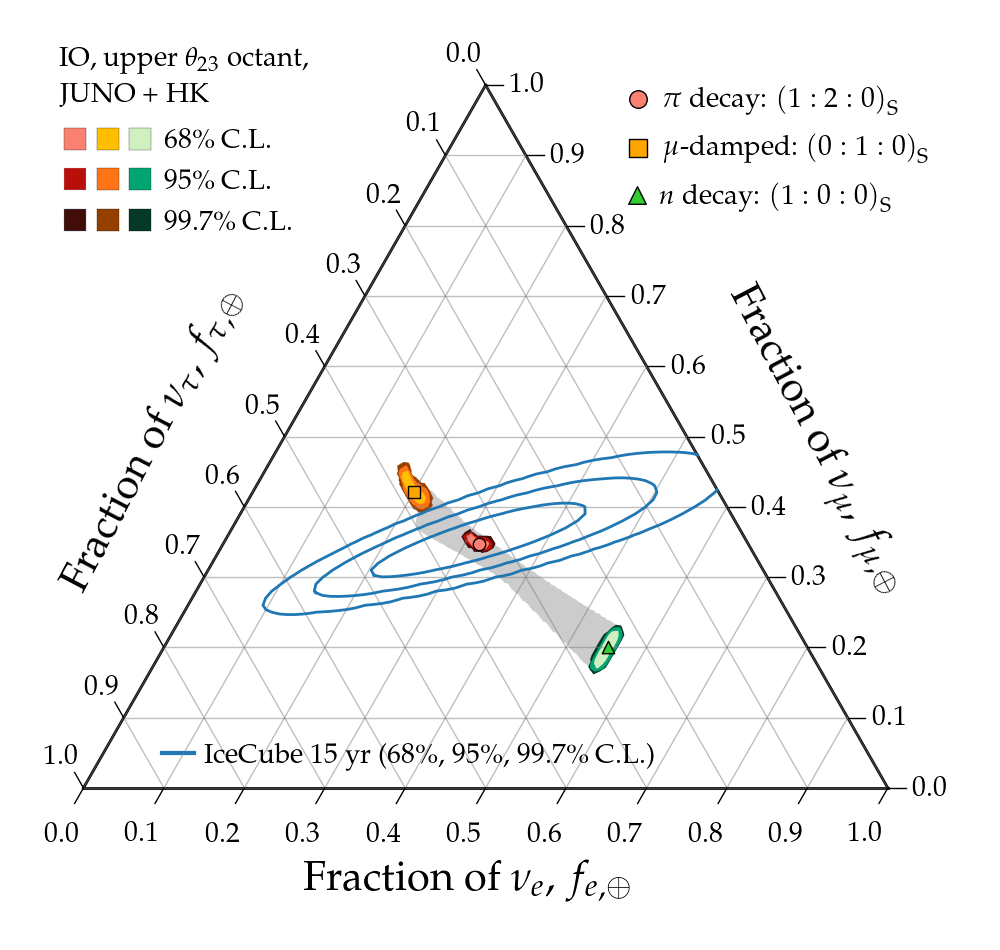}
  \includegraphics[trim=0 0.5cm 0 0, clip, width=0.49\textwidth]{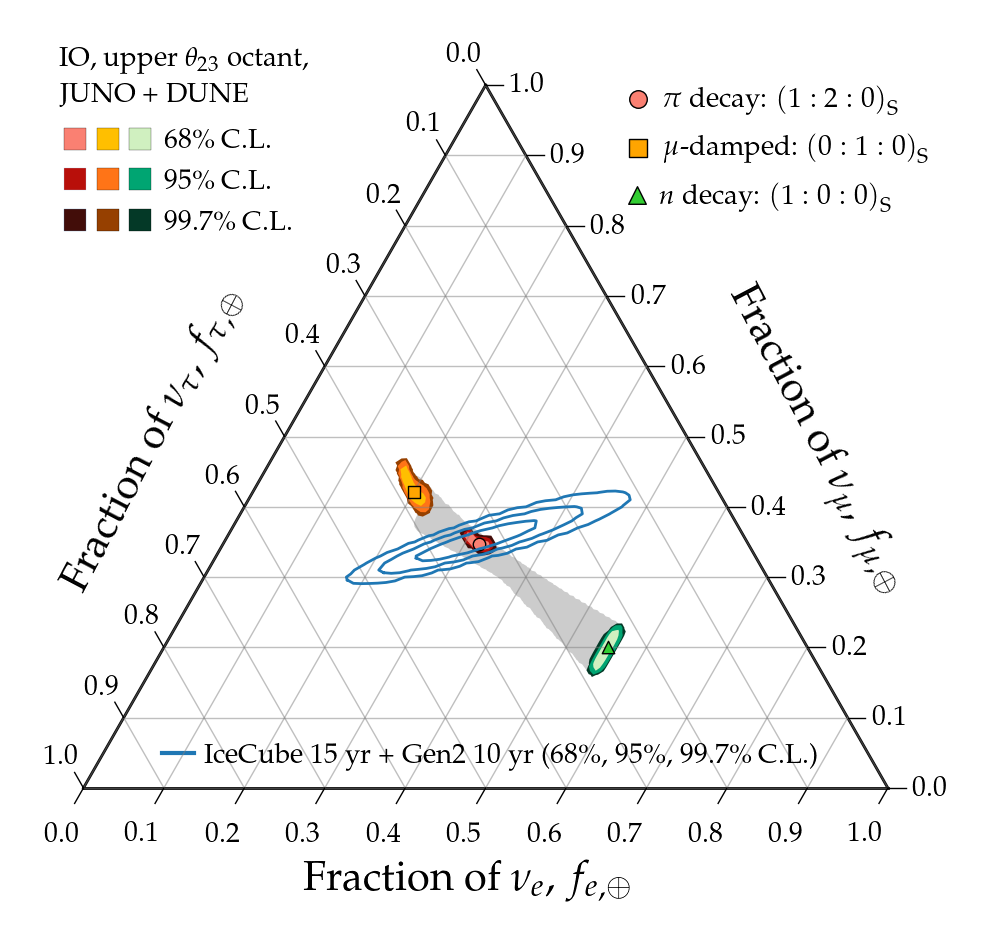}
%  \internallinenumbers
  \caption{Standard oscillation regions, for benchmark flavor compositions $f_{\alpha, {\rm S}}$ at the source: inverted ordering (IO), upper $\theta_{23}$ octant.}
  \label{fig:triangle_sm_upper_io_fixed}
\end{figure*}

%%%%%%%%%%%%%%%%%%%%%%%%%%%%%%%%%%%%%%%%%%%%%%%%%%%%%%%%
%%%%%%%%%%%%%%%%%%%%%%%%%%%%%%%%%%%%%%%%%%%%%%%%%%%%%%%%

% SM, NO, upper th23 octant, all fS
\begin{figure*}
  \centering
  \includegraphics[trim=0 0.5cm 0 0, clip, width=0.49\textwidth]{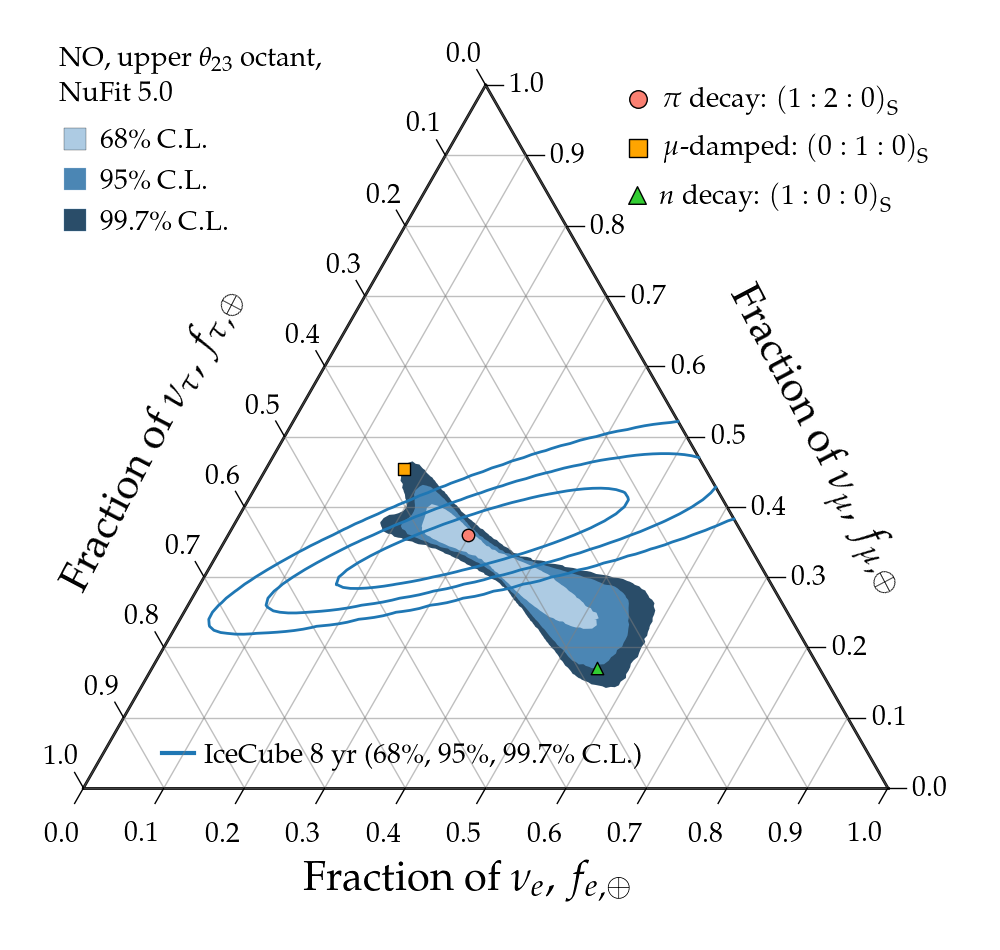}
  \includegraphics[trim=0 0.5cm 0 0, clip, width=0.49\textwidth]{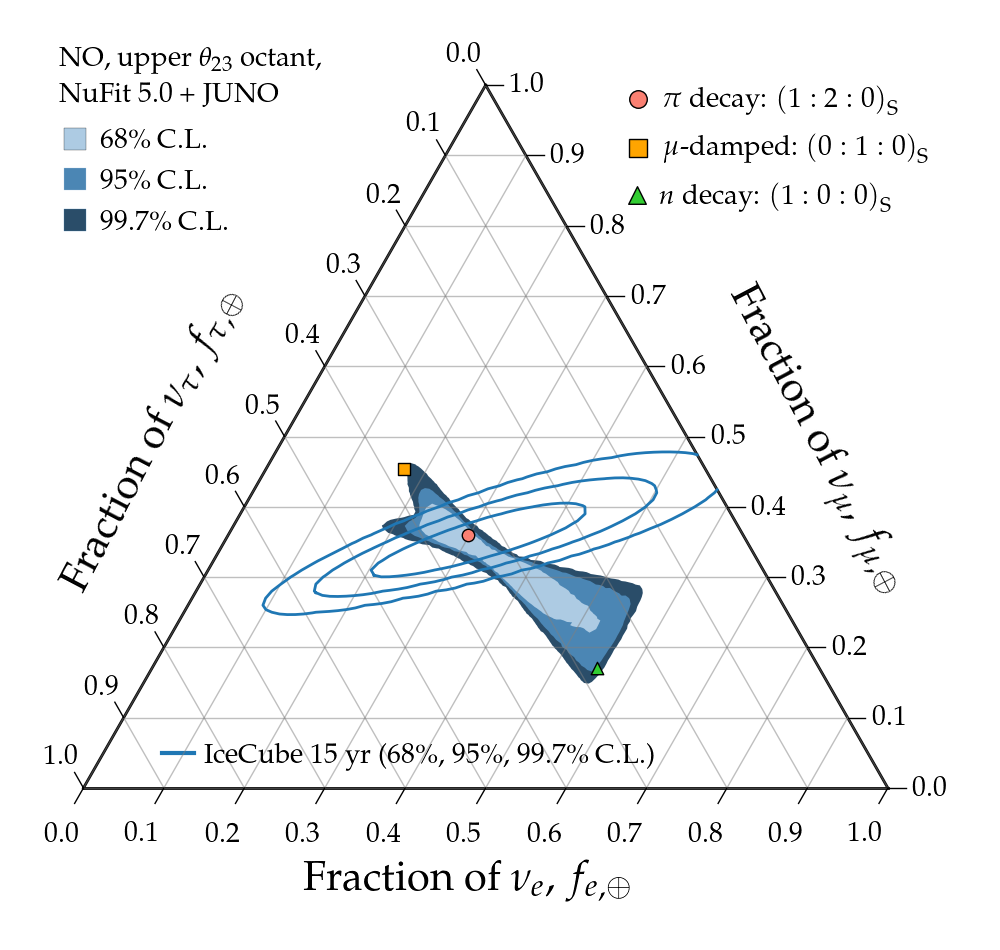}
  \includegraphics[trim=0 0.5cm 0 0, clip, width=0.49\textwidth]{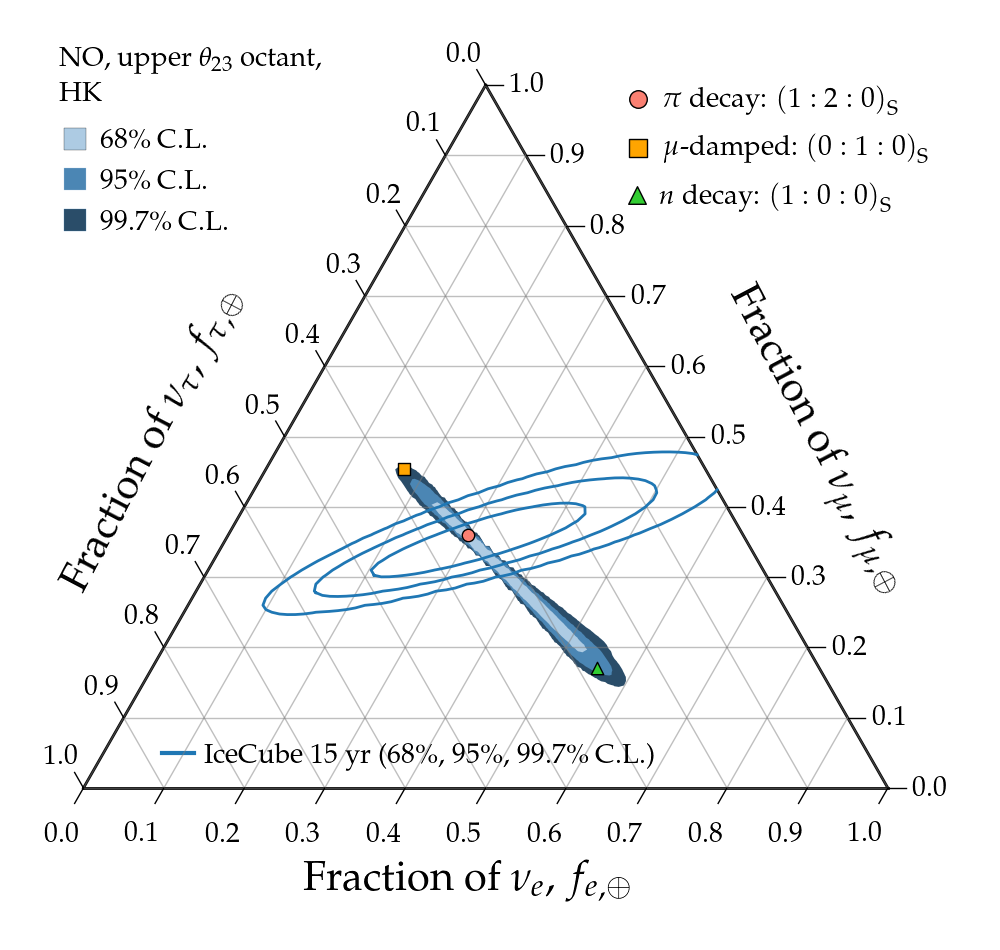}
  \includegraphics[trim=0 0.5cm 0 0, clip, width=0.49\textwidth]{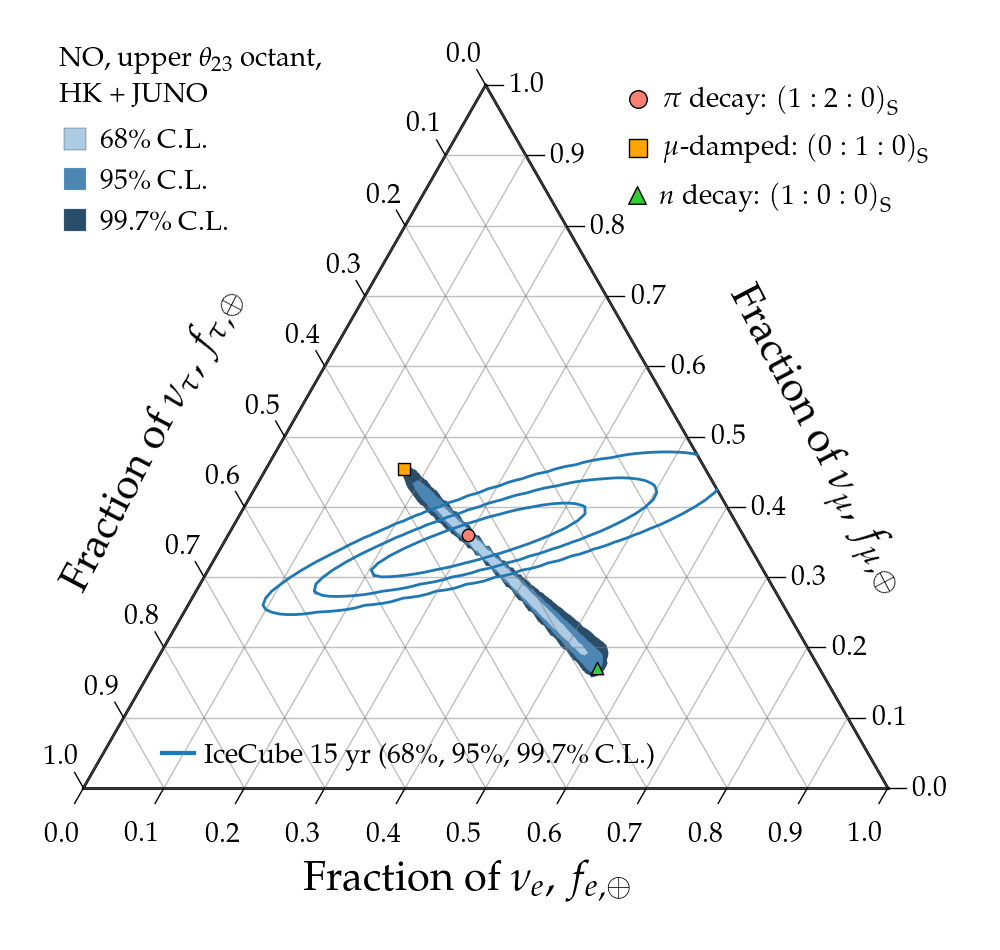}
  \includegraphics[trim=0 0.5cm 0 0, clip, width=0.49\textwidth]{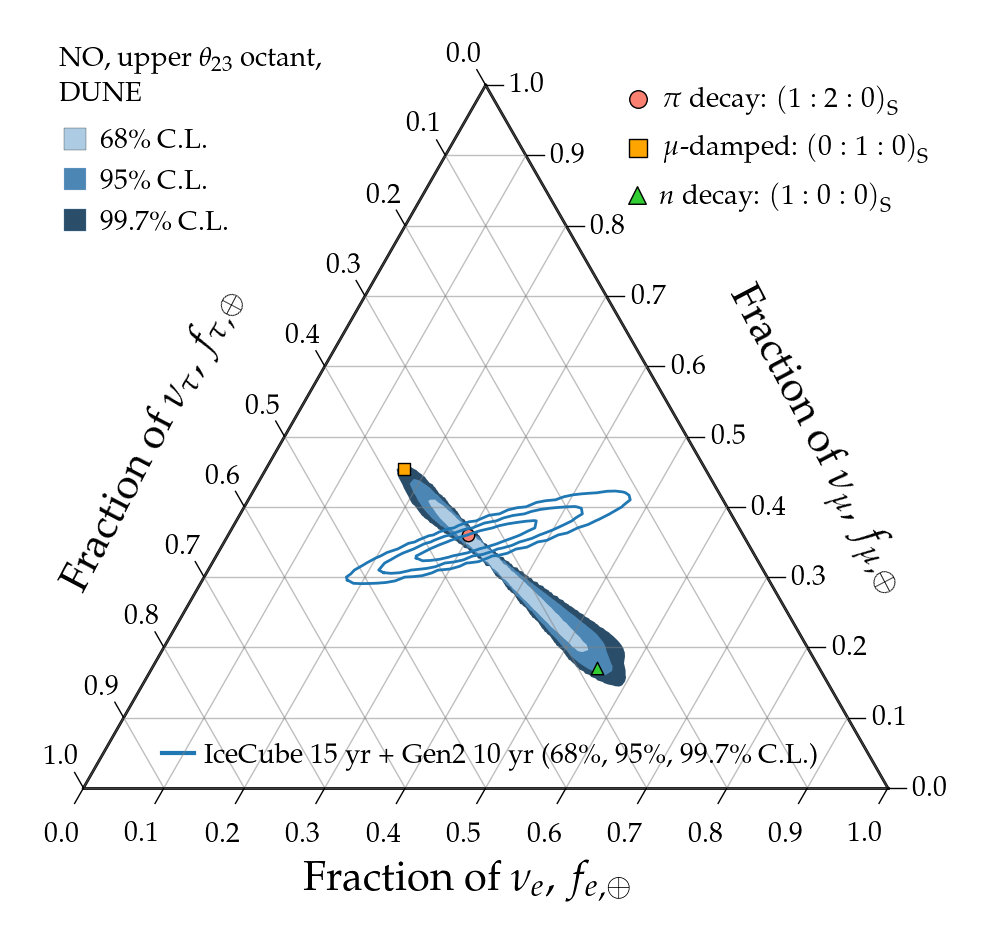}
    \includegraphics[trim=0 0.5cm 0 0, clip, width=0.49\textwidth]{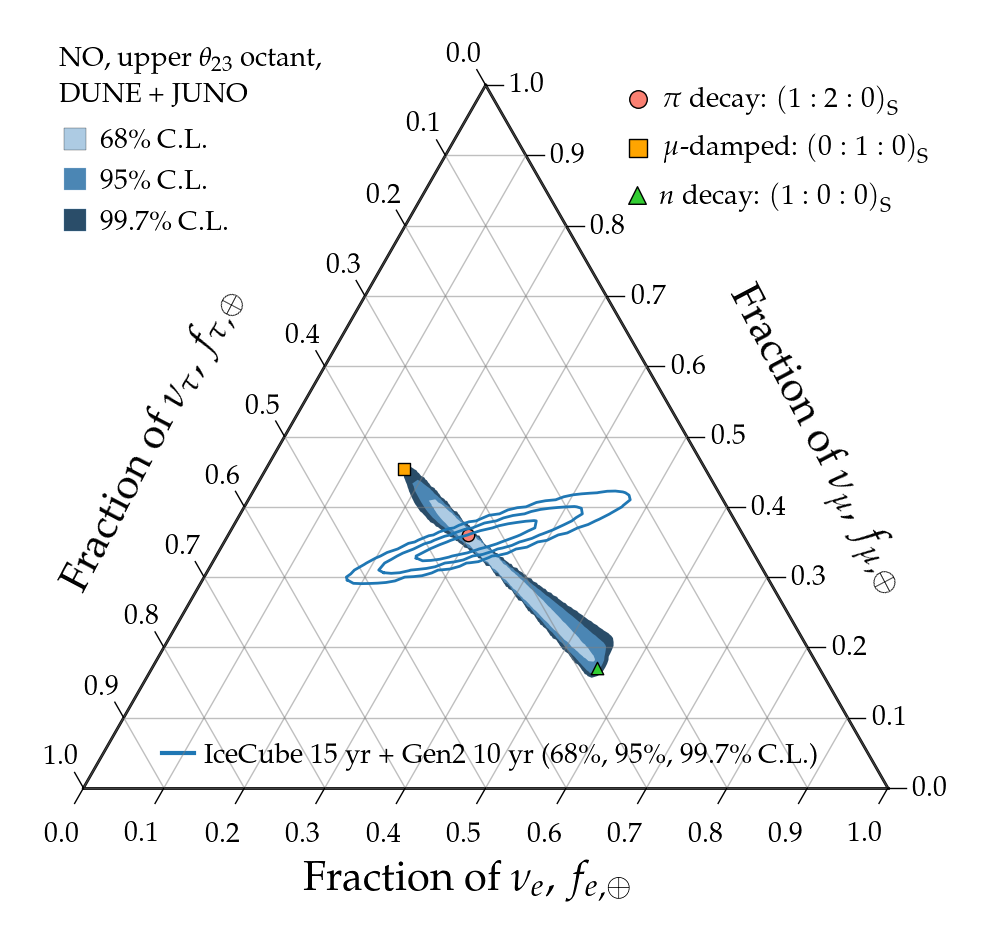}
%  \internallinenumbers
  \caption{Standard oscillation regions, varying over all flavor compositions $f_{\alpha, {\rm S}}$ at the source: normal ordering (NO), upper $\theta_{23}$ octant.}
  \label{fig:triangle_sm_upper_no}
\end{figure*}

% SM, NO, upper th23 octant, fixed fS
\begin{figure*}
  \centering
  \includegraphics[trim=0 0.5cm 0 0, clip, width=0.49\textwidth]{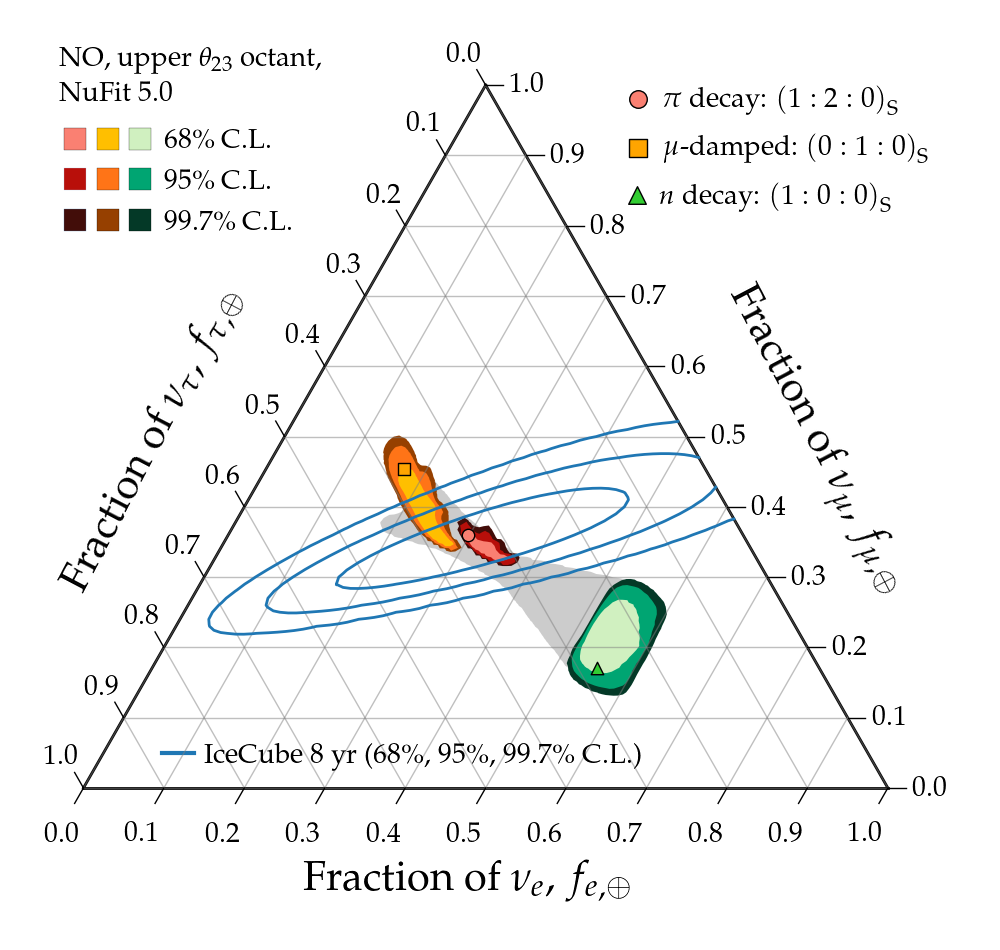}
  \includegraphics[trim=0 0.5cm 0 0, clip, width=0.49\textwidth]{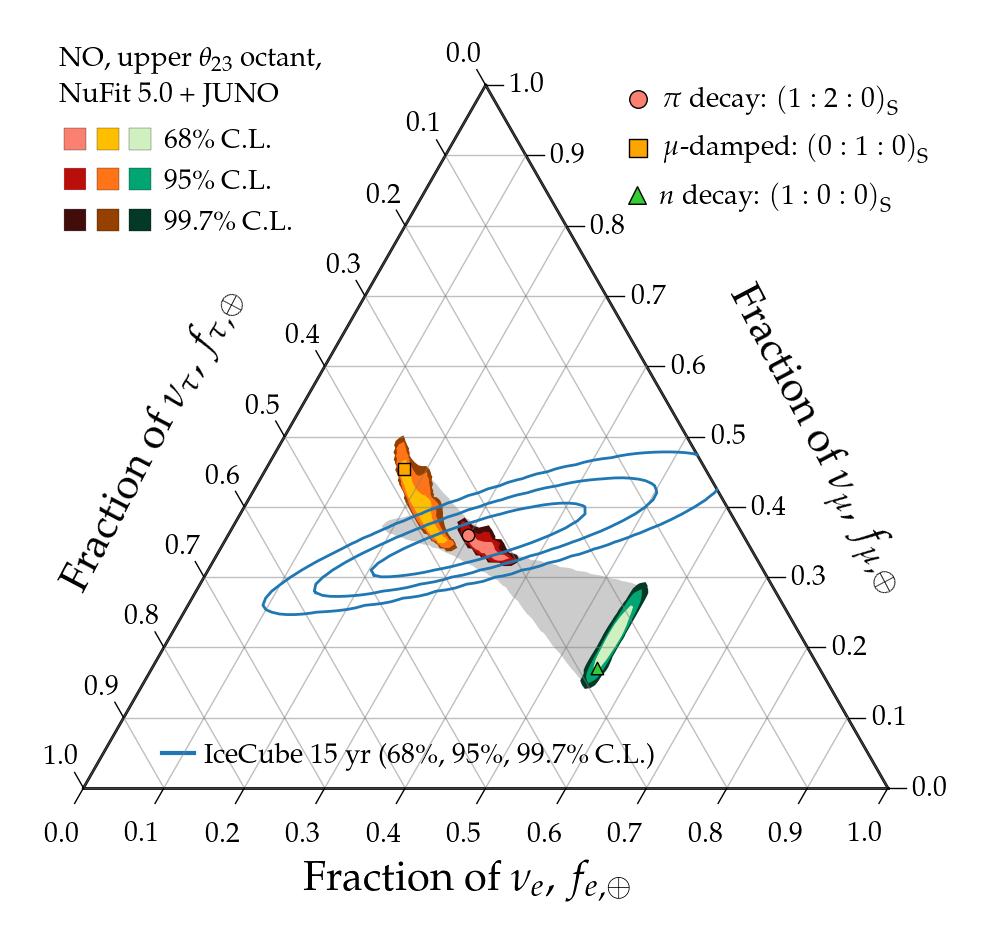}
  \includegraphics[trim=0 0.5cm 0 0, clip, width=0.49\textwidth]{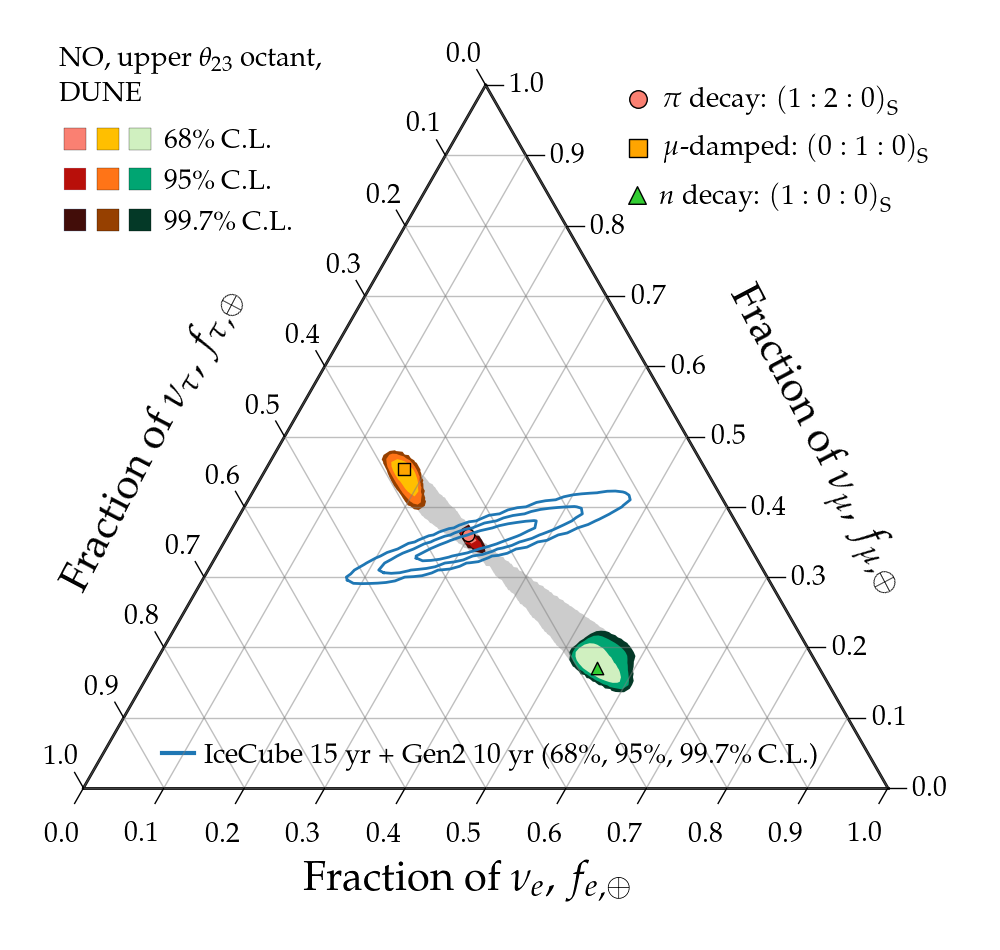}
  \includegraphics[trim=0 0.5cm 0 0, clip, width=0.49\textwidth]{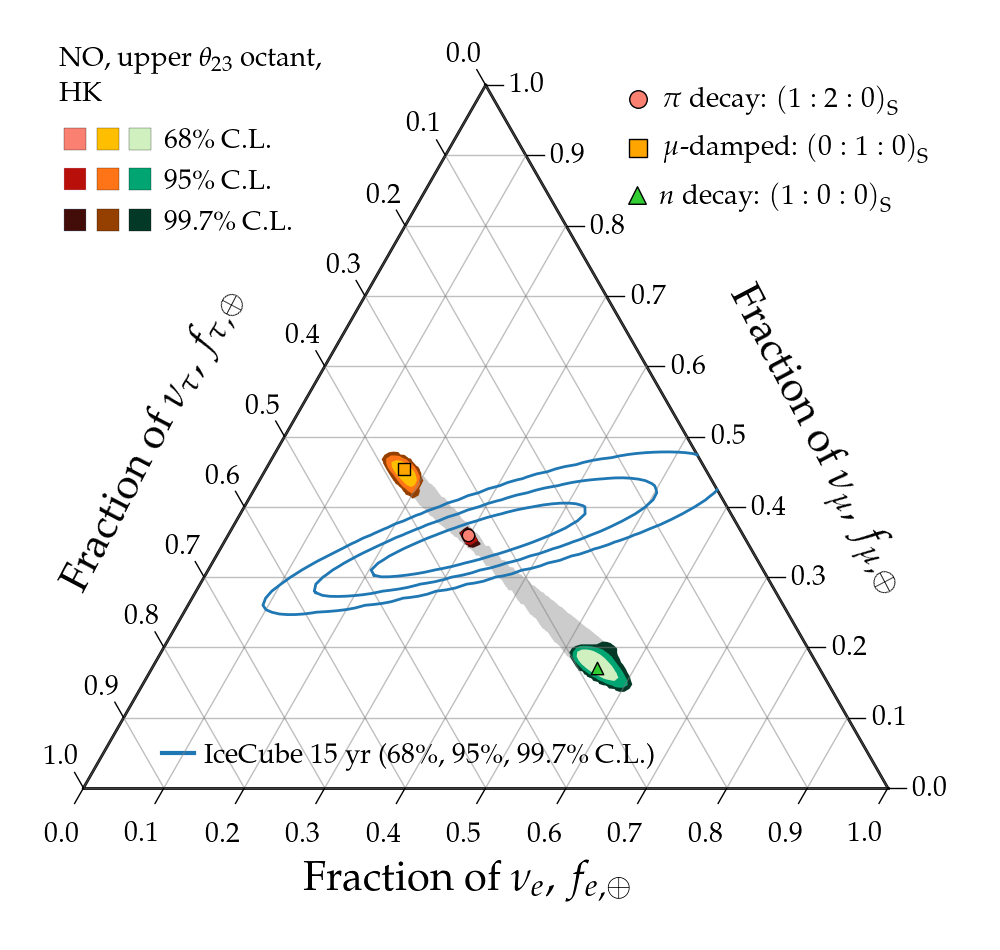}
  \includegraphics[trim=0 0.5cm 0 0, clip, width=0.49\textwidth]{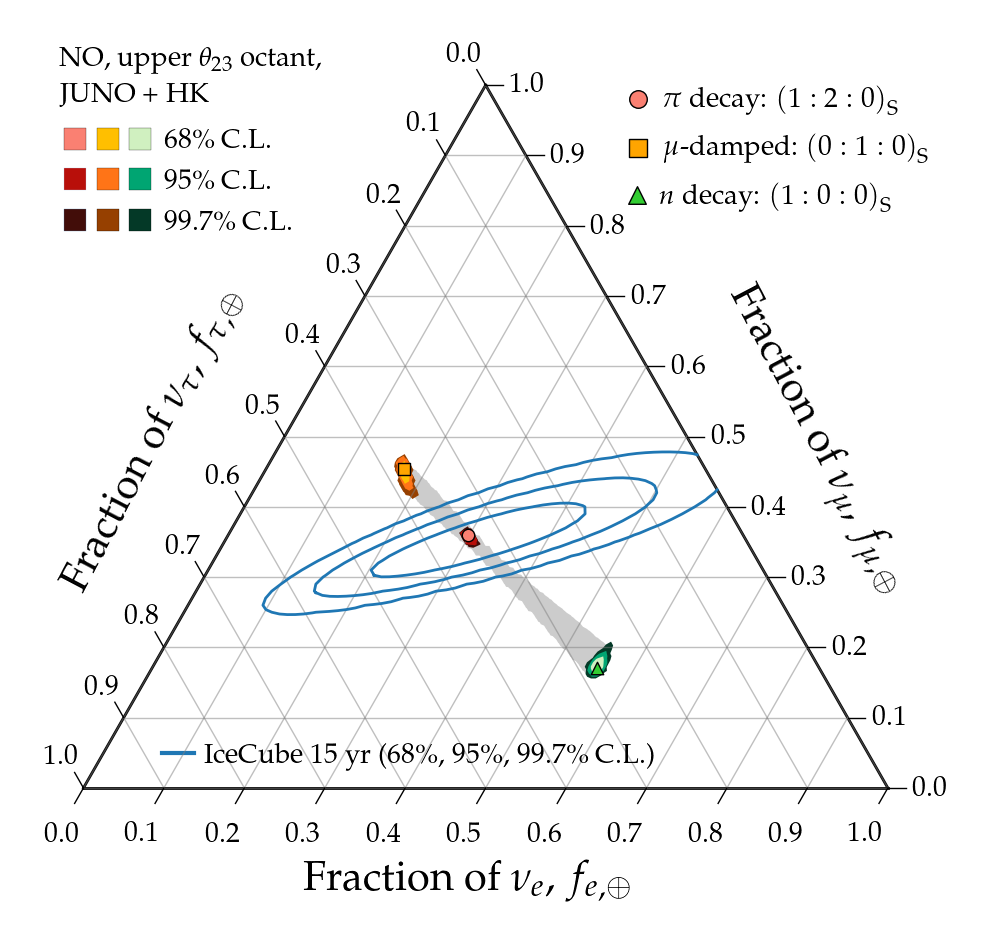}
  \includegraphics[trim=0 0.5cm 0 0, clip, width=0.49\textwidth]{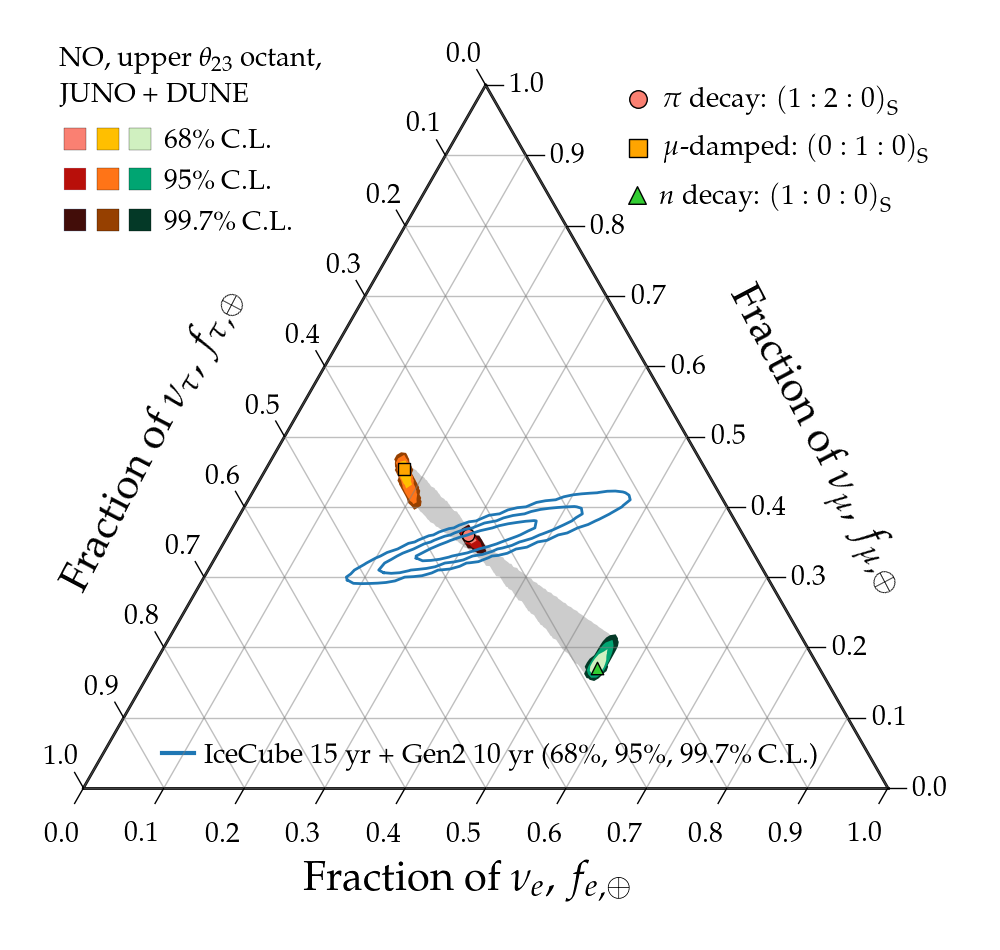}
%  \internallinenumbers
  \caption{Standard oscillation regions, for benchmark flavor compositions $f_{\alpha, {\rm S}}$ at the source: normal ordering (NO), upper $\theta_{23}$ octant.}
  \label{fig:triangle_sm_upper_no_fixed}
\end{figure*}

%%%%%%%%%%%%%%%%%%%%%%%%%%%%%%%%%%%%%%%%%%%%%%%%%%%%%%%%
%%%%%%%%%%%%%%%%%%%%%%%%%%%%%%%%%%%%%%%%%%%%%%%%%%%%%%%%

% SM, NO, lower th23 octant, all fS
\begin{figure*}
  \centering
  \includegraphics[trim=0 0.5cm 0 0, clip, width=0.49\textwidth]{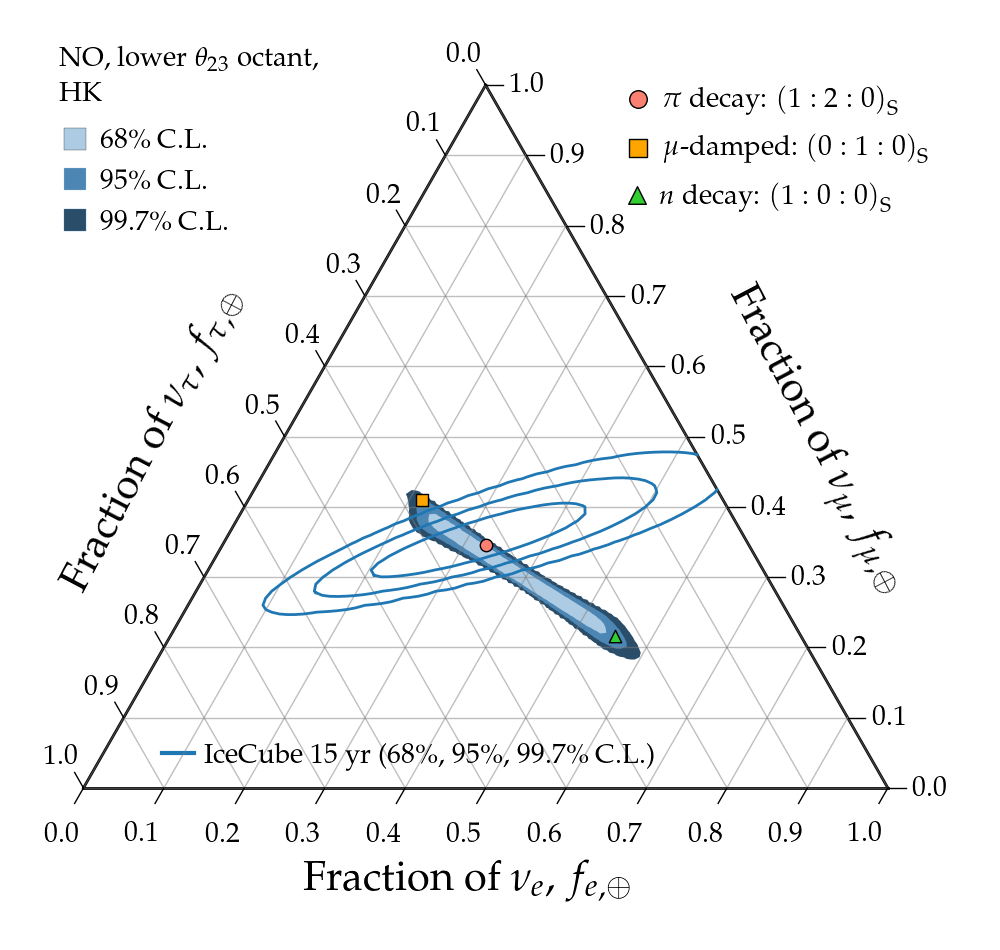}
  \includegraphics[trim=0 0.5cm 0 0, clip, width=0.49\textwidth]{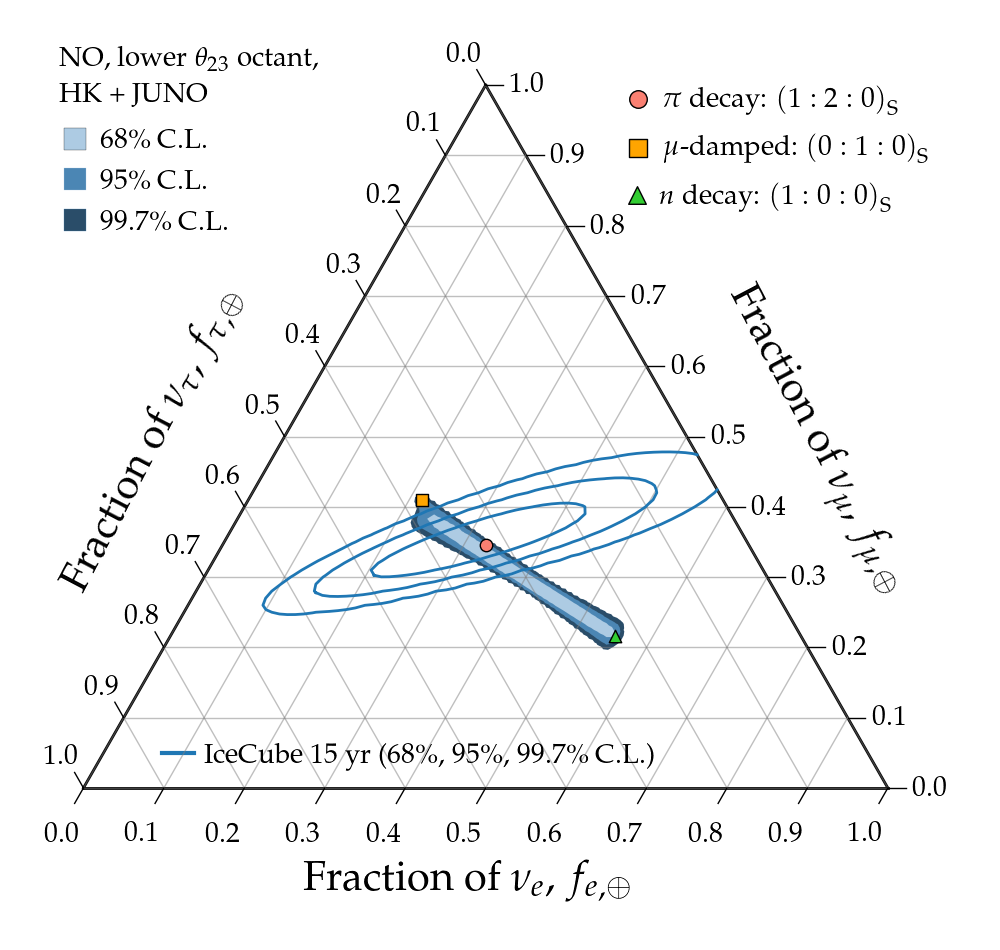}
  \includegraphics[trim=0 0.5cm 0 0, clip, width=0.49\textwidth]{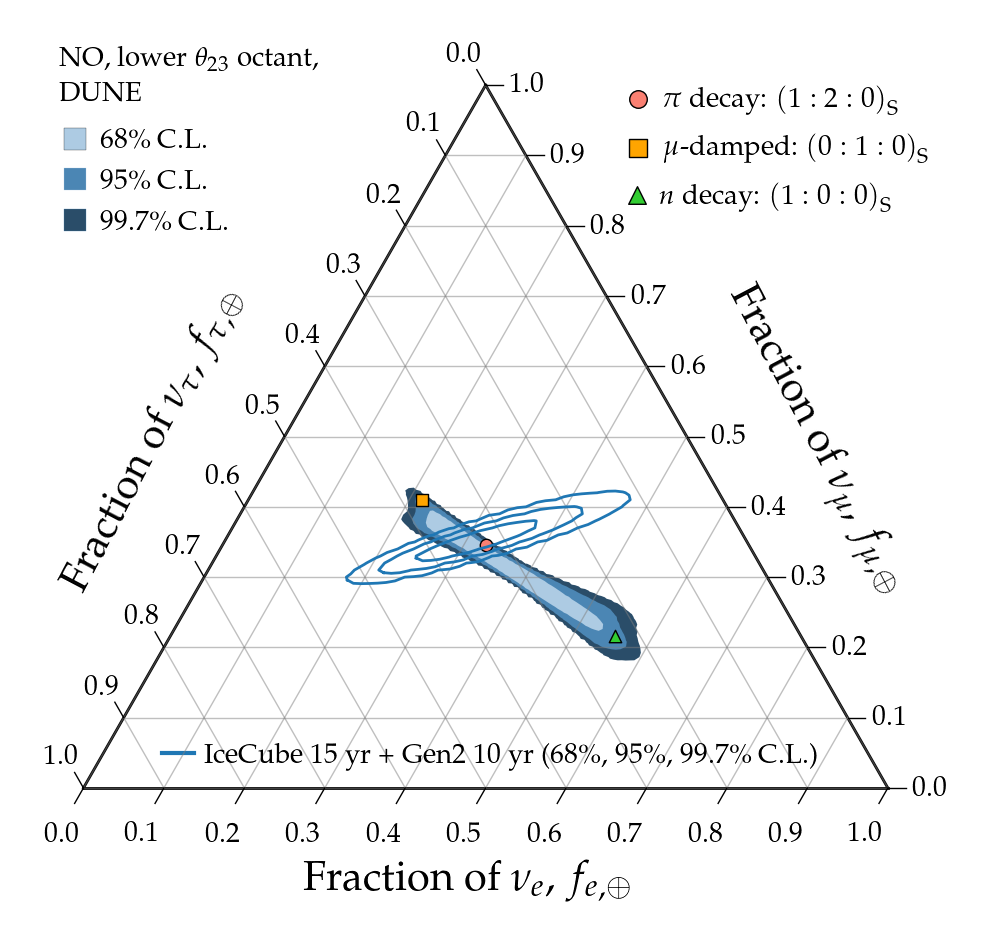}
    \includegraphics[trim=0 0.5cm 0 0, clip, width=0.49\textwidth]{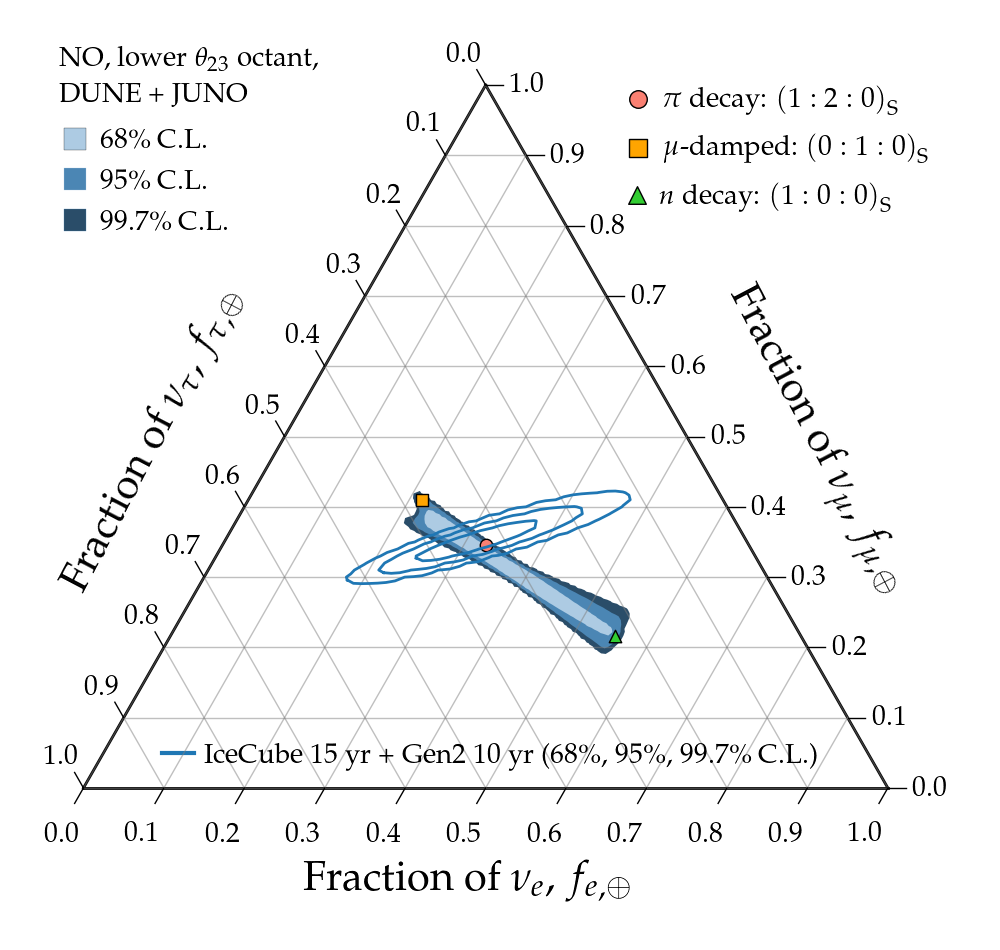}
%  \internallinenumbers
  \caption{Standard oscillation regions, varying over all flavor compositions $f_{\alpha, {\rm S}}$ at the source: normal ordering (NO), lower $\theta_{23}$ octant.}
  \label{fig:triangle_sm_lower_no}
\end{figure*}

% SM, NO, lower th23 octant, fixed fS
\begin{figure*}
  \centering
  \includegraphics[trim=0 0.5cm 0 0, clip, width=0.49\textwidth]{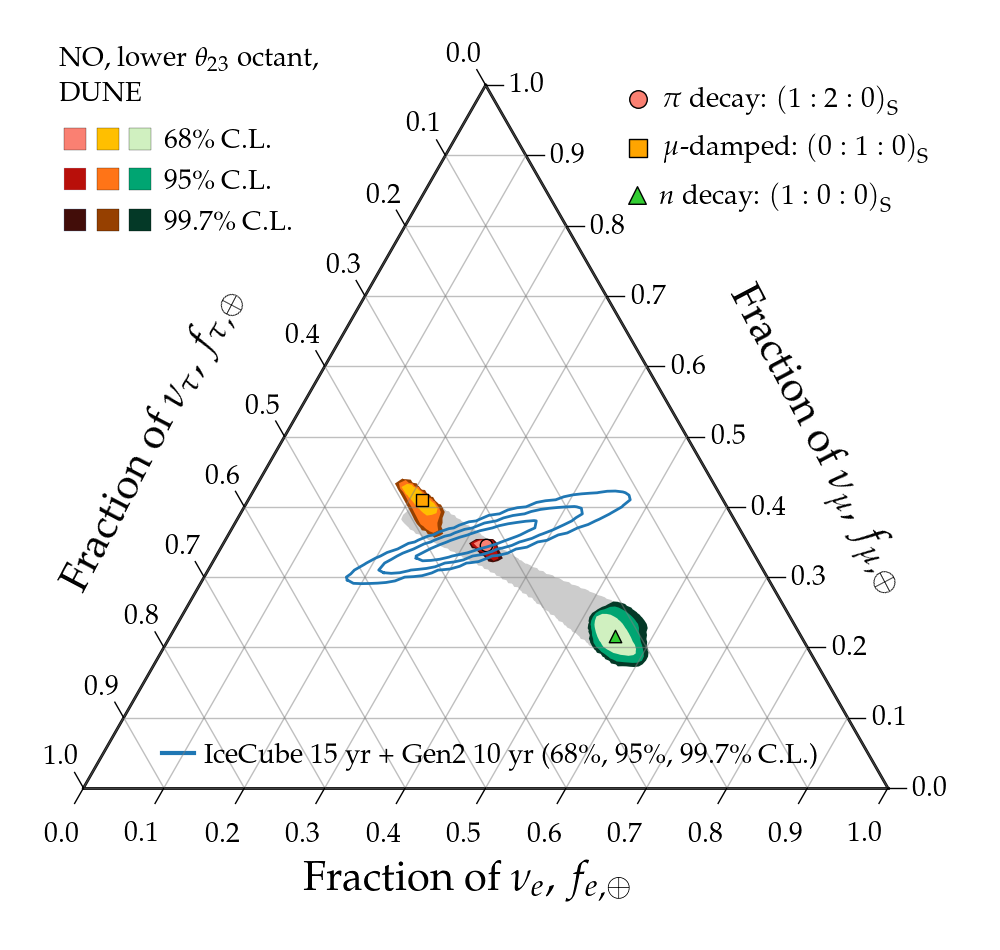}
  \includegraphics[trim=0 0.5cm 0 0, clip, width=0.49\textwidth]{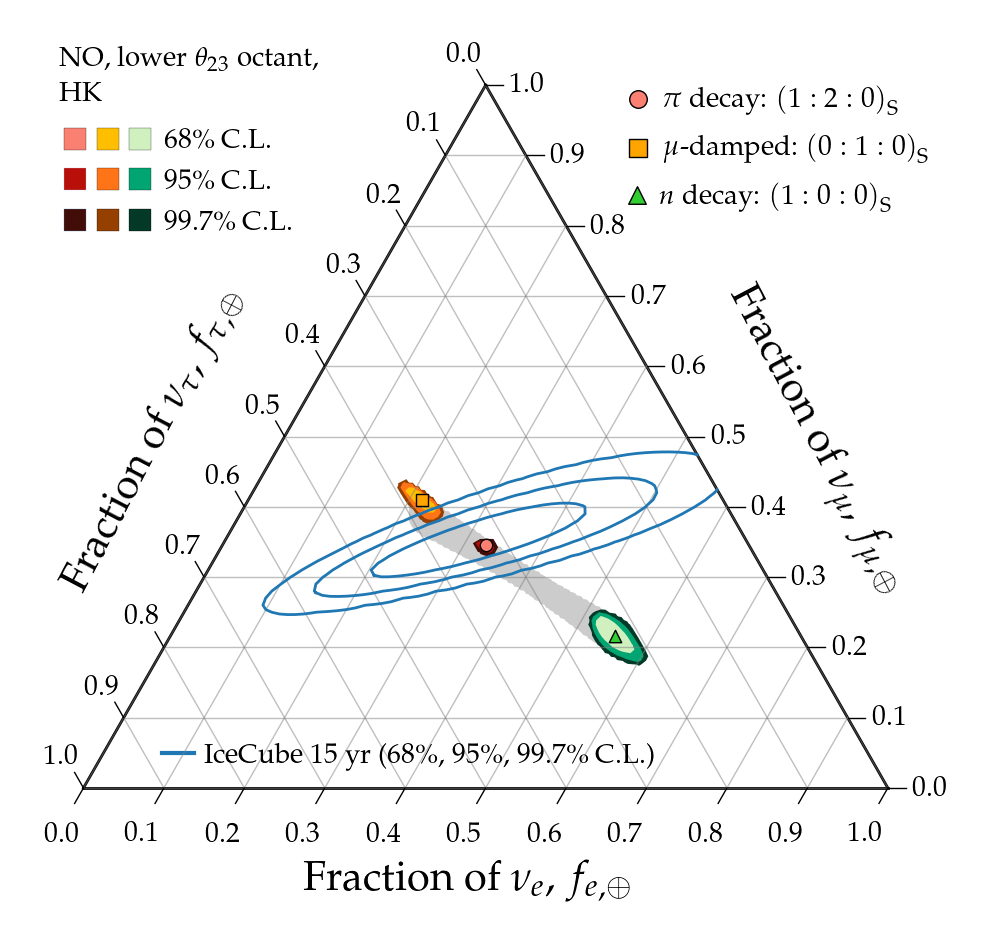}
  \includegraphics[trim=0 0.5cm 0 0, clip, width=0.49\textwidth]{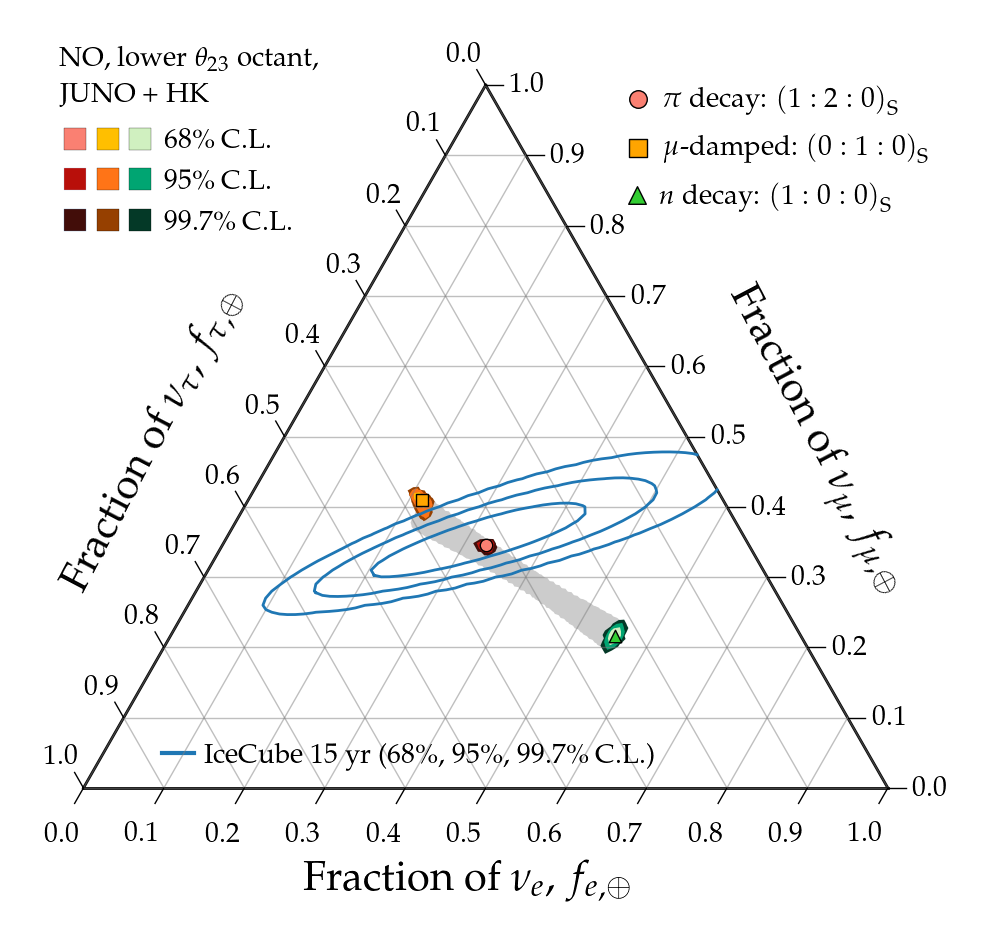}
  \includegraphics[trim=0 0.5cm 0 0, clip, width=0.49\textwidth]{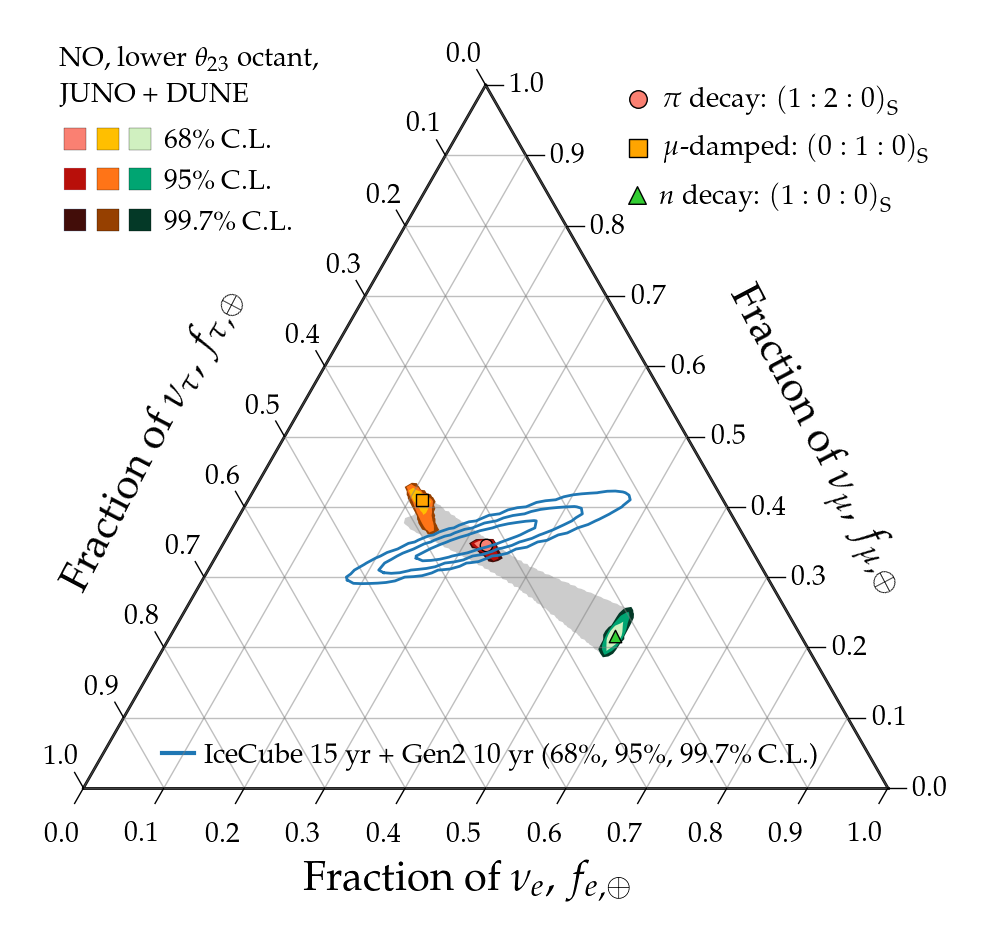}
%  \internallinenumbers
  \caption{Standard oscillation regions, for benchmark flavor compositions $f_{\alpha, {\rm S}}$ at the source: normal ordering (NO), lower $\theta_{23}$ octant.}
  \label{fig:triangle_sm_lower_no_fixed}
\end{figure*}

%%%%%%%%%%%%%%%%%%%%%%%%%%%%%%%%%%%%%%%%%%%%%%%%%%%%%%%%
\end{document}